\documentclass[
reprint,
superscriptaddress,
nofootinbib,
 amsmath,amssymb,
 aps,
showkeys,
a4paper,twocolumn, floatfix
]{revtex4-2}
\usepackage{dsfont}
\usepackage{graphicx}
\usepackage{dcolumn}
\usepackage{bm}
\usepackage[hidelinks]{hyperref}
\usepackage[normalem]{ulem}

\hypersetup{
    colorlinks,
    linkcolor={red!50!black},
    citecolor={blue!50!black},
    urlcolor={blue!80!black}
}
\usepackage{comment}
\usepackage{physics}

\usepackage{tikz}
\usetikzlibrary{matrix,backgrounds,quantikz}
\pgfdeclarelayer{myback}
\pgfsetlayers{myback,background,main}

\tikzset{mycolor/.style = {line width=1bp,color=#1}}%
\tikzset{myfillcolor/.style = {draw,fill=#1}}%

\NewDocumentCommand{\highlight}{O{blue!40} m m}{%
\draw[mycolor=#1] (#2.north west)rectangle (#3.south east);
}
\usepackage{booktabs}
\usepackage{multirow}

\usepackage{mathtools}

\usepackage{amsthm}
\theoremstyle{plain}

\newtheorem*{theorem*}{Theorem}

\newtheorem*{observation*}{Observation}
\newtheorem*{conject*}{Conjecture}
\newtheorem*{definition*}{Definition}

\begin{document}

\preprint{APS/123-QED}

\title{Noise-Induced Equalization in quantum learning models}

\author{Francesco Scala}
 \email{francesco.scala@unibas.ch}
\affiliation{Department of Mathematics and Computer Science,
University of Basel (Switzerland)}

\author{Giacomo Guarnieri}
 \affiliation{Dipartimento di Fisica ``A. Volta,'' Universit\`a di Pavia, via Bassi 6, 27100 Pavia (Italy)
}
\affiliation{INFN Sezione di Pavia, Via Agostino Bassi 6, I-27100, Pavia, Italy}

\author{Aurelien Lucchi}
\affiliation{Department of Mathematics and Computer Science,
University of Basel (Switzerland)}

\date{\today}

\begin{abstract}
Quantum noise is known to strongly affect quantum computation, thus potentially limiting the performance of currently available quantum processing units. 
Even learning models based on variational quantum algorithms, which were designed to cope with the limitations of state-of-the art noisy hardware capabilities, are affected by noise-induced barren plateaus, arising when the noise level becomes too strong. However, the generalization performances of such quantum machine learning algorithms can also be positively influenced by a proper level of noise, despite its generally detrimental effects. 
Here, we propose a pre-training procedure to determine the quantum noise level leading to desirable optimisation landscape properties. 
We show that an optimized level of quantum noise induces an ``equalization'' of the directions in the Riemannian manifold, flattening(/enhancing) the initially steep(/shallow) ones by redistributing sensitivity across its principal eigen-directions.
We analyse this noise-induced equalization through the lens of the Quantum Fisher Information Matrix, thus providing a recipe that allows to estimate the noise level inducing the strongest equalization.
We finally benchmark these conclusions with extensive numerical simulations providing evidence of the beneficial noise effects in the neighborhood of the best equalization, often leading to improved generalization. 
\end{abstract}

\maketitle

\section{Introduction\label{sec:intro}}

\begin{figure*}
    \centering
    \includegraphics[trim={1cm 8.9cm .5cm 8.9cm}, clip,width=.99\linewidth]{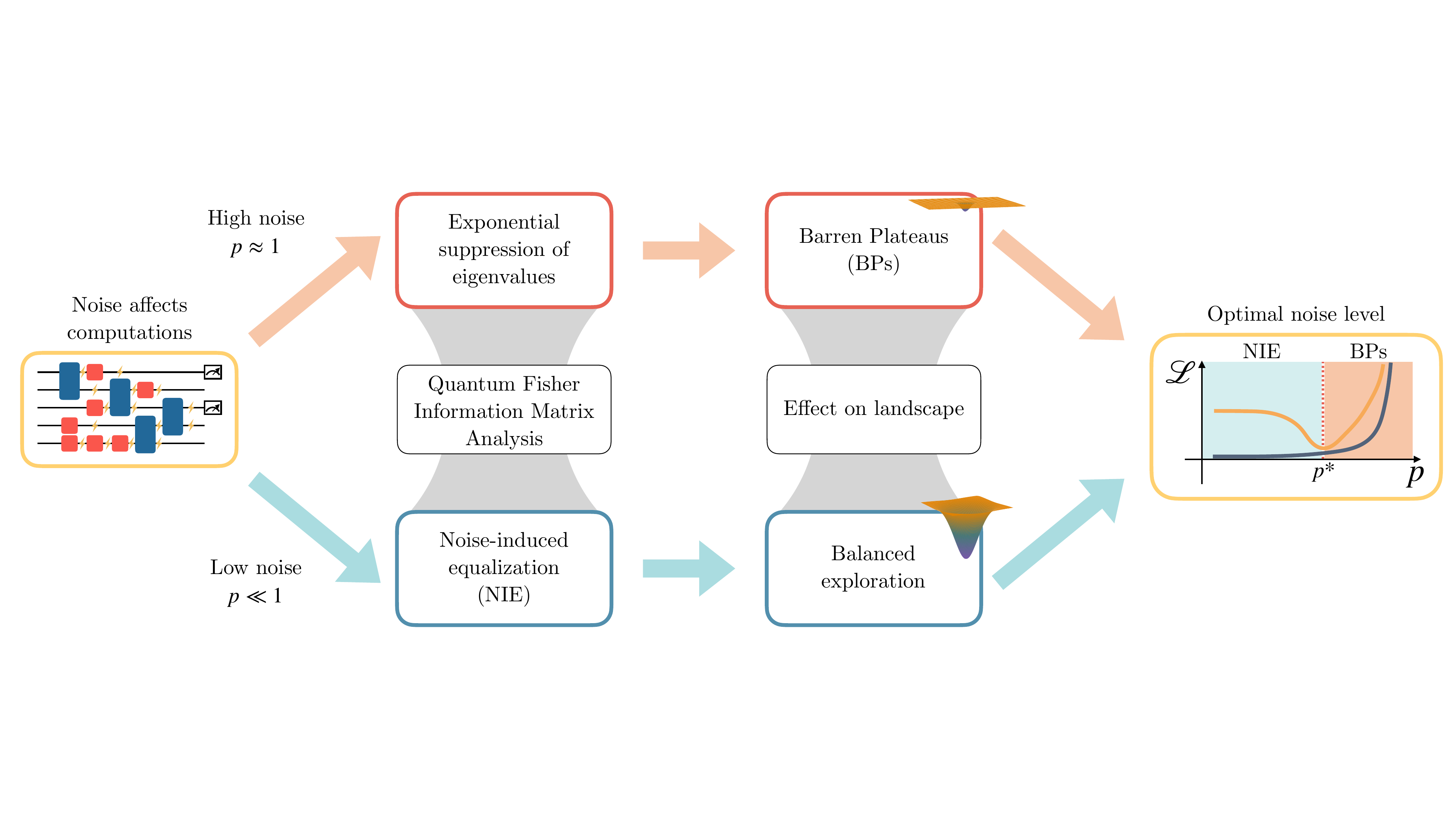}
    \caption{Schematic illustration of the noise analysis proposed in this work. The action of noise on the QNN algorithmic performances can be analyzed via the Quantum Fisher Information Matrix. If noise is too strong, the eigenvalues of the QFIM are exponentially suppressed, which is known to lead to the BPs issue. However, small levels of noise induce an \emph{equalization}
    process, in which smaller eigenvalues gain more and more importance, thus allowing a more balanced exploration of the cost landscape. We argue that an optimal level of noise exists, $p^*$, which gives the best equalization and leads to improved generalization performances of the QNN.}
    \label{fig:enter-label}
\end{figure*}

In the noisy intermediate-scale quantum era~\cite{Preskill_2018}, quantum processing units are inherently affected by various sources of noise~\cite{nielsen00}. 
Accurately modeling and understanding quantum noise is essential for the advancement of quantum computing, quantum information, and quantum machine learning (QML)~\cite{2014QMLWittek,Biamonte2017quantum,cerezo2022challenges, preskill2025nisqmegaquopmachine}. 
A significant research effort is currently focusing on mitigating the detrimental effects of noise to preserve the reliability of quantum algorithms~\cite{Temme2017prl,Kandala2019nature, cai2023error_mitigation}. 
Among the several areas in which Quantum Computing research is currently pursued, QML has attracted considerable interest due to its integration of quantum computation with classical optimization techniques in the so-called variational quantum algorithms~\cite{Cerezo2021VQA}. 
However, noise can practically hinder the potential advantages of QML over classical machine learning models~\cite{stilck2021limitations,depalma2023limitations}, and severely impact the trainability~\cite{McClean2018barren, Arrasmith2022equivalence} of parameterized quantum circuits~\cite{Peruzzo2014VQE,Benedetti2019pqc,Cerezo2021VQA}, including Quantum Neural Networks (QNNs)~\cite{Mangini2021qnns,abbas2021power}. In particular, noise-induced barren plateaus (BPs)~\cite{wang2021noise, schumann2023emergence,garciamartin2023effects} can cause the gradient of the loss function to vanish, thus making classical optimization ineffective. As a consequence, researchers have been investigating the trainability conditions of QNNs in various scenarios~\cite{oliv2022evaluating,borras2023impact}, and novel techniques have been developed to cope with these challenges~\cite{gu2021adaptive, ito2023latencyaware}.  

Among the different origins of BPs, noise stands out due to its distinctive nature. A recent study~\cite{garciamartin2023effects} explores the impact of noise on overparametrization~\cite{kiani2020learning,haug2021capacity,Larocca2023theory}, providing valuable insights into how a considerable amount of realistic noise can induce BPs. 
In essence, noise induces an exponential suppression of the model's ability to explore the Hilbert space, thus severely limiting its expressivity. 
This phenomenon can be analyzed through the Quantum Fisher Information Matrix (QFIM)~\cite{liu2014fidelity, liu2020fisher, meyer2021fisher, pezze2025advancesmultiparameterquantumsensing, scandi2023quantum}, a central quantity in quantum parameter estimation theory that quantifies the sensitivity of a quantum state to multi-parameter variations. The rank of the QFIM determines the number of informative directions in parameter space that are accessible for optimization~\cite{haug2021capacity, Larocca2023theory}. Recent work has also shown that the presence of noise can, in specific regimes, enhance parameter estimation rather than degrade it~\cite{peng2024enhanced, chen2024quantum, garcia2025noise}.

Classical machine learning models~\cite{mohri2018foundations} are known to exploit noise to induce good generalization properties with different techniques, such as noise injection~\cite{camuto2020explicit,li2020adaptive, levi2022noise,orvieto2022anticorrelated,orvieto2023explicit}, stochastic gradient descent~\cite{smith2020generalization,sclocchi2023dissecting}, data augmentation~\cite{bishop1995training} and dropout~\cite{wan2013dropconnect, srivastava2014dropout}.
In fact, even if a given model is very well optimized on training data, there is no guarantee that it will perform just as well on unseen data. 
Overfitting occurs when a model memorizes random noise in the training data, rather than learning the underlying patterns that allow it to make accurate predictions on new, unseen examples.~\cite{hawkins2004problem, ying2019overview, banchi2021generalization}. 
Moreover, a careful analysis of the Fisher Information Matrix (FIM) of learning models leads to the conclusion that well-conditioned FIMs at initialization are often associated with a more favourable optimization landscape (more uniform curvature)~\cite{Pennington2018Fisher}, thus setting the stage for better generalization~\cite{Baldassi2021Unveiling}. Recently in QML, there has also been an increasing interest in understanding how incoherent processes may help, for example, to avoid BPs~\cite{sannia2024engineered}, or to escape saddle points~\cite{liu2023stochastic}, to provide generative modelling~\cite{parigi2024quantum} or reinforcement learning~\cite{oliveraatencio2025impactamplitudephasedamping}, or having better generalization capabilities~\cite{scala2023dropout,Heyraud2022noisy,hu2023tackling, Khanal2025_data_JSuperC, zhu2025optimizer, yang2025stability}. 

In this work, we provide a novel procedure to identify a beneficial quantum noise level, $p^*$, before the onset of noise-induced BPs, positively reshaping the optimization landscape and in which vicinity generalization capabilities may be enhanced. 
While quantum noise on average dampens the sensitivity to parameters variations leading to BPs, our results show that modest, optimized noise levels reshape the Riemannian manifold associated to the quantum model of interest, making its curvature the curvature of the landscape more uniform across different directions. 
We refer to this phenomenon as noise-induced “equalization” (NIE), which we investigate by analysing the eigenspectrum of the QFIM by means of a newly introduced spectrally-resolved measure. 
Furthermore, we conjecture and numerically verify that in the neighborhood of the noise level yielding the best equalization, superior generalization is promoted, as the reshaping of the optimization landscape favors the parameter space exploration over its exploitation. 
We notice that the improvement in generalization performances is not a direct property of the QFIM spectrum at initialization; rather, it is a knock-on effect induced by enabling a smoother training dynamics, which in turn tends to land the optimization in flatter — and thus often more generalizable — parameter space regions.
The protocol presented in this work is applied before training, and it only depends on the model design, which makes it applicable to various settings and datasets.
Finally, we corroborate the quality of the proposed procedure by comparing its estimated optimal noise level to the one obtained by means of a recently proposed generalization bound that also depends on the QFIM spectrum~\cite{Khanal2025_data_JSuperC}. 
We show that our procedure allows for a better estimate of useful noisy regimes as compared to the mentioned generalization bound. 

The manuscript is organized as follows: in Sec.~\ref{sec:background} we introduce the Quantum Fisher Information Matrix, QNNs as QML models and how to describe quantum noise; in Sec.~\ref{sec:noise impact} we discuss the impact of noise on QNNs, and we define the concept of noise-induced equalization; in Sec.~\ref{sec:results} we validate our theories by numerical simulations and in Sec.~\ref{sec:discussion} we discuss our findings and their potential impact; finally, in Sec.~\ref{sec:methods} we describe the details of our implementations.

\section{Background}
\label{sec:background}

In this section, we provide an introduction to the general concept of ``information matrix'', and in particular to the QFIM, as well as to the quantum algorithms known as QNNs. For completeness, we also provide an outline of the theoretical description of QNN architectures and their overparametrization. Finally, we summarize the noise channel formalism as well as the different kinds of prototypical quantum noise models.

\subsection{Quantum Fisher Information}
\label{subsec:qfim}
The optimisation of a parameterized quantum circuits corresponds to adjusting a set of circuit parameters so as to prepare a desired target quantum state. 

The natural framework of this problem is the so-called quantum parameter‑estimation theory~\cite{pezze2025advancesmultiparameterquantumsensing}, in which the Quantum Fisher Information Matrix (QFIM) quantifies how sensitive the quantum state is upon changes of the parameters, in analogy with the classical Fisher Information Matrix (FIM)~\cite{liu2014fidelity,liu2020fisher,meyer2021fisher,pezze2025advancesmultiparameterquantumsensing,scandi2023quantum}. 
This quantity plays a foundational role in quantum metrology (via the Quantum Cramér–Rao bound) and has recently also been applied to analyse overparameterization of parameterized quantum circuits~\cite{haug2021capacity,Larocca2023theory}. Thus the QFIM provides a powerful and principled entry point for studying how circuit parameterisation maps to state space and ultimately to algorithmic performance.
However, it is important to recognise that, in the multi‐parameter quantum regime, the QFIM is not the only tool and additional bounds become relevant. For instance, the Holevo Cramér–Rao bound gives the most general lower bound on the covariance of unbiased estimators when measurement incompatibilities or collective strategies matter~\cite{albarelli2019evaluating}.

For pure states, the QFIM can be derived from the quantum fidelity, a contrast function quantifying the overlap between two quantum states. For a parameterized family of pure states, $\ket{\psi(\boldsymbol{\theta})}$, where $\boldsymbol{\theta}$ represents the parameters array, one of the possible measures of quantum fidelity  between two states is defined as the squared overlap:  
\begin{equation}
f(\ket{\psi(\boldsymbol{\theta})}, \ket{\psi(\boldsymbol{\theta'})}) = |\langle \psi(\boldsymbol{\theta'}) \ket{\psi(\boldsymbol{\theta})}|^2 \,.
\end{equation}  
One can then define a fidelity-based distance, $d_f = 2(1 - f) $, from which the QFIM is derived as the Hessian of this distance~\cite{meyer2021fisher}:  
\begin{equation}
\mathcal{F}_{ij} = \frac{\partial^2}{\partial \delta\theta_i \partial \delta\theta_j} d_f(\ket{\psi(\boldsymbol{\theta})}, \ket{\psi(\boldsymbol{\theta} + \delta \boldsymbol{\theta})})\, ,
\end{equation}  
where $\delta \boldsymbol{\theta}$ represents a small shift of parameters. Hence, the QFIM elements are explicitly given by:  
\begin{align}
[\mathcal{F}(\boldsymbol{\theta})]_{ij} = 4 \mathrm{Re}
\large\{&\langle \partial_i \psi(\boldsymbol{\theta}) \ket{\partial_j \psi(\boldsymbol{\theta})} + \nonumber\\
&-\langle \partial_i \psi(\boldsymbol{\theta}) \ket{\psi(\boldsymbol{\theta})} \langle \psi(\boldsymbol{\theta}) \ket{\partial_j \psi(\boldsymbol{\theta})} \large\} \, ,
\end{align}  
in which $\ket{\partial_i \psi(\boldsymbol{\theta})} = \partial \ket{\psi(\boldsymbol{\theta})} / \partial \theta_i $. 

For mixed states, the QFIM generalizes using the Bures distance and Uhlmann fidelity. The Bures distance provides a measure of the dissimilarity between two density matrices $\rho$ and $\sigma$ as
\begin{equation}
d_B(\rho, \sigma) = \sqrt{2 \big( 1 - \sqrt{F(\rho, \sigma)} \big)} \, ,
\end{equation}  
in which $F(\rho, \sigma) $ is the so called Uhlmann fidelity 
\begin{equation}
    F(\rho, \sigma) = \bigl(\Tr\sqrt{\sqrt{\rho}\sigma\sqrt{\rho}}\bigr)^2\, ,
\end{equation}
which quantifies the maximal overlap between their purifications.  
Hence, the QFIM for mixed states can be derived by considering the spectral decomposition of the density matrix $\rho = \sum_k \lambda_k \ket{\psi_k}\bra{\psi_k}$, where $\lambda_k$ are the eigenvalues and $\ket{\psi_k}$ the corresponding eigenstates. Then, the QFIM incorporates contributions from diagonal and off-diagonal terms, respectively~\cite{liu2014fidelity, meyer2021fisher}:  
\begin{align}
[\mathcal{F}(\boldsymbol{\theta})]_{ij} &= \sum_{\substack{k\\\lambda_k\neq0}} \left[\frac{(\partial_i \lambda_k )(\partial_j \lambda_k)}{\lambda_k} +4\lambda_k\mathrm{Re}\left\{\langle  \partial_i\psi_k| \partial_j \psi_k  \rangle\right\}\right]+\nonumber\\
&- \sum_{\substack{k,l\\\lambda_k,\lambda_l\neq0}} \frac{8 \lambda_k \lambda_l}{\lambda_k + \lambda_l} \mathrm{Re}\left\{\langle \partial_i \psi_l | \psi_k \rangle \langle  \psi_k| \partial_j \psi_l  \rangle \right\} \, .
\end{align} 
We notice that the latter matrix is positive semidefinite, real, and symmetric thus inducing a proper metric onto the parameterized manifold.

Extending the concept of the QFIM to non-pure quantum states allows capturing the sensitivity to parameter variations in real-world scenarios where quantum noise is also present. 
A crucial property for what follows is the general contractivity of the QFIM under the action of a quantum channel $\Lambda_t[\cdot]$(i.e. a completely positive and trace preserving dynamical map acting on the state space) 
\begin{equation}\label{eq:DPI}
    \mathcal{F}(\Lambda_t[\rho(\boldsymbol\theta)])\preccurlyeq  \mathcal{F}(\rho(\boldsymbol\theta)),
\end{equation} 
where $\preccurlyeq$ represents the L\"{o}wner ordering of matrices.
Eq.\eqref{eq:DPI}, known as Data-Processing Inequality, physically implies that the ability to discriminate two quantum states can only be degraded under the action of noise. 
A special case of the Data-Processing Inequality implies also that the classical Fisher information $\mathcal{I}$ extracted after a quantum measurement process is always upper bounded by the quantum Fisher information~\cite{meyer2021fisher}:
\begin{equation}
    \mathcal{I}(\Tr[\rho(\boldsymbol\theta) O])\preccurlyeq \mathcal{F}(\rho(\boldsymbol\theta)) \,  ,
\end{equation}
implying that the information that can be extracted from a state is always smaller than the information contained in the state itself. 

Since the QFIM provides at least the same amount of information as the classical FIM, in the following, we will analyze QNNs, defined as parameterized quantum circuits, by using the QFIM instead of the classical FIM.

The analysis we propose is based on the QFIM eigenspectrum. As just introduced, the QFIM describes the sensitivity to parameter variations of a certain quantum model. This sensitivity can also be interpreted as the importance of different parameters in the computation. Hence, analyzing the eigenspectrum of the QFIM corresponds to conducting the study in a rotated framework, in which the QFIM is diagonal. 
Notice that this approach considers only linearly independent directions in the parameters space, which may be obtained by linear combinations of the directions defined by the variational parameters.
As a last remark, since the QFIM is computed as the Hessian of the fidelity between quantum states, it can be considered as an indicator of the flatness/steepness of the Riemannian manifold of quantum states.
Consequently, its eigenvalues can also be related to the steepness in the ``state landscape'' with respect to the eigen-directions.

\subsection{Quantum Neural Networks}
\label{subsec:qnns}

Given a dataset $\left\{\mathbf{x}_i, y_i\right\}_{i=1}^{M}$, 
where data points $\left\{\mathbf{x}_i\right\}_{i=1}^{M}$ are sampled from a distribution $\mathcal{D}$ with corresponding labels $\left\{y_i\right\}_{i=1}^{M}$, 
we define a QNN model by selecting an observable (i.e., a Hermitian operator) $O$ and computing its expectation value with respect to a $L$-layer parameterized quantum state, represented as a density matrix $\rho_L(\mathbf{x}_i, \boldsymbol{\theta})$:
\begin{equation}
    f(\mathbf{x}_i, \boldsymbol{\theta}) = \mathrm{Tr}\left[ O \rho(\mathbf{x}_i, \boldsymbol{\theta}) \right] \, . 
\end{equation}
The density matrix can be represented as   
\begin{equation}
    \rho_L(\mathbf{x}_i, \boldsymbol{\theta}) = U_L(\mathbf{x}_i, \boldsymbol{\theta}) \rho_0 U_L^\dagger(\mathbf{x}_i, \boldsymbol{\theta}) \, ,
\end{equation}
in which $\rho_0 = (\ket{0}\bra{0})^{\otimes n}$ is the initial quantum register state. The evolution operator $U_L(\mathbf{x}_i, \boldsymbol{\theta})$ is then explicitly given by  
\begin{equation}
    U_L(\mathbf{x}_i, \boldsymbol{\theta}) = \prod_{l=0}^{L} U_l(\boldsymbol{\theta}) S_l(\mathbf{x}_i) \, ,
\end{equation}
where $U_l(\boldsymbol{\theta})$ represent the trainable parameterized unitaries, and $S_l(\mathbf{x}_i)$ are encoding operations that embed the classical data points, $\mathbf{x}_i$. Here, $L$ denotes the number of layers (i.e., the depth) of the QNN. While we focus on scalar output functions for simplicity, this model can be readily extended to produce a vector-valued output, $\boldsymbol{f}$, by assigning different observables $O_j$ to each component $f_j$.

In practical QNN implementations, quantum computations are inevitably subject to noise. This can be modeled by the action of a quantum channel $\mathcal{N}_l$~\cite{nielsen00}, which may affect the system after each encoding operation $ S_l $ and after each trainable unitary $ U_l $. 
Analitically, any quantum channel $\mathcal{N}$ acting on a density matrix $\rho$ can be described using the Kraus decomposition as
\begin{equation}
    \mathcal{N}(\rho)=\sum_k K_k\rho K_k^\dagger \, ,
\end{equation}
where ${K_k}$ are the Kraus operators associated with the channel and satisfy the completeness relation $\sum_k K_k^\dagger K_k = I$ to ensure trace preservation.
This representation provides a convenient and general framework to model various noise processes, such as depolarization, dephasing, or amplitude damping, by specifying the appropriate set of $K_k$ (see Appendix~\ref{appendix:noise chan}). Using this framework, noisy models can then be trained according to the known variational quantum algorithm scheme~\cite{Cerezo2021VQA}.
In this work, we investigate the impact of these different noise channels on QNN training and generalization. 

Noisy models can then be trained according to the known variational quantum algorithm scheme~\cite{Cerezo2021VQA}.

A QNN is said to be overparameterized when the rank of the QFIM reaches its maximal value for all the elements in the training set~\cite{Larocca2023theory}. 
This corresponds to the situation in which adding more parameters to the model does not increase the rank of its associated QFIM. 
Notice that this definition of overparametrization only depends on features of the QNN model, while it is independent from the particular loss function employed, or the given variational problem to be solved. 
In particular, the rank of the QFIM can be considered as one of the possible definitions of the effective dimension of the QNN~\cite{haug2021capacity, abbas2021effective,Khanal2025_data_JSuperC}:
\begin{equation}
    \label{eq:d_eff}
    d_{eff} = \mathrm{rank}\left[\mathcal{F}(\boldsymbol\theta)\right] \, .
\end{equation}
The maximal achievable dimension 
for pure states is $d_{eff}^{max}=2^{n+1}-2$, which corresponds to the number of independent real parameters in the state vector describing the quantum state. For mixed states this value is enhanced to $d_{eff}^{max}=2^{2n+1}$, since their full description requires defining the corresponding density matrices.
Alternatively, the effective dimension can also be defined as the number of QFIM eigenvalues that are above a given threshold.

\section{The impact of noise on QNNs}
\label{sec:noise impact}

After having introduced the QFIM and the main models employed to approximate the description of the different quantum noise channels, here we start diving into the effects of such sources of noise on the QNN performances. First, we briefly report some preliminary results, mostly known from the literature~\cite{garciamartin2023effects}, highlighting the effects of noise on overparameterized QNNs and motivating the analysis conducted in this work. Then, the main contribution of the present work will be introduced, i.e., the NIE procedure and its relationship with the performance of the QNN learning model. As already anticipated, the QFIM will be the main tool employed to analyse the effects of noise.

\subsection{Motivation of the study}

Computing the eigenspectrum of the QFIM associated to a specific QNN provides valuable insights into how variations in the parameter space may impact the output quantum state.
In the case of overparameterized quantum models, only a subset of all the parameters ``actively'' contributes to changing the quantum state, while the other parameters act redundantly in the same directions within the parameter space. 
Mathematically, such a redundancy results in null eigenvalues and a saturated rank of the QFIM.
Building upon previous theoretical knowledge~\cite{muller2016relative}, Ref.~\cite{garciamartin2023effects} studied the effect of depolarizing noise on overparameterized QNNs by means of the QFIM. 

The main outcome of this latter analysis is that the QFIM entries (as well as the eigenvalues) undergo an exponential suppression in the case of depolarizing noise, either global or local, combined with general unital Pauli channels. In particular, for the latter case, the scaling is exponential in both the probability of applying depolarizing noise ($p$) and the total number of noisy gates in the circuit. 
This exponential decay of the QFIM implies that the QNN becomes insensitive to parameter variations as the level of depolarizing noise increases, thus resulting in a flat loss landscape. This behaviour explains the rise of noise-induced barren plateaus.

Moreover, it was also noticed that small local depolarizing noise (possibly combined with unital Pauli channels) may increase the QFIM rank in overparameterized QNNs. This corresponds to an increase in the value of previously null eigenvalues, even though the QFIM, and hence the average of all the eigenvalues, is exponentially suppressed overall. We stress that the null eigenvalues increase only for small noise intensities. This effect stops when these eigenvalues become comparable to the non-zero ones. After this threshold in noise level, they start to be exponentially suppressed too. 
This leads to two different regimes. First, the null eigenvalues become non-trivial, and the QNN can be considered as \emph{quasi-overparameterized}, meaning that noise enables the exploration of new directions, effectively reducing the level of overparametrization. However, as the noise level gets higher and higher, \textit{all} the eigenvalues are exponentially suppressed, which ultimately results in noise-induced barren plateaus. 

In practice, this implies that in the overparameterized regime, the noiseless QFIM becomes ill-conditioned due to its smallest eigenvalue vanishing, whereas the presence of moderate quantum noise improves its conditioning.
This phenomenon strongly resembles what was noticed in Ref.~\cite{Pennington2018Fisher} for classical NNs: linear networks have ill-conditioned FIM, while nonlinear NNs do not possess null FIM eigenvalues.
This change in FIM conditioning was shown to improve the effectiveness of first-order optimization methods.
Given that quantum noise can be considered as a quantum nonlinearity, this motivates the study of how this phenomenon actually affects the performance of QNN learning models.
In what follows, we are going to argue that this quasi-overparameterized regime is part of a more general phenomenon in which the least important parameters gain relevance with respect to the most important ones. We name this process the ``noise-induced equalization'', as detailed in the following. 

\subsection{Noise-induced equalization}

Inspired by these previous results on overparameterized QNN, here we aim at giving an answer to the following question: ``\emph{what are the consequences of this change in the relative importance of the parameters?}'' by further analyzing the QFIM eigenspectrum.

Before presenting the numerical results in the next Section, we introduce a new, useful QFIM-based framework to better interpret the NIE phenomenon.
As anticipated in Sec.~\ref{subsec:qfim}, the eigenvalues of the QFIM can be interpreted as the steepness of the Riemannian manifold associated with the learning model under analysis.
Sorting the eigenspectrum by intensity would correspond to assigning a \emph{steepness rank} $r$ to all the different directions. 
It has to be noticed that such a rank (and in general eigenvalues) would depend on the specific QFIM under analysis, which changes for different input data $\mathbf{x}$, variational parameters $\boldsymbol{\theta}$ and, if present, the quantum noise level $p$, which implies that $r$ depends on all these variables as well. 
This means that the QFIM is a \emph{local} object sensitive to changes in 
the underlying data and model parameters, as well as quantum noise. For this reason, here we are going to define a direction-blind, but spectrally resolved, information measure which will then be evaluated at multiple points in order to draw some useful insights into the \emph{global} landscape.
Hence, given a QNN with $P$ parameters and with an associated QFIM, let us denote with $\{\lambda_r\}_{r=1}^P$ the ordered eigenspectrum of the QFIM such that $\lambda_1\leq\lambda_2\leq\dots\leq\lambda_P$. 
We note that in cases of degenerate eigenvalues, we manually and arbitrarily assign an relative rank between them, e.g., if $\lambda_a = \lambda_b$, we set $r(a) > r(b)$. After fully sorting the eigenspectrum, the original directions no longer matter and we focus solely on the magnitudes of the ordered eigenvalues. In other words, our goal is to compare the overall distortion of the manifold without regard to which specific directions in the space are more or less distorted.
Suppose that each operation in the quantum model is noisy, i.e., that every gate is followed by a quantum channel with a noise level quantified by $p$.
To understand the effect of $p$ on the given model, we can study the ``rank-wise'' change in the importance of each direction in the parameter space, $I_r(p)$, with respect to the noiseless case ($p=p_0=0$), which is explicitly defined as: 
\begin{equation}
    \label{eq:rel_change_lambda}
    I_r(p)=\frac{\lambda_r(p)}{\lambda_r(p_0)} \, .
\end{equation}
For the above reasons, Eq.~\eqref{eq:rel_change_lambda} can be interpreted as a change in steepness in the quantum state space, allowing us to identify how distorted the manifold is in different directions, thus offering a finer characterization at the single eigenvalue level. 
In this sense, $I_r(p)$ provides a novel, spectrally-resolved perspective on the degradation of quantum information, going beyond previously studied aggregated measures~\cite{haug2021capacity, abbas2021effective,garciamartin2023effects,Khanal2025_data_JSuperC} and enabling a more detailed understanding of how information geometry deforms with increasing noise.

The choice for this specific measure is guided by multiple factors. First, from the Data-Processing Inequality and the L\"{o}wner order relation we know that the trace of the QFIM can only decrease under the action of quantum noise (for depolarizing noise, we know that it is moreover exponentially suppressed~\cite{garciamartin2023effects}). The same argument also excludes the average of eigenvalues (and other global quantities like the determinant~\cite{Khanal2025_data_JSuperC}) among the candidates for measuring the effect of noise on the spectrum. Hence, we may think of splitting the eigenspectrum into two parts, i.e., large and small eigenvalues, then excluding the largest eigenvalues, and computing the average of the small eigenvalues in order to characterise their behaviour. While this approach may seem a valuable path, how to determine a proper splitting of the eigenspectrum remains unclear for a generic QNN. In fact, while such a splitting may appear quite evident for overparameterized QNNs, generalizing this procedure is non-trivial. By considering the measure defined in Eq.~\eqref{eq:rel_change_lambda}, instead, we can compare each noisy eigenvalue to its noiseless version, thus giving a clear quantification of how much the specific element of the QFIM eigenspectrum is affected by noise, in particular, whether it is increased or lowered, respectively. Notice that the possibility of some of the smallest eigenvalues increasing does not contradict the L\"{o}wner ordering relation between the noiseless and noisy QFIM. In fact, assuming that the noise will decrease the trace of the QFIM, the noiseless QFIM only \emph{weakly majorize}~\cite{horn1994matrix_analysis} the noisy QFIM, meaning that the following inequality holds for the partial sums of decreasingly ordered eigenvalues
\begin{equation}
    \sum_{r=P}^k\lambda_r(p)\leq \sum_{r=P}^k\lambda_r(p_0) \quad \text{for }k=P,\dots, 1 \, .
\end{equation}

We now define the concept of equalization and specify the conditions under which it is considered optimal, a regime in which improved exploration is expected. The equalization definition is based on empirical deductions derived from extensive numerical simulations (see the next Section). 
In particular, we numerically observe that certain levels of noise increase the least important eigenvalues, while the most relevant eigenvalues are damped. In fact, this is the process we define ``noise-induced equalization'' (NIE), which we formalize as follows:
\begin{observation*}[Noise-induced equalization]
    Given a QNN model whose associated QFIM ordered eigenspectrum is $\{\lambda_r\}_{r=1}^P$, for all noise levels $p\geq 0$ there exists an integer $R\in [0,P]$ such that
    \begin{equation}
    \label{eq:equalization}
        \begin{cases}
              I_r(p)> 1 \ &\forall r\leq R \, , \\
              I_r(p)\leq 1 &\forall r> R  \, .
        \end{cases}
    \end{equation}
\end{observation*}
Here, notice that $R=0$ is also included as a possibility, to account for the situation in which all the eigenvalues lose importance (i.e., they are exponentially suppressed). The latter is the case of high noise levels.  As a consequence of the former Observation we can define the level of noise inducing the best equalization:
\begin{definition*}[Best NIE]
\label{th:conjecture}
    Given a QNN model, the best noise-induced equalization arises when subject to quantum noise of level $p^*$:
    \begin{equation}
    \label{eq:conjecture}
        p^*=\frac{1}{R_{max}}\sum_{r=1}^{R_{max}}\mathbb{E}_{\mathbf{x},\theta}\left[p_r^* \right]\, ,
    \end{equation}
    with $R_{max}=\max_p R$ and $p^*_r=\arg\max_p I_{r}(p)$.
\end{definition*}
\begin{figure}
    \centering
    \includegraphics[width=\linewidth]{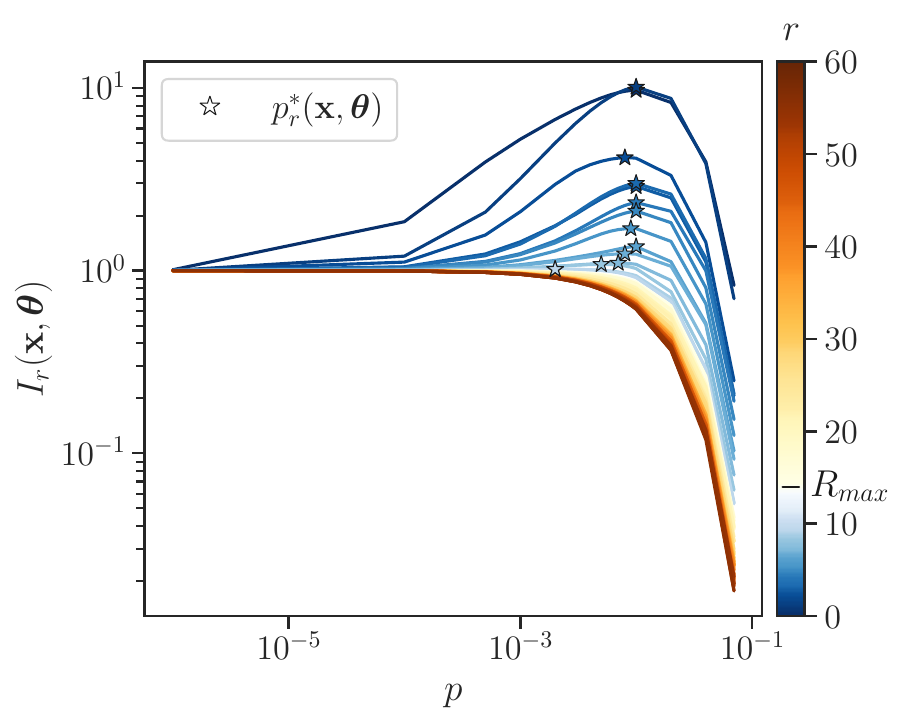}
    \caption{Change in steepness $I_r(\mathbf{x}, \boldsymbol{\theta})$ for the for input $\mathbf{x}$ and parameter vector $\boldsymbol{\theta}$, for an underparametrized QNN of $n=5$ qubits subject to dephasing noise.}
    \label{fig:I_r}
\end{figure}
Here, $R_{max}$ is taken to consider all and only the eigenvalues that can acquire importance by changing the noise level. 
Then, $p^*$ is obtained by averaging the noise values leading to the maximal gain ($p_r^*=\arg\max_p I_{r}(p)$) per each eigenvalue that can acquire importance ($r\leq R_{max}$). 
In Fig.~\ref{fig:I_r}, we show how our measure varies for different eigenvalues as a function of the quantum noise. 
The averaging over inputs and parameters is taken in order to cancel the dependency on inputs and specific points in the landscape. 
While this is necessary to determine a unique noise level over the landscape, getting rid of data dependence may limit the power of the method in the context of generalization, as it is known that such performances are influenced by the data distribution~\cite{langford2002pac, dziugaite2017computing,rivasplata2020pac}. 

It is worth emphasizing that Eq.~\eqref{eq:conjecture} is different from using the $\arg\max$ of $\bar{{I}}_{R_{max}}(p)=\sum_r^{R_{max}}\mathbb{E}_{\mathbf{x},\theta}({I}_{r})/R_{max}$, as it would allow to only select noise levels among the tested ones. 

Following this definition, we conjecture that around the noise level $p^*$ the QNN should experience improved performances. The intuition behind this conjecture is that noise equalization allows to reduce the hyper-specialization of the most influential directions in parameter space, while reinforcing the weakest ones, i.e., reducing exploitation in favor of exploration. 

In Appendix~\ref{appendix:noise dla}, we approach the effect of quantum noise from a theoretical perspective. 
In particular, after introducing the necessary background knowledge on Dynamical Lie Algebras (DLAs) in closed and open quantum systems, we give insights into how noise can induce equalization in two different situations: 
when the generators of the unitary dynamics either commute with the noise superoperators or when they do not.
When noise introduces new generators in the quantum dynamics, new directions are explored if they do not commute with the ones already present in the noiseless setting. 
This reduces the ratio between the number of variational parameters and the number of directions that can be explored, effectively reducing the exploitation of redundant directions. 
This is the reason why the null eigenvalues are activated when we apply noise to overparameterized QNNs. 
Anyway, this does not explain how equalization takes place in underparameterized QNNs, where less important eigenvalues are strengthened without any activation of new eigenvalues. 
To provide a partial explanation to this phenomenon, inspired by Ref.~\cite{garcia2025noise}, we analytically show in Appendix~\ref{appendix:analytical toys} how the QFIM eigenvalues are affected by noise for two toy models with noise operators commuting with the generators of the DLA associated with the quantum circuit, thus not enlarging the DLA. This allows us to prove the existence of an optimal level of noise $p^*$ and demonstrate how information flows in directions that were not active in the noiseless setting.

Hence, there can be situations in which noise enhances expressivity, potentially enabling more complex transformations, even in the presence of decoherence.
Noise-induced equalization could then be regarded as the sweet spot between enhanced expressivity (strenghtened low eigenvalues) and detrimental noise effects given by loss of coherence (weakened high eigenvalues). 
Ultimately, this could allow the improvement of generalization capabilities for QNN models.

From the point of view of space curvature, this can also be intuitively seen as a reshaping of the Riemannian manifold where extremely steep directions are smoothened, while flatter directions gain some steepness. This may result in a landscape in which minima are wider and flatter, a property which has been associated to better generalization performance~\cite{Baldassi2021Unveiling}. 
Hence, we conjecture that such an \emph{equalization} in the QFIM eigenvalues related to QNNs (not necessarily overparameterized) might be the reason behind the improved generalization properties that have been numerically observed, e.g., in Ref.~\cite{somogyi2024niose_reg}, and analytically predicted with generalization bounds in Refs.~\cite{Khanal2025_data_JSuperC,zhu2025optimizer,yang2025stability}.

In Sec.~\ref{sec:results}, we provide numerical evidence to confirm our anticipated conjecture, demonstrating its applicability to both overparameterized and underparameterized QNNs. Our results show that NIE of the eigenspectrum occurs in both cases, thus providing a key pre-training insight into the behavior of QNNs. Through this phenomenon, we derive an estimate for $p^*$ based on our conjecture, and validate that such a noise level actually often leads to desirable generalization performance across various use cases.

\subsection{Comparison with generalization bound}
\label{sec:gen bound}

As already mentioned, the connection between noise and generalization of QNN has started to appear in the recent literature. In particular, in the case of Ref.~\cite{somogyi2024niose_reg} empirical conclusions have been drawn after numerical evidence in selected cases, while in Refs.~\cite{Khanal2025_data_JSuperC,zhu2025optimizer,yang2025stability} the authors have tried to quantify the regularizing power of noise by means of generalization bounds. 
In contrast to these approaches, in the present work we study the origin of such a phenomenon, which allows us to give an estimate of a noise level, $p^*$, before actually training the QNN model, where we argue that improved performances may be observed. 
In fact, to further strengthen our claim, in Sec.~\ref{sec:results} we apply our estimation procedure to the very same dataset considered in Ref.~\cite{somogyi2024niose_reg} to check whether we have compatible findings.
Moreover, Ref.~\cite{Khanal2025_data_JSuperC} provides a generalization bound for noisy quantum circuits depending on the QFIM spectrum among other quantities, but the dependency of these elements on noise and the specific role of QFIM is not discussed.
Understanding how the quantum noise explicitly enters the generalization bound could potentially lead to an alternative method to estimate $p^*$. 

In Appendix~\ref{appendix:explicit bound}, we explicitly report the expression of this bound, grouping together in the quantity $B(p)$ all the noise-dependent terms. 
The main quantities affected by noise are: (i) the effective dimension ($d_{eff}$), (ii) the square root of the QFIM determinant ($m$), and (iii) the Lipschitz constant of the model ($L_f$). Specifically, we realize how the first two quantities 
are affected by noise through their dependence on the QFIM. In fact, we know that the QFIM eigenvalues undergo an equalization process, from which they are exponentially damped by noise. This implies that both the effective dimension (derived by the rank of the QFIM) and the determinant will change under the action of noise. The noise progressively smoothens the optimization landscape until it is completely flat (noise-induced BPs). As the Lipschitz function $L_f$ dominates the gradient norm, increasing the noise will reduce the gradient norm and, consequently, $L_f$.
In Appendix~\ref{appendix:numerical genbound}, we are going to monitor the variation of $B(p)$ on increasing levels of noise, explicitly showing after numerical simulations that its minimum occurs for a non-trivial level of noise. As shown in Sec.~\ref{sec:results}, this noise level could be used as a \emph{broad estimate} of $p^*$, since it is obtained from a bound. More specifically, this kind of theoretical tool associates good generalization performances with a small generalization gap. However, the latter may also result from underfitting, making it rather vacuous.

To conclude this section, we remark that since the noise levels on current hardware are lower than the ones required for regularization~\cite{somogyi2024niose_reg}, noisy dynamics can be obtained by exploiting ancillary qubits via Stinespring dilation~\cite{Stinespring1955}. With the present framework, such noise levels could be obtained by simulations anticipating the training procedure, after which incoherent dynamics can be artificially introduced in the quantum circuits for regularization purposes in QML applications.

\section{Numerical results}
\label{sec:results}

\begin{figure*}
    \centering
    \includegraphics[width=\linewidth]{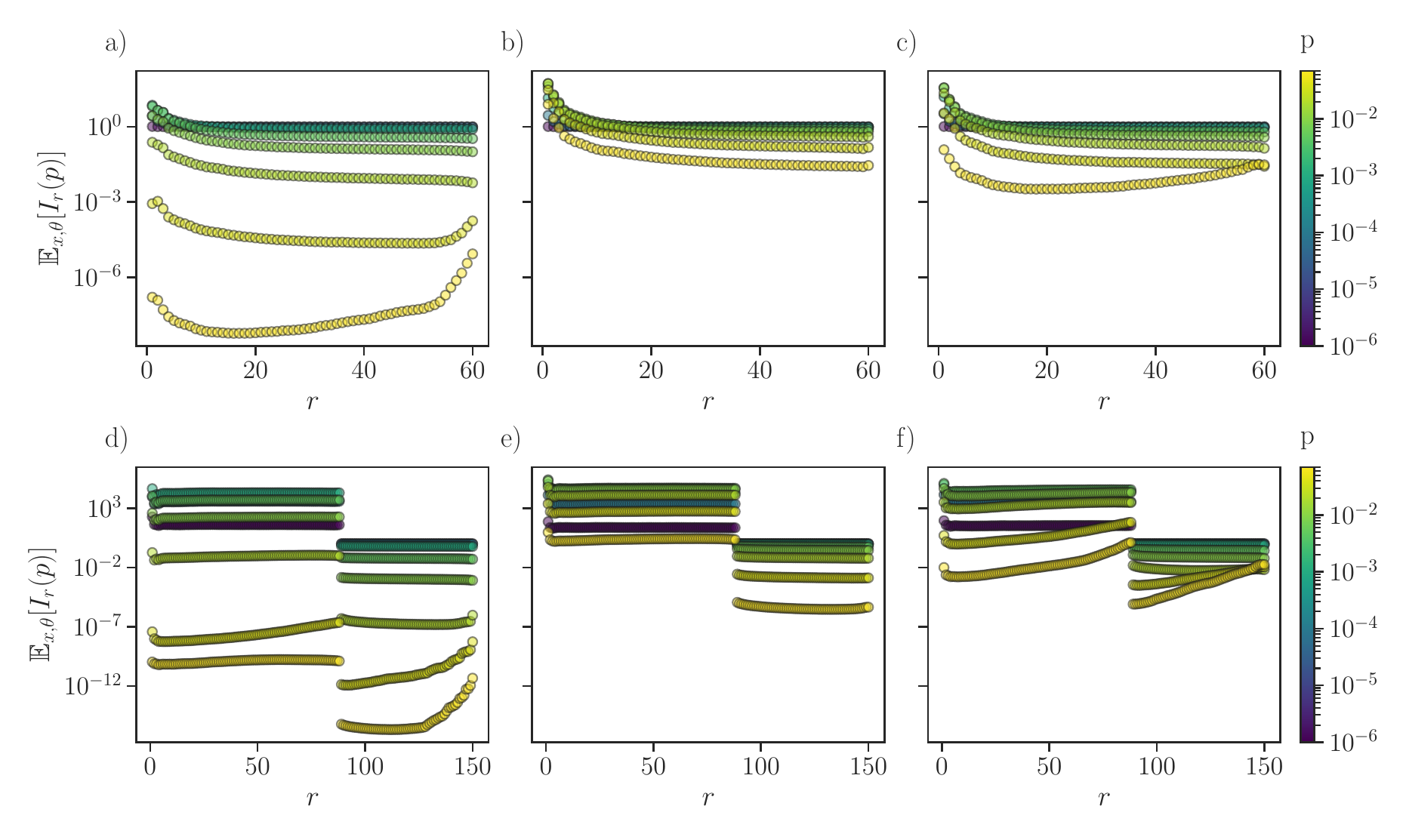}
    \caption{Relative change of the averaged QFIM eigenvalues $\lambda_m$ under different levels of noise $p$ with respect to the noiseless case ($p=0$) for a), d) depolarizing, b), e) dephasing and c), f) amplitude-damping noise. The first row represents the relative change for an underparameterized model ($60$ parameters), while the second for an overparameterized one ($150$ parameters).
    The average is computed over all the inputs of the training set (noisy sinusoidal dataset) and 5 different parameter vectors.}
    \label{fig:relative change sin}
\end{figure*}

\begin{figure*}
    \centering
    \includegraphics[trim={.5cm .5cm .5cm .5cm},clip,width=0.95\linewidth]{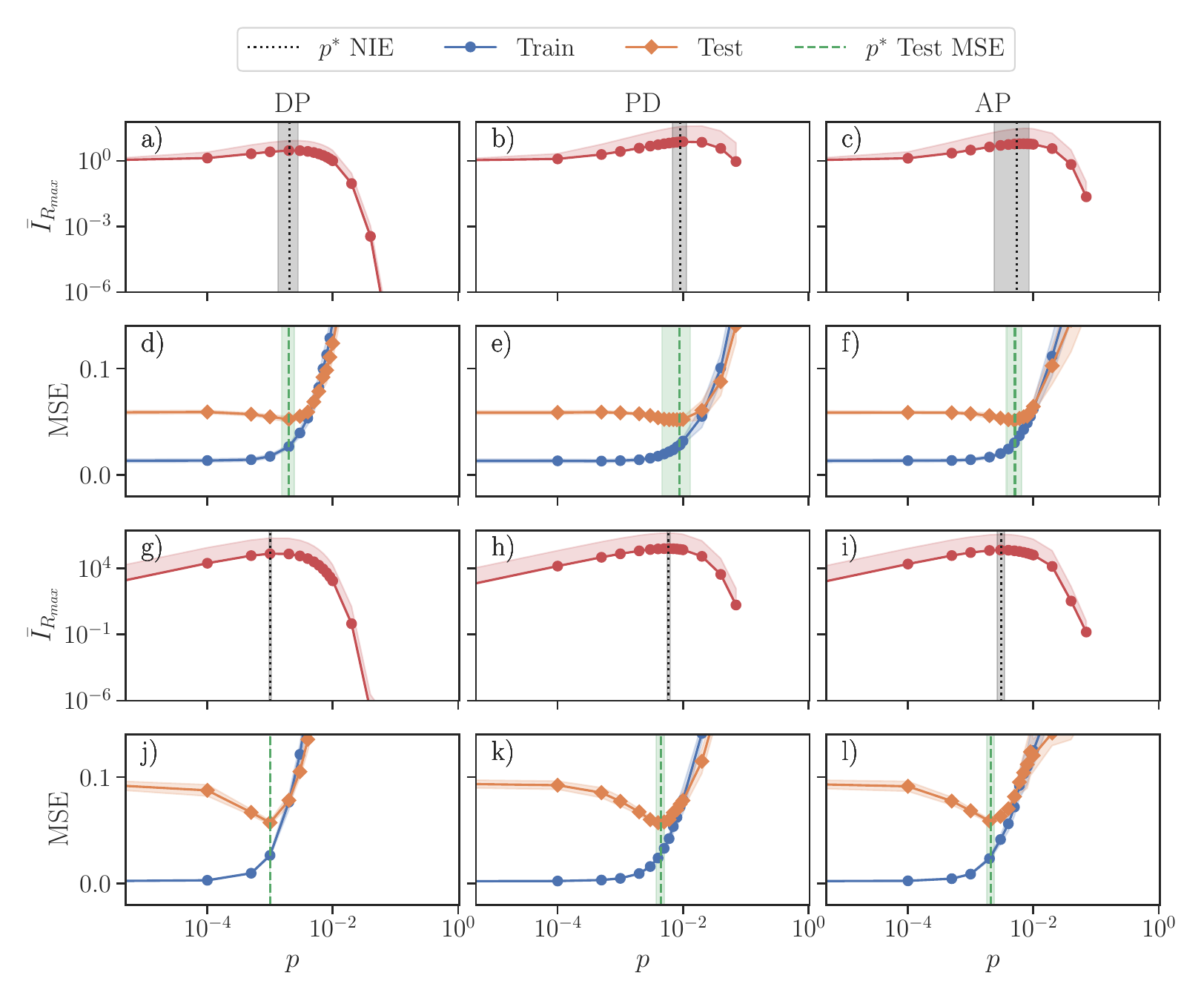}
    \caption{\textbf{Sinusoidal} Comparison of optimal noise level $p^*$ obtained from NIE and test MSE on noisy sinusoidal dataset. Columns correspond to different types of noise: depolarizing (DP), phase damping (PD) and amplitude damping (AD). First and second rows refer to the underparameterized model, while third and fourth to the overparameterized one. We can see that $p^*$ estimated with the two methods (vertical lines) are in good agreement. Shaded regions represent one standard deviation above and below the mean, computed over 10 independent runs, and are plotted throughout, though not always visible (except for $\bar{I}_{R_{\max}}$, where only the upper standard deviation is shown for clarity).} 
    \label{fig:generalization p* sin}
\end{figure*}

In this section, we provide numerical evidence that specific levels of quantum noise, in the neighbourhood of a certain $p^*$, can induce regularization in QNNs suffering from overfitting. In particular, via a pre-training analysis of the QFIM eigenspectrum, we are able to estimate the noise level $p^*$, typical of the particular noise under investigation and model design, which will correspond to the regime where the optimization landscape is smoother. This strategy is advantageous, as it allows one to find a noisy good operating regime without extensive repetitions of the training procedure using hyperparameter grid search.

We start by presenting how noise affects the QFIM eigenspectrum in both underparameterized and overparameterized QNNs. 
In particular, we show that different noise levels have different effects on the eigenvalues and from this changing behavior we can estimate the noise level $p^*$ inducing the strongest equalization. 
This is studied for depolarizing, dephasing and amplitude-damping noise.
Once we have the optimal level $p^*$, we train the quantum model for different noise values and show that the equalization can induce a regularizing effect, effectively reducing overfitting.
Eventually, we analogously try to find a good noisy regime by means of the generalization bound given in Eq.~\eqref{eq:gen_bound}. 

As a use case, we select a regression task with a noisy sinusoidal dataset. It is built by drawing with uniform probability 50 data samples $x$, which is then divided into 30\% training samples and 70\% test samples. The analytical expression describing the labels that we assign to these points is the following:
\begin{equation}
    y = \sin(\pi x) + \epsilon\,,
\end{equation}
where $x \in [-1,1]$ and $\epsilon$ is an additive white Gaussian noise with amplitude equal to 0.4, zero mean and a standard deviation of 0.5. 
In addition, we also verify the applicability of our method for the diabetes dataset analysed in Ref.~\cite{somogyi2024niose_reg} and another two-dimensional regression dataset. 
In the Appendix we provide additional details about the datasets (Appendix~\ref{appendix:datasets}), quantum models (Appendix~\ref{appendix: qnn models}) and additional experiments and analysis (Appendices~\ref{appendix:diabetes dataset},~\ref{appendix:additional exp} and~\ref{appendix:numerical genbound}).

\subsection{QFIM suppression}

As introduced in Sec.~\ref{subsec:qfim} the QFIM is a powerful tool to describe how sensitive a parametric quantum model is to small variations of its parameters. 
It is then reasonable to use this instrument to gain a deeper understanding of how noise affects the action of a QNN, as already done for overparameterized QNNs in Ref.~\cite{garciamartin2023effects}. 
In particular, we focus our attention on how much the eigenspectrum of the QFIM changes under the action of different kinds of quantum noise with respect to the noiseless case ($p=0$) for both underparameterized and overparameterized QNNs. This analysis is carried out using our novel spectrally-resolved measure $I_r(p)$, defined in Eq.~\eqref{eq:rel_change_lambda}. 
More in detail, we are interested in seeing the effects on average on the landscapes. For this purpose, we would need the expectation of $I_r(p)$ with respect to both the data and parameters distribution. 
To approximate this, $\mathbb{E}_{x,\theta}$ is calculated as the average over training samples and 5 parameters initializations. One could also retain some data samples and perform this analysis based on a validation set.

In Fig.~\ref{fig:relative change sin}, we show the relative change of the QFIM eigenvalues $\lambda_m$ under different levels of noise $p$ for depolarizing, dephasing and amplitude-damping noise for both underparameterized and overparameterized models of $n=5$ qubits with maximal expressivity (i.e. overparametrization threshold at $2^{5+1}-2=62$~\cite{haug2021capacity}). 
The dataset is a noisy sinuoidal and the circuit is a Hardware Efficient Ansatz (HEA)~\cite{leone2024HEA} (see Appendix~\ref{appendix:datasets} and Appendix~\ref{appendix: qnn models} for more details). The eigenvalues are indexed in increasing order.
In all these plots, we can notice that there is a first phase where some of the least relevant eigenvalues are increased by the presence of noise, while eigenvalues with higher index $m$ either decrease or stay stable. 
Since the QFIM eigenvalues are associated with a linearly independent direction in the state space, this means that directions (in the diagonalized framework) with a growing eigenvalue are gaining importance in the computation with respect to the noiseless case. 
After a certain threshold value $p^*$ then the whole eigenspectrum is suppressed exponentially, as already shown in Ref.~\cite{garciamartin2023effects}. 
Here we want to stress that the increasing importance of the least relevant eigen-directions takes place not only in overparameterized QNNs, but is a more general phenomenon occurring also in underparameterized models. 
This opens the question of whether such a phenomenon could also happen in other settings/systems outside QML and what the consequences could be.

\subsection{Noise regularization}
\label{subsec:noise regular}

\begin{table}[]
\begin{tabular}{l|c|c|c|}
\cline{2-4}
& \multicolumn{1}{c|}{DP} & \multicolumn{1}{c|}{PD} & \multicolumn{1}{c|}{AD} \\ \hline
\multicolumn{1}{|l|}{NIE} & $(2.1 \pm 0.7) 10^{-3}$                   & $(8.9 \pm 2.5) 10^{-3}$                  & $(5.5 \pm 3.0) 10^{-3}$                   \\ \hline
\multicolumn{1}{|l|}{Test MSE}            & $(2.0 \pm 0.4) 10^{-3}$    & $(8.8 \pm 4.1) 10^{-3}$       & $(5.1 \pm 1.4) 10^{-3}$   \\ \hline
\multicolumn{1}{|l|}{$B(p)$}               & $(7.7 \pm 0.5) 10^{-3}$                   & $(2.3 \pm 0.7) 10^{-2}$                   & $(1.2 \pm 0.4) 10^{-2}$                   \\ \hline
\multicolumn{1}{|l|}{Gen. gap}            & $(5.5 \pm 1.0) 10^{-3}$    & $(2.2 \pm 0.6) 10^{-2}$       & $(1.24 \pm 0.5) 10^{-2}$     
\\ \hline
\end{tabular}
\caption{\textbf{Sinusoidal} Comparison of different values of $p^*$ for depolarizing (DP), phase damping (PD) and amplitude damping (AD) for sinusoidal dataset estimated with different methods for underparameterized QNN.}
\label{tab:p* underparam}
\end{table}

\begin{table}[]
\begin{tabular}{l|c|c|c|}
\cline{2-4}
                         & \multicolumn{1}{c|}{DP} & \multicolumn{1}{c|}{PD} & \multicolumn{1}{c|}{AD} \\ \hline
\multicolumn{1}{|l|}{NIE} & $(1.0 \pm 0.0) 10^{-3}$ & $(5.9 \pm 0.3) 10^{-3}$ & $(3.1 \pm 0.4) 10^{-3}$ \\ \hline
\multicolumn{1}{|l|}{Test MSE}               & $(1.0 \pm 0.0) 10^{-3}$ & $(4.4 \pm 0.7) 10^{-3}$ & $(2.1 \pm 0.3) 10^{-3}$  \\ \hline
\multicolumn{1}{|l|}{$B(p)$}                 & $(1.8 \pm 0.4) 10^{-3}$ & $(1.0 \pm 0.01) 10^{-2}$ & $(6.0 \pm 0.6) 10^{-3}$ \\ \hline
\multicolumn{1}{|l|}{Gen. gap}            & $(2.0 \pm 0.0) 10^{-3}$    & $(9.2 \pm 0.6) 10^{-3}$       & $(8.7 \pm 1.6) 10^{-3}$   \\ \hline
\end{tabular}
\caption{\textbf{Sinusoidal} Comparison of different values of $p^*$ for depolarizing (DP), phase damping (PD) and amplitude damping (AD) for sinusoidal dataset estimated with different methods for overparameterized QNN. }
\label{tab:p* overparam}
\end{table}

We now proceed with the verification of our conjecture, i.e. the critical point $p^*$, where the least important directions in the Riemannian manifold have the maximal relative increase compared to the noiseless case, is also the noise level determining desirable generalization performances achievable by applying quantum noise only. 
The noise level $p^*$ is determined by averaging the noise values leading to the maximal gain per each eigenvalue that can acquire importance as defined in Eq.~\eqref{eq:conjecture}.
At this point, we need to specify what we mean by \emph{desirable generalization properties}, as there is no simple unique measure to evaluate it. 
In fact, one may consider desirable either the situation where the test error is minimal or the one where the generalization gap (the difference between training and test error) closes. 
While looking at these situations, one still has to take into account the value of the error on the training set, if this becomes too large, a low test error (compared to the training one) or a closing gap might be related to underfitting the data, which is not beneficial in the end. 
For this reason, we compare $p^*$ obtained by the NIE analysis with both the noise level leading to the minimal test error, measured in terms of Mean Squared Error (MSE) (see Sec.~\ref{sec:methods}),  and the one associated with a closing generalization gap.  
The comparison of these estimates for different quantum noise channels and QNN depth is presented both in Fig.~\ref{fig:generalization p* sin} for a visual understanding and in Tabs.~\ref{tab:p* underparam}-\ref{tab:p* overparam} for a more compact and precise assessment. 
We would like to remark that we do not make use of any technique that might induce regularization (as stochastic gradient descent, L2 regularization or shot noise) other than quantum noise in order to identify its influence. 

More in detail, Figs.~\ref{fig:generalization p* sin}a-f display results for an underparameterized QNN, while Figs.~\ref{fig:generalization p* sin}g-l gather the same information for the same model in the overparameterized regime. 
As anticipated, in the first and third row of Fig.~\ref{fig:generalization p* sin} we show how $\bar{I}_{R_{max}}(p)$ changes upon variation of the noise level. 
In particular, here we plot the average trend over different QFIM matrices (five random parameter vectors per training data), and the shade represents one standard deviation. 
To have a better estimate, the optimal level of $p^*$ is computed individually on different QFIMs, averaging over different inputs and variational parameters as described in Eq.~\eqref{eq:conjecture}. The vertical dotted line highlights our estimate of the optimal level of noise $p^*$, while the shaded band represent one standard deviation. 

Similarly, in the second and fourth row, we show the value of the final MSE on training and test data, averaged over 10 different initializations, together with the estimate of the best noise level according to such values. Specifically, $p^*$ is determined for the single runs and then averaged. Different columns in Fig.~\ref{fig:generalization p* sin} report results for different kinds of quantum noise, depolarizing, phase and amplitude damping, respectively.

We can notice a good agreement between the two different estimates in all the configurations. This confirms that the NIE-based procedure allows an estimation of the optimal noise level $p^*$ with a neighbourhood corresponding to a dip in the test MSE. In particular, our approximation of $p^*$ in most cases exactly coincides with what is found to be the level of noise inducing the minimal test MSE (see Tab.~\ref{tab:p* underparam} and Tab.~\ref{tab:p* overparam}). When the two levels do not exactly match, this could be due to multiple factors, such as the finite number of test samples and initializations with which we evaluate the test MSE, the finite number of training samples and parameters configurations used to compute the average eigenspectrum and the discrete set of noise levels studied. All these factors contribute to the size of the error bar, allowing for compatibility between the approximations. We note that for the overparameterized QNN, the error in the NIE-based determination of $p^*$ is much smaller with respect to the underparameterized case. This is strictly due to the noise enabling new directions to be explored, which happens irrespective of the different training samples or parameters configurations.

We also investigate the feasibility of determining $ p^* $ using the generalization bound given in Appendix~\ref{appendix:explicit bound}. In particular, we focus on a single term in the bound $B(p)$, which is the only noise-dependent term. 
This approach proves viable since $B(p)$ (and hence the bound) exhibits a minimum (see numerical experiments in Appendix~\ref{appendix:numerical genbound}) for nontrivial values of $ p $ whose value is reported in  Tabs.~\ref{tab:p* underparam}-\ref{tab:p* overparam}. 
Nonetheless, several limitations arise due to the quantities required for computing the bound. First, the determinant becomes problematic in models with a large number of parameters, as many eigen-directions are associated with eigenvalues smaller than $1$. 
This situation yields a determinant that is nearly zero, causing the bound to diverge to infinity and violating one of the theorem's assumptions ($ \sqrt{\det(\mathcal{F}(\theta))} \geq m > 0 $).  
In contrast, when the determinant remains significantly different from zero, as observed in QNNs with fewer parameters, increasing noise induces barren plateaus that suppress the gradient of the model. This suppression results in the Lipschitz constant $ L_f $ approaching zero, thereby biasing the minimum toward $ p = 1 $.  

Furthermore, generalization bounds assess the gap between training and optimal performance, where a closed gap is ideally indicative of optimal generalization. 
However, the gap may also narrow as a result of deteriorating training loss, rather than an improvement in generalization performance. 
By looking at Tabs.~\ref{tab:p* underparam}-\ref{tab:p* overparam} it is possible to see a fair agreement between the values of $p^*$ estimated from the closing generalization gap and the one from generalization bound. 
Unfortunately, most of the times this happens for values of training MSE that are way worse than the initial ones. 
These inherent limitations become particularly evident when analyzing the diabetes dataset in Appendix \ref{appendix:diabetes dataset}.  

Since the proposed approach relies solely on a subset of the eigenspectrum, it circumvents these issues while requiring fewer computational resources, rendering it suitable for a wide range of QNN architectures.

\section{Discussion}
\label{sec:discussion}

In this work, we have investigated the impact of quantum noise on the generalization properties of QNNs. 
In particular, we have shown a correlation between the improvement of generalization performance of a QNN and the noise-induced equalization (NIE) effect in the eigenspectrum of the Quantum Fisher Information Matrix (QFIM), whereby the least important eigen-directions gain relevance while the most relevant ones lose it. 
This result can be intuitively explained by combining previous results on quantum noise and overparametrization~\cite{garciamartin2023effects}, conditioning of neural networks~\cite{Pennington2018Fisher} and the relationship between wide minima and generalization~\cite{Baldassi2021Unveiling}.
Since the noise level inducing the best equalization can be deduced based on the previously mentioned QFIM analysis, we propose this as an effective protocol allowing to determine the noise level that leads to enhanced exploration and consequently desirable generalization performance for the given QNN model.

More in detail, we have numerically showcased the NIE effect by introducing the spectrally-resolved measure $I_r(p)$, which quantifies the relative change of importance in the directions of the Riemannian manifold as a function of noise. Then, we identified an optimal noise level, $p^*$, obtained by averaging the noise values leading to the maximal increase of importance ($\arg \max_p I_r (p)$) per each eigenvalue that can acquire importance ($r \leq R_{max}$), via Eq.~\eqref{eq:conjecture}. 
Remarkably, $p^*$ is, in most of the analyzed cases, compatible with the noise level that yields the most beneficial generalization performance, reinforcing the idea that noise can play a constructive role in improving generalization. 
This method has a significant advantage over other common regularization techniques, as it allows for the optimal noise level (i.e., $p^*$) to be determined without extensive hyperparameter grid search requiring repetitions of the training procedure. 
Furthermore, our results are in agreement with the noise levels experimentally determined in Ref.~\cite{somogyi2024niose_reg} for the diabetes dataset, suggesting that the NIE effect is not a merely theoretical curiosity, but it could actually be observed in practical implementations of QML models.

A comparison with existing generalization bounds highlights the limitations of previous theoretical results. Specifically, Refs.~\cite{zhu2025optimizer,yang2025stability} provide bounds that suggest an improvement in generalization with increasing noise. 
However, these bounds are only applicable in the case of Stochastic Gradient Descent (SGD) optimizer, and they do not provide a method to determine an optimal noise level ($p^*$). 
In fact, their predictions become vacuous when noise-induced barren plateaus impede model training. In contrast, the bound introduced in Ref.~\cite{Khanal2025_data_JSuperC} provides an interesting dependence on the determinant of the QFIM. 
However, such a dependence is left implicit, while the explicit dependence on the noise level is not discussed. 
By numerically computing this bound as a function of noise, we observed that it exhibits a minimum, thus suggesting the existence of an optimal noise level. 
Nevertheless, this approach is hindered by intrinsic limitations related to the determinant of the QFIM, as well as to the effect of noise-induced barren plateaus, which prevent an accurate estimation of $p^*$. 
We stress that our procedure only depends on the QNN design, making it extremely versatile and applicable to disparate datasets and optimizers.

Looking forward, several directions remain open for future research. 
Given the known suppression of the eigenspectrum under noise, an interesting avenue would be to analytically describe the initial growth of the least important eigenvalues. As a first step, in this work, we analytically showed that the enhancement of low eigenvalues is indeed possible for small toy problems.
A combination of this growth with the known exponential decay mechanisms~\cite{garciamartin2023effects} could potentially lead to an analytical determination of $p^*$. 

Computational improvements for this pre-training analysis could be achieved in the QFIM calculation by leveraging techniques such as SPSA~\cite{Gacon2021simultaneous}, Stein's identity~\cite{halla2025estimationQFI} or classical shadows~\cite{Huang2020shadows} similarly to what has been done in Ref.~\cite{shi2025weightedapproxQNG}. 
Moreover, it would be interesting to apply a similar investigation leveraging the weighted approximate metric tensor~\cite{shi2025weightedapproxQNG} instead of the QFIM. 
This would take into account additional information coming from the observable possibly leading to more accurate estimate of the optimal noise level $p^*$, especially for shallower circuits where the locality of the hamiltonian could imply a limited light-cone contribution.

An intrinsic limitation of the introduced technique lies in the fact that we seek better QFIM conditioning on average over the optimization landscape, but then learning is usually initialized at random points that might have different conditioning with respect to the previously analyzed points. A potential enhancement could stem from the combination with meta-learning techniques, similar to the ones employed in Ref.~\cite{ait2025sculpting}.

Finally, a broader perspective concerns the implications of noise-induced equalization beyond the scope of quantum machine learning. In fact, it was recently shown in Ref.~\cite{garcia2025noise} that incoherent dynamics can lead to metrological advantages in quantum sensing. This is the same principle at the heart of the NIE. It remains an open question whether similar effects could manifest in other quantum paradigms related to quantum sensing like, for example, quantum thermodynamics. Investigating these aspects in different fields could further enable a deeper understanding of the interplay between noise, optimization, and generalization in quantum models.

\section{Methods}
\label{sec:methods}

Here we provide some insights into practical details related to this work, leaving the most technical part in the appendices. 
Numerical simulations of quantum circuits are performed in Python with Pennylane~\cite{pennylane} in combination with JAX~\cite{jax2018github}. 
For noisy simulations, we execute circuits in the density matrix formalism, applying a noise channel after each gate, while for noiseless we rely on statevector simulations. The QFIM matrix is derived from quantum circuits by leveraging JAX as well.

For what concerns the discrete set of noise levels studied, different settings have been between sinusoidal and diabetes datasets. In particular, for the sinusoidal dataset the noise levels sampled are: 
\begin{align*}
p_{list}^{sin}=\{&10^{-6}, 10^{-4}, 5\cdot10^{-4}, 10^{-3},2\cdot10^{-3},\\
                        & 3\cdot10^{-3}, 4\cdot10^{-3},5\cdot10^{-3}, 6\cdot10^{-3},\\
                        & 7\cdot10^{-3}, 8\cdot10^{-3}, 9\cdot10^{-3}, 10^{-2},\\
                        &2\cdot 10^{-2}, 4\cdot 10^{-2}, 7\cdot 10^{-2}\}. 
\end{align*}
While for diabetes dataset we have added additional noise levels that we studied in Ref.~\cite{somogyi2024niose_reg}: 
\begin{align*}
    p_{list}^{diab}=p_{list}^{sin}\cup\{
                        & 10^{-2.75}, 10^{-2.5} ,10^{-2.25}, 10^{-1.75},3\cdot 10^{-2},\\
                        &   10^{-1.5} , 5\cdot 10^{-2},10^{-1.25}, 8\cdot 10^{-2},9\cdot 10^{-2},\\
                        &10^{-1}, 10^{-0.75}, 10^{-0.5} , 10^{-0.25},  1\}
\end{align*}
The analysis of NIE is conducted on multiple QFIM. In particular, per each training data, we compute the QFIM at 5 random points of the parameter space. 
In Sec.~\ref{subsec:noise regular}, we propose to compute $p^*$ as the mean of $\arg\max_p I_r(p) \quad \forall r\leq R_{max}$ on single runs. 
We point out that $R_{max}$ is not unique across different QFIMs and depends on the specific input $\mathbf{x}$ and parameter vector $\theta$.
The selection of $R_{max}$ as the minimum over the different runs is done for convenience, as when averaging and computing the standard deviation, the number of eigenvalues taken into account would be the same. One could have also set $R_{max}$ as the average of the different $R_{max}^{(j)}$, or the closest integer to the mean of the various $R_{max}^{(j)}$, anyway, this might lead to include some eigenvalues that are not increased in all the analyzed QFIMs. In addition, the denominator of $I_r(p)$ is $\max\{10^{-10}, \lambda_r(p_0)\}$ to avoid numerical issues. We invite the interested reader to check out the available code~\cite{githubrepo} for further technical details. 

We perform numerical simulations for training quantum 
machine learning models using the Adam optimizer~\cite{kingma2014adam} with hyperparameters $\eta = 0.01$, $\beta_1 = 0.9$, and $\beta_2 = 0.999$. An important difference from Ref.~\cite{somogyi2024niose_reg} is that we do not employ batch gradient as this is known to have a positive influence on generalization~\cite{smith2020generalization}, while our goal is to show the genuine regularizing effect of quantum noise. The cost function employed is the Mean Squared Error (MSE), which can be written as:
\begin{equation}
\label{eq:mse}
\mathcal{L}(\boldsymbol{\theta})=\text{MSE}(\mathbf{x},y) = \frac{1}{M} \sum_{i=1}^{M} \left(f(\mathbf{x}_i, \boldsymbol{\theta}\right) - y_i)^2
\end{equation}
where $f(\mathbf{x}_i, \boldsymbol{\theta})$ are the predicted outputs, $y_i$ are the true outputs (labels), and $M$ is the number of samples. We trained the models on 10 different initializations of the parameters, which were chosen to be unrelated to the 5 initializations used in our pre-training analysis of the QFIM. This approach allows for a more general assessment of the optimal level of noise $p^*$ leading to the best regime in terms of generalization. To determine this optimal level, we estimated the mean of the argmin of the final test MSE (FTMSE) determined on single runs:
\begin{equation}
    p_j^* =\arg\min_p\text{FTMSE}_j \, ,
\end{equation}
while the error on the estimation is computed as the standard deviation.

Ultimately, to estimate the generalization bound given in Eq.~\eqref{eq:gen_bound}, we take the effective dimension $d_{eff}$ as average the rank of the QFIM (as prescribed by Eq.~\eqref{eq:d_eff}) over training data each evaluated at 5 random points in the parameter space. We approximate the Lipschitz required by the bound
constant as the maximum gradient over these same configurations.

\section*{Code availability statement}
Code to reproduce the results and to create all figures and tables presented in this manuscript is available at Github repository~\cite{githubrepo}.

\section*{Aknowledgments}
FS and GG thanks Davide Cugini, Davide Nigro, Francesco Ghisoni, Dario Gerace and Sabri Meyer for insightful scientific discussions and feedback.
FS and AL acknowledges the support from SNF grant No. 214919. G.G. kindly acknowledges support from the Ministero dell’Università e della Ricerca (MUR) under the “Rita Levi-Montalcini” grant and to INFN.

\section*{AI tools disclaimer}
ChatGPT was used to improve the readability of parts of the paper. No new content was created by the AI tool. The authors have checked all texts and take full responsibility for the result. 

\bibliography{bibliography}

@article{Temme2017prl,
  title = {Error Mitigation for Short-Depth Quantum Circuits},
  author = {Temme, Kristan and Bravyi, Sergey and Gambetta, Jay M.},
  journal = {Phys. Rev. Lett.},
  volume = {119},
  issue = {18},
  pages = {180509},
  numpages = {5},
  year = {2017},
  month = {Nov},
  publisher = {American Physical Society},
  doi = {10.1103/PhysRevLett.119.180509},
  url = {https://link.aps.org/doi/10.1103/PhysRevLett.119.180509}
}

@article{Kandala2019nature,
	author = {Kandala, Abhinav and Temme, Kristan and C{\'o}rcoles, Antonio D. and Mezzacapo, Antonio and Chow, Jerry M. and Gambetta, Jay M.},
	date = {2019/03/01},
	doi = {10.1038/s41586-019-1040-7},
	id = {Kandala2019},
	isbn = {1476-4687},
	journal = {Nature},
	number = {7749},
	pages = {491--495},
	title = {Error mitigation extends the computational reach of a noisy quantum processor},
	url = {https://doi.org/10.1038/s41586-019-1040-7},
	volume = {567},
	year = {2019}}

@article{cai2023error_mitigation,
  title = {Quantum error mitigation},
  author = {Cai, Zhenyu and Babbush, Ryan and Benjamin, Simon C. and Endo, Suguru and Huggins, William J. and Li, Ying and McClean, Jarrod R. and O'Brien, Thomas E.},
  journal = {Rev. Mod. Phys.},
  volume = {95},
  issue = {4},
  pages = {045005},
  numpages = {37},
  year = {2023},
  month = {Dec},
  publisher = {American Physical Society},
  doi = {10.1103/RevModPhys.95.045005},
  url = {https://link.aps.org/doi/10.1103/RevModPhys.95.045005}
}

@book{nielsen00,
  title={Quantum computation and quantum information},
  author={Nielsen, Michael A and Chuang, Isaac L},
  year={2010},
  publisher={Cambridge university press}
}

@article{Stinespring1955,
  title = {Positive Functions on C-Algebras},
  volume = {6},
  ISSN = {0002-9939},
  url = {http://dx.doi.org/10.2307/2032342},
  DOI = {10.2307/2032342},
  number = {2},
  journal = {Proceedings of the American Mathematical Society},
  publisher = {JSTOR},
  author = {Stinespring,  W. Forrest},
  year = {1955},
  month = apr,
  pages = {211}
}

@article{Preskill_2018,
   title={Quantum Computing in the NISQ era and beyond},
   volume={2},
   journal={Quantum},
   publisher={Verein zur Forderung des Open Access Publizierens in den Quantenwissenschaften},
   author={Preskill, John},
   year={2018},
   
   pages={79}
}

@misc{preskill2025nisqmegaquopmachine,
      title={Beyond NISQ: The Megaquop Machine}, 
      author={John Preskill},
      year={2025},
      eprint={2502.17368},
      archivePrefix={arXiv},
      primaryClass={quant-ph},
      url={https://arxiv.org/abs/2502.17368}, 
}

@article{Khanal2025_data_JSuperC,
	author = {Khanal, Bikram and Rivas, Pablo},
	date = {2025/03/13},
	doi = {10.1007/s11227-025-06966-9},
	id = {Khanal2025},
	isbn = {1573-0484},
	journal = {The Journal of Supercomputing},
	number = {4},
	pages = {611},
	title = {Data-dependent generalization bounds for parameterized quantum models under noise},
	url = {https://doi.org/10.1007/s11227-025-06966-9},
	volume = {81},
	year = {2025}}

@article{banchi2021generalization,
  title={Generalization in quantum machine learning: A quantum information standpoint},
  author={Banchi, Leonardo and Pereira, Jason and Pirandola, Stefano},
  journal={PRX Quantum},
  volume={2},
  number={4},
  pages={040321},
  year={2021},
  publisher={APS}
}

@article{somogyi2024niose_reg,
  title={Method for noise-induced regularization in quantum neural networks},
  author={Somogyi, Wilfrid and Pankovets, Ekaterina and Kuzmin, Viacheslav and Melnikov, Alexey},
  journal={arXiv preprint arXiv:2410.19921},
  year={2024}
}

@article{yang2025stability,
  title={Stability and Generalization of Quantum Neural Networks},
  author={Yang, Jiaqi and Xie, Wei and Xu, Xiaohua},
  journal={arXiv preprint arXiv:2501.12737},
  year={2025}
}

@article{zhu2025optimizer,
  title={Optimizer-Dependent Generalization Bound for Quantum Neural Networks},
  author={Zhu, Chenghong and Yao, Hongshun and Liu, Yingjian and Wang, Xin},
  journal={arXiv preprint arXiv:2501.16228},
  year={2025}
}

@article{haug2021capacity,
	doi = {10.1103/prxquantum.2.040309},
	url = {https://doi.org/10.1103\%2Fprxquantum.2.040309},
	year = 2021,
	month = {oct},
	publisher = {American Physical Society ({APS})},
	volume = {2},
	number = {4},
	author = {Tobias Haug and Kishor Bharti and M.S. Kim},
	title = {Capacity and Quantum Geometry of Parametrized Quantum Circuits},
	journal = {{PRX} Quantum}
}

@article{abbas2021effective,
  title={Effective dimension of machine learning models},
  author={Abbas, Amira and Sutter, David and Figalli, Alessio and Woerner, Stefan},
  journal={arXiv preprint arXiv:2112.04807},
  year={2021}
}

@article{Larocca2023theory,
	doi = {10.1038/s43588-023-00467-6},
	url = {https://doi.org/10.1038\%2Fs43588-023-00467-6},
	year = 2023,
	month = {jun},
	publisher = {Springer Science and Business Media {LLC}},
	volume = {3},
	number = {6},
	pages = {542--551},
	author = {Mart{\'{\i}}n Larocca and Nathan Ju and Diego Garc{\'{\i}}a-Mart{\'{\i}}n and Patrick J. Coles and Marco Cerezo},
	title = {Theory of overparametrization in quantum neural networks},
	journal = {Nat Comput Sci}
}

@misc{kiani2020learning,
      title={Learning Unitaries by Gradient Descent}, 
      author={Bobak Toussi Kiani and Seth Lloyd and Reevu Maity},
      year={2020},
      eprint={2001.11897},
      archivePrefix={arXiv},
      primaryClass={quant-ph}
}

@article{Cerezo2021VQA,
	doi = {10.1038/s42254-021-00348-9},
	url = {https://doi.org/10.1038\%2Fs42254-021-00348-9},
	year = 2021,
	month = {aug},
	publisher = {Springer Science and Business Media {LLC}},
	volume = {3},
	number = {9},
	pages = {625--644},
	author = {M. Cerezo and Andrew Arrasmith and Ryan Babbush and Simon C. Benjamin and Suguru Endo and Keisuke Fujii and Jarrod R. McClean and Kosuke Mitarai and Xiao Yuan and Lukasz Cincio and Patrick J. Coles},
	title = {Variational quantum algorithms},
	journal = {Nature Reviews Physics}
}

@article{Peruzzo2014VQE,
    year = 2014,
    publisher = {Springer Science and Business Media {LLC}},
	volume = {5},
	number = {1},
	author = {Alberto Peruzzo and Jarrod McClean and Peter Shadbolt and Man-Hong Yung and Xiao-Qi Zhou and Peter J. Love and Al{\'{a}}n Aspuru-Guzik and Jeremy L. O'Brien},
	title = {A variational eigenvalue solver on a photonic quantum processor},
	journal = {Nat. Commun.}
}

@article{Biamonte2017quantum,
  title={Quantum machine learning},
  author={Biamonte, Jacob and Wittek, Peter and Pancotti, Nicola and Rebentrost, Patrick and Wiebe, Nathan and Lloyd, Seth},
  journal={Nature},
  volume={549},
  number={7671},
  pages={195--202},
  year={2017},
  publisher={Nature Publishing Group UK London}
}

@article{cerezo2022challenges,
  title={Challenges and opportunities in quantum machine learning},
  author={Cerezo, Marco and Verdon, Guillaume and Huang, Hsin-Yuan and Cincio, Lukasz and Coles, Patrick J},
  journal={Nature Computational Science},
  volume={2},
  number={9},
  pages={567--576},
  year={2022},
  publisher={Nature Publishing Group US New York}
}

@book{2014QMLWittek,
	address = {Boston},
	title = {Quantum Machine Learning},
	author = {Peter Wittek},
	publisher = {Academic Press},
	year = {2014}}

@article{Benedetti2019pqc,
	year = 2020,
	journal={Quantum Sci. Technol.},
	publisher = {{IOP} Publishing},
	volume = {5},
	number = {1},
	pages = {019601},
	author = {Marcello Benedetti and Erika Lloyd and Stefan Sack and Mattia Fiorentini},
	title = {Parameterized quantum circuits as machine learning models}
}

@article{Mangini2021qnns,
	doi = {10.1209/0295-5075/134/10002},
	url = {https://doi.org/10.1209\%2F0295-5075\%2F134\%2F10002},
    year = 2021,
	month = {apr},
  	publisher = {{IOP} Publishing},
  	volume = {134},
  	number = {1},
  	pages = {10002},
  	author = {S. Mangini and F. Tacchino and D. Gerace and D. Bajoni and C. Macchiavello},
  	title = {Quantum computing models for artificial neural networks},
  	journal = {Europhysics Letters}
}

@article{abbas2021power,
	doi = {10.1038/s43588-021-00084-1},
	url = {https://doi.org/10.1038\%2Fs43588-021-00084-1},
	year = 2021,
	month = {jun},
	publisher = {Springer Science and Business Media {LLC}},
	volume = {1},
	number = {6},
	pages = {403--409},
	author = {Amira Abbas and David Sutter and Christa Zoufal and Aurelien Lucchi and Alessio Figalli and Stefan Woerner},
	title = {The power of quantum neural networks},
	journal = {Nat Comput Sci}
}

@article{leone2024HEA,
  title={On the practical usefulness of the hardware efficient ansatz},
  author={Leone, Lorenzo and Oliviero, Salvatore FE and Cincio, Lukasz and Cerezo, M},
  journal={Quantum},
  volume={8},
  pages={1395},
  year={2024},
  publisher={Verein zur F{\"o}rderung des Open Access Publizierens in den Quantenwissenschaften}
}

@article{garciamartin2023effects,
  title = {Effects of noise on the overparametrization of quantum neural networks},
  author = {Garc\'{\i}a-Mart\'{\i}n, Diego and Larocca, Mart\'{\i}n and Cerezo, M.},
  journal = {Phys. Rev. Res.},
  volume = {6},
  issue = {1},
  pages = {013295},
  numpages = {17},
  year = {2024},
  month = {Mar},
  publisher = {American Physical Society},
  doi = {10.1103/PhysRevResearch.6.013295},
  url = {https://link.aps.org/doi/10.1103/PhysRevResearch.6.013295}
}

@article{muller2016relative,
  title={Relative entropy convergence for depolarizing channels},
  author={M{\"u}ller-Hermes, Alexander and Stilck Fran{\c{c}}a, Daniel and Wolf, Michael M},
  journal={Journal of Mathematical Physics},
  volume={57},
  number={2},
  year={2016},
  publisher={AIP Publishing}
}

@article{du2023problem,
  title={Problem-dependent power of quantum neural networks on multiclass classification},
  author={Du, Yuxuan and Yang, Yibo and Tao, Dacheng and Hsieh, Min-Hsiu},
  journal={Physical Review Letters},
  volume={131},
  number={14},
  pages={140601},
  year={2023},
  publisher={APS}
}

@article{McClean2018barren,
	doi = {10.1038/s41467-018-07090-4},
	url = {https://doi.org/10.1038%2Fs41467-018-07090-4},
	year = 2018,
	month = {nov},
	publisher = {Springer Science and Business Media {LLC}},
	volume = {9},
	number = {1},
	author = {Jarrod R. McClean and Sergio Boixo and Vadim N. Smelyanskiy and Ryan Babbush and Hartmut Neven},
	title = {Barren plateaus in quantum neural network training landscapes},
	journal = {Nat Commun}
}

@article{Arrasmith2022equivalence, 
title={Equivalence of quantum barren plateaus to cost concentration and narrow gorges}, 
volume={7}, 
ISSN={2058-9565}, 
url={http://dx.doi.org/10.1088/2058-9565/ac7d06}, 
DOI={10.1088/2058-9565/ac7d06}, 
number={4}, 
journal={Quantum Science and Technology}, 
publisher={IOP Publishing}, 
author={Arrasmith, Andrew and Holmes, Zoë and Cerezo, M and Coles, Patrick J}, 
year={2022}, month=aug, pages={045015} }

@article{sannia2024engineered,
  title={Engineered dissipation to mitigate barren plateaus},
  author={Sannia, Antonio and Tacchino, Francesco and Tavernelli, Ivano and Giorgi, Gian Luca and Zambrini, Roberta},
  journal={npj Quantum Information},
  volume={10},
  number={1},
  pages={81},
  year={2024},
  publisher={Nature Publishing Group UK London}
}

@article{wang2021noise,
	doi = {10.1038/s41467-021-27045-6},
	url = {https://doi.org/10.1038%2Fs41467-021-27045-6},
	year = 2021,
	month = {nov},
	publisher = {Springer Science and Business Media {LLC}},
	volume = {12},
	number = {1},
	author = {Samson Wang and Enrico Fontana and M. Cerezo and Kunal Sharma and Akira Sone and Lukasz Cincio and Patrick J. Coles},
	title = {Noise-induced barren plateaus in variational quantum algorithms},
	journal = {Nat Commun}
}

@misc{schumann2023emergence,
      title={Emergence of noise-induced barren plateaus in arbitrary layered noise models}, 
      author={Marco Schumann and Frank K. Wilhelm and Alessandro Ciani},
      year={2023},
      eprint={2310.08405},
      archivePrefix={arXiv},
      primaryClass={quant-ph}
}

@misc{oliv2022evaluating,
      title={Evaluating the impact of noise on the performance of the Variational Quantum Eigensolver}, 
      author={Marita Oliv and Andrea Matic and Thomas Messerer and Jeanette Miriam Lorenz},
      year={2022},
      eprint={2209.12803},
      archivePrefix={arXiv},
      primaryClass={quant-ph}
}

@article{borras2023impact,
doi = {10.1088/1742-6596/2438/1/012093},
url = {https://dx.doi.org/10.1088/1742-6596/2438/1/012093},
year = {2023},
month = {feb},
publisher = {IOP Publishing},
volume = {2438},
number = {1},
pages = {012093},
author = {Kerstin Borras and Su Yeon Chang and Lena Funcke and Michele Grossi and Tobias Hartung and Karl Jansen and Dirk Kruecker and Stefan Kühn and Florian Rehm and Cenk Tüysüz and Sofia Vallecorsa},
title = {Impact of quantum noise on the training of quantum Generative Adversarial Networks},
journal = {Journal of Physics: Conference Series},
}

@misc{gu2021adaptive,
  author = {Gu, Andi and Lowe, Angus and Dub, Pavel A. and Coles, Patrick J. and Arrasmith, Andrew},
  eprint={2108.10434},
  title = {Adaptive shot allocation for fast convergence in variational quantum algorithms},
  publisher = {arXiv},
  year = {2021},
  archivePrefix={arXiv},
  primaryClass={quant-ph}
}

@misc{ito2023latencyaware,
      title={Latency-aware adaptive shot allocation for run-time efficient variational quantum algorithms}, 
      author={Kosuke Ito},
      year={2023},
      eprint={2302.04422},
      archivePrefix={arXiv},
      primaryClass={quant-ph}
}

@article{stilck2021limitations,
	doi = {10.1038/s41567-021-01356-3},
	url = {https://doi.org/10.1038%2Fs41567-021-01356-3},
	year = 2021,
	month = {oct},
	publisher = {Springer Science and Business Media {LLC}},
	volume = {17},
	number = {11},
	pages = {1221--1227},
	author = {Daniel Stilck Fran{\c{c}}a and Raul Garc{\'{\i}}a-Patr{\'{o}}n},
	title = {Limitations of optimization algorithms on noisy quantum devices},
	journal = {Nat. Phys.}
}

@article{depalma2023limitations,
  title = {Limitations of Variational Quantum Algorithms: A Quantum Optimal Transport Approach},
  author = {De Palma, Giacomo and Marvian, Milad and Rouz\'e, Cambyse and Fran\ifmmode \mbox{\c{c}}\else \c{c}\fi{}a, Daniel Stilck},
  journal = {PRX Quantum},
  volume = {4},
  issue = {1},
  pages = {010309},
  numpages = {30},
  year = {2023},
  month = {Jan},
  publisher = {American Physical Society},
  doi = {10.1103/PRXQuantum.4.010309},
  url = {https://link.aps.org/doi/10.1103/PRXQuantum.4.010309}
}

@article{hu2023tackling,
  title = {Tackling Sampling Noise in Physical Systems for Machine Learning Applications: Fundamental Limits and Eigentasks},
  author = {Hu, Fangjun and Angelatos, Gerasimos and Khan, Saeed A. and Vives, Marti and T\"ureci, Esin and Bello, Leon and Rowlands, Graham E. and Ribeill, Guilhem J. and T\"ureci, Hakan E.},
  journal = {Phys. Rev. X},
  volume = {13},
  issue = {4},
  pages = {041020},
  numpages = {34},
  year = {2023},
  month = {Oct},
  publisher = {American Physical Society},
  doi = {10.1103/PhysRevX.13.041020},
  url = {https://link.aps.org/doi/10.1103/PhysRevX.13.041020}
}

@misc{liu2023stochastic,
      title={Stochastic noise can be helpful for variational quantum algorithms}, 
      author={Junyu Liu and Frederik Wilde and Antonio Anna Mele and Liang Jiang and Jens Eisert},
      year={2023},
      eprint={2210.06723},
      archivePrefix={arXiv},
      primaryClass={quant-ph}
}

@article{parigi2024quantum,
  title={Quantum-Noise-Driven Generative Diffusion Models},
  author={Parigi, Marco and Martina, Stefano and Caruso, Filippo},
  journal={Advanced Quantum Technologies},
  pages={2300401},
  year={2024},
  publisher={Wiley Online Library}
}

@article{Heyraud2022noisy,
  title = {Noisy quantum kernel machines},
  author = {Heyraud, Valentin and Li, Zejian and Denis, Zakari and Le Boit\'e, Alexandre and Ciuti, Cristiano},
  journal = {Phys. Rev. A},
  volume = {106},
  issue = {5},
  pages = {052421},
  numpages = {16},
  year = {2022},
  month = {Nov},
  publisher = {American Physical Society},
  doi = {10.1103/PhysRevA.106.052421},
  url = {https://link.aps.org/doi/10.1103/PhysRevA.106.052421}
}

@article{scala2023dropout,
  title={A General Approach to Dropout in Quantum Neural Networks},
  author={Scala, Francesco and Ceschini, Andrea and Panella, Massimo and Gerace, Dario},
  journal={Advanced Quantum Technologies},
  pages={2300220},
  year={2023},
  publisher={Wiley Online Library},
  url={https://advanced.onlinelibrary.wiley.com/doi/full/10.1002/qute.202300220}
}

@misc{oliveraatencio2025impactamplitudephasedamping,
      title={Impact of Amplitude and Phase Damping Noise on Quantum Reinforcement Learning: Challenges and Opportunities}, 
      author={María Laura Olivera-Atencio and Lucas Lamata and Jesús Casado-Pascual},
      year={2025},
      eprint={2503.24069},
      archivePrefix={arXiv},
      primaryClass={quant-ph},
      url={https://arxiv.org/abs/2503.24069}, 
}

@inproceedings{Pennington2018Fisher,
 author = {Pennington, Jeffrey and Worah, Pratik},
 booktitle = {Advances in Neural Information Processing Systems},
 editor = {S. Bengio and H. Wallach and H. Larochelle and K. Grauman and N. Cesa-Bianchi and R. Garnett},
 pages = {},
 publisher = {Curran Associates, Inc.},
 title = {The Spectrum of the Fisher Information Matrix of a Single-Hidden-Layer Neural Network},
 url = {https://proceedings.neurips.cc/paper_files/paper/2018/file/18bb68e2b38e4a8ce7cf4f6b2625768c-Paper.pdf},
 volume = {31},
 year = {2018}
}

@article{scandi2023quantum,
  title={Quantum Fisher Information and its dynamical nature},
  author={Scandi, Matteo and Abiuso, Paolo and Surace, Jacopo and De Santis, Dario},
  journal={Reports on Progress in Physics},
  year={2023}
}

@article{liu2020fisher,
doi = {10.1088/1751-8121/ab5d4d},
url = {https://dx.doi.org/10.1088/1751-8121/ab5d4d},
year = {2019},
month = {dec},
publisher = {IOP Publishing},
volume = {53},
number = {2},
pages = {023001},
author = {Jing Liu and Haidong Yuan and Xiao-Ming Lu and Xiaoguang Wang},
title = {Quantum Fisher information matrix and multiparameter estimation},
journal = {Journal of Physics A: Mathematical and Theoretical},
}

@article{meyer2021fisher,
  doi = {10.22331/q-2021-09-09-539},
  url = {https://doi.org/10.22331/q-2021-09-09-539},
  title = {Fisher {I}nformation in {N}oisy {I}ntermediate-{S}cale {Q}uantum {A}pplications},
  author = {Meyer, Johannes Jakob},
  journal = {{Quantum}},
  issn = {2521-327X},
  publisher = {{Verein zur F{\"{o}}rderung des Open Access Publizierens in den Quantenwissenschaften}},
  volume = {5},
  pages = {539},
  month = sep,
  year = {2021}
}

@article{liu2014fidelity,
title = {Fidelity susceptibility and quantum Fisher information for density operators with arbitrary ranks},
journal = {Physica A: Statistical Mechanics and its Applications},
volume = {410},
pages = {167-173},
year = {2014},
issn = {0378-4371},
doi = {https://doi.org/10.1016/j.physa.2014.05.028},
url = {https://www.sciencedirect.com/science/article/pii/S0378437114003926},
author = {Jing Liu and Heng-Na Xiong and Fei Song and Xiaoguang Wang},
}

@misc{pezze2025advancesmultiparameterquantumsensing,
      title={Advances in multiparameter quantum sensing and metrology}, 
      author={Luca Pezzè and Augusto Smerzi},
      year={2025},
      eprint={2502.17396},
      archivePrefix={arXiv},
      primaryClass={quant-ph},
      url={https://arxiv.org/abs/2502.17396}, 
}

@article{Gacon2021simultaneous,
  doi = {10.22331/q-2021-10-20-567},
  url = {https://doi.org/10.22331/q-2021-10-20-567},
  title = {Simultaneous {P}erturbation {S}tochastic {A}pproximation of the {Q}uantum {F}isher {I}nformation},
  author = {Gacon, Julien and Zoufal, Christa and Carleo, Giuseppe and Woerner, Stefan},
  journal = {{Quantum}},
  issn = {2521-327X},
  publisher = {{Verein zur F{\"{o}}rderung des Open Access Publizierens in den Quantenwissenschaften}},
  volume = {5},
  pages = {567},
  month = oct,
  year = {2021}
}

@misc{halla2025estimationQFI,
      title={Estimation of Quantum Fisher Information via Stein's Identity in Variational Quantum Algorithms}, 
      author={Mourad Halla},
      year={2025},
      eprint={2502.17231},
      archivePrefix={arXiv},
      primaryClass={quant-ph},
      url={https://arxiv.org/abs/2502.17231}, 
}

@misc{shi2025weightedapproxQNG,
      title={Weighted Approximate Quantum Natural Gradient for Variational Quantum Eigensolver}, 
      author={Chenyu Shi and Vedran Dunjko and Hao Wang},
      year={2025},
      eprint={2504.04932},
      archivePrefix={arXiv},
      primaryClass={quant-ph},
      url={https://arxiv.org/abs/2504.04932}, 
}

@article{albarelli2019evaluating,
  title={Evaluating the Holevo Cram{\'e}r-Rao bound for multiparameter quantum metrology},
  author={Albarelli, Francesco and Friel, Jamie F and Datta, Animesh},
  journal={Physical review letters},
  volume={123},
  number={20},
  pages={200503},
  year={2019},
  publisher={APS}
}

@book{horn1994matrix_analysis,
  title={Topics in matrix analysis},
  author={Horn, Roger A and Johnson, Charles R},
  year={1994},
  publisher={Cambridge university press}
}

@book{mohri2018foundations,
  title={Foundations of machine learning},
  author={Mohri, Mehryar and Rostamizadeh, Afshin and Talwalkar, Ameet},
  year={2018},
  publisher={MIT press}
}

@article{hawkins2004problem,
  title={The problem of overfitting},
  author={Hawkins, Douglas M},
  journal={Journal of chemical information and computer sciences},
  volume={44},
  number={1},
  pages={1--12},
  year={2004},
  publisher={ACS Publications}
}

@inproceedings{ying2019overview,
  title={An overview of overfitting and its solutions},
  author={Ying, Xue},
  booktitle={Journal of physics: Conference series},
  volume={1168},
  number={2},
  pages={022022},
  year={2019},
  organization={IOP Publishing}
}

@article{nakkiran2021deep,
  title={Deep double descent: Where bigger models and more data hurt},
  author={Nakkiran, Preetum and Kaplun, Gal and Bansal, Yamini and Yang, Tristan and Barak, Boaz and Sutskever, Ilya},
  journal={Journal of Statistical Mechanics: Theory and Experiment},
  volume={2021},
  number={12},
  pages={124003},
  year={2021},
  publisher={IOP Publishing}
}

@article{bishop1995training,
  title={Training with noise is equivalent to Tikhonov regularization},
  author={Bishop, Chris M},
  journal={Neural computation},
  volume={7},
  number={1},
  pages={108--116},
  year={1995},
  publisher={MIT Press}
}

@article{camuto2020explicit,
  title={Explicit regularisation in gaussian noise injections},
  author={Camuto, Alexander and Willetts, Matthew and Simsekli, Umut and Roberts, Stephen J and Holmes, Chris C},
  journal={Advances in Neural Information Processing Systems},
  volume={33},
  pages={16603--16614},
  year={2020}
}

@inproceedings{li2020adaptive,
  title={Adaptive Gaussian noise injection regularization for neural networks},
  author={Li, Yinan and Liu, Fang},
  booktitle={International Symposium on Neural Networks},
  pages={176--189},
  year={2020},
  organization={Springer}
}

@article{levi2022noise,
  title={Noise injection node regularization for robust learning},
  author={Levi, Noam and Bloch, Itay M and Freytsis, Marat and Volansky, Tomer},
  journal={arXiv preprint arXiv:2210.15764},
  year={2022}
}

@inproceedings{orvieto2022anticorrelated,
  title={Anticorrelated noise injection for improved generalization},
  author={Orvieto, Antonio and Kersting, Hans and Proske, Frank and Bach, Francis and Lucchi, Aurelien},
  booktitle={International Conference on Machine Learning},
  pages={17094--17116},
  year={2022},
  organization={PMLR}
}

@inproceedings{orvieto2023explicit,
  title={Explicit regularization in overparametrized models via noise injection},
  author={Orvieto, Antonio and Raj, Anant and Kersting, Hans and Bach, Francis},
  booktitle={International Conference on Artificial Intelligence and Statistics},
  pages={7265--7287},
  year={2023},
  organization={PMLR}
}

@inproceedings{wan2013dropconnect,
  title={Regularization of neural networks using dropconnect},
  author={Wan, Li and Zeiler, Matthew and Zhang, Sixin and Le Cun, Yann and Fergus, Rob},
  booktitle={International conference on machine learning},
  pages={1058--1066},
  year={2013},
  organization={PMLR}
}

@article{srivastava2014dropout,
  author  = {Nitish Srivastava and Geoffrey Hinton and Alex Krizhevsky and Ilya Sutskever and Ruslan Salakhutdinov},
  title   = {Dropout: A Simple Way to Prevent Neural Networks from Overfitting},
  journal = {Journal of Machine Learning Research},
  year    = {2014},
  volume  = {15},
  number  = {56},
  pages   = {1929--1958},
  url     = {http://jmlr.org/papers/v15/srivastava14a.html}
}

@inproceedings{smith2020generalization,
  title={On the generalization benefit of noise in stochastic gradient descent},
  author={Smith, Samuel and Elsen, Erich and De, Soham},
  booktitle={International Conference on Machine Learning},
  pages={9058--9067},
  year={2020},
  organization={PMLR}
}

@article{zhang2021understanding,
  title={Understanding deep learning (still) requires rethinking generalization},
  author={Zhang, Chiyuan and Bengio, Samy and Hardt, Moritz and Recht, Benjamin and Vinyals, Oriol},
  journal={Communications of the ACM},
  volume={64},
  number={3},
  pages={107--115},
  year={2021},
  publisher={ACM New York, NY, USA}
}

@inproceedings{sclocchi2023dissecting,
  title={Dissecting the effects of SGD noise in distinct regimes of deep learning},
  author={Sclocchi, Antonio and Geiger, Mario and Wyart, Matthieu},
  booktitle={International Conference on Machine Learning},
  pages={30381--30405},
  year={2023},
  organization={PMLR}
}

@article{Baldassi2021Unveiling,
title={Unveiling the structure of wide flat minima in neural networks},
author={Carlo Baldassi and Clarissa Lauditi and E. Malatesta and Gabriele Perugini and R. Zecchina},
journal={Physical review letters},
year={2021},
volume={127 27},
pages={ 278301 },
doi={10.1103/PhysRevLett.127.278301}
}

@article{Huang2020shadows,
   title={Predicting many properties of a quantum system from very few measurements},
   volume={16},
   ISSN={1745-2481},
   url={http://dx.doi.org/10.1038/s41567-020-0932-7},
   DOI={10.1038/s41567-020-0932-7},
   number={10},
   journal={Nature Physics},
   publisher={Springer Science and Business Media LLC},
   author={Huang, Hsin-Yuan and Kueng, Richard and Preskill, John},
   year={2020},
   month=jun, pages={1050–1057} }

@article{langford2002pac,
  title={PAC-Bayes \& margins},
  author={Langford, John and Shawe-Taylor, John},
  journal={Advances in neural information processing systems},
  volume={15},
  year={2002}
}

@article{dziugaite2017computing,
  title={Computing nonvacuous generalization bounds for deep (stochastic) neural networks with many more parameters than training data},
  author={Dziugaite, Gintare Karolina and Roy, Daniel M},
  journal={arXiv preprint arXiv:1703.11008},
  year={2017}
}

@article{rivasplata2020pac,
  title={PAC-Bayes analysis beyond the usual bounds},
  author={Rivasplata, Omar and Kuzborskij, Ilja and Szepesv{\'a}ri, Csaba and Shawe-Taylor, John},
  journal={Advances in Neural Information Processing Systems},
  volume={33},
  pages={16833--16845},
  year={2020}
}

@article{sklearn,
  title={Scikit-learn: Machine Learning in {P}ython},
  author={Pedregosa, F. and Varoquaux, G. and Gramfort, A. and Michel, V.
          and Thirion, B. and Grisel, O. and Blondel, M. and Prettenhofer, P.
          and Weiss, R. and Dubourg, V. and Vanderplas, J. and Passos, A. and
          Cournapeau, D. and Brucher, M. and Perrot, M. and Duchesnay, E.},
  journal={Journal of Machine Learning Research},
  volume={12},
  pages={2825--2830},
  year={2011}
}

@misc{pennylane,
  doi = {10.48550/ARXIV.1811.04968},
  url = {https://arxiv.org/abs/1811.04968},
  author = {Bergholm, Ville and Izaac, Josh and Schuld, Maria and others
            },
  title = {PennyLane: Automatic differentiation of hybrid quantum-classical computations},
  publisher = {arXiv},
  year = {2018},
  copyright = {arXiv.org perpetual, non-exclusive license}
}

@software{jax2018github,
  author = {James Bradbury and Roy Frostig and Peter Hawkins and Matthew James Johnson and Chris Leary and Dougal Maclaurin and George Necula and Adam Paszke and Jake Vander{P}las and Skye Wanderman-{M}ilne and Qiao Zhang},
  title = {{JAX}: composable transformations of {P}ython+{N}um{P}y programs},
  url = {http://github.com/google/jax},
  version = {0.3.13},
  year = {2018},
}

@article{kingma2014adam,
  title={Adam: A method for stochastic optimization},
  author={Kingma, Diederik P and Ba, Jimmy},
  journal={arXiv preprint arXiv:1412.6980},
  year={2014}
}

@misc{kempkes2025doubledescentquantummachine,
      title={Double descent in quantum kernel methods}, 
      author={Marie Kempkes and Aroosa Ijaz and Elies Gil-Fuster and Carlos Bravo-Prieto and Jakob Spiegelberg and Evert van Nieuwenburg and Vedran Dunjko},
      year={2025},
      eprint={2501.10077},
      archivePrefix={arXiv},
      primaryClass={quant-ph},
      url={https://arxiv.org/abs/2501.10077}, 
}

@article{garcia2025noise,
  title={Noise-enhanced quantum clocks and global field sensors},
  author={Garc{\'\i}a-Pintos, Luis Pedro},
  journal={arXiv preprint arXiv:2507.02071},
  year={2025}
}

@article{chen2024quantum,
  title={Quantum metrology enhanced by leveraging informative noise with error correction},
  author={Chen, Hongzhen and Chen, Yu and Liu, Jing and Miao, Zibo and Yuan, Haidong},
  journal={Physical Review Letters},
  volume={133},
  number={19},
  pages={190801},
  year={2024},
  publisher={APS}
}

@article{peng2024enhanced,
  title={Enhanced quantum metrology with non-phase-covariant noise},
  author={Peng, Jia-Xin and Zhu, Baiqiang and Zhang, Weiping and Zhang, Keye},
  journal={Physical Review Letters},
  volume={133},
  number={9},
  pages={090801},
  year={2024},
  publisher={APS}
}

@article{ait2025sculpting,
  title={Sculpting Quantum Landscapes: Fubini-Study Metric Conditioning for Geometry-Aware Learning in Parameterized Quantum Circuits},
  author={Ait Haddou, Marwan and Bennai, Mohamed},
  year={2025},
  journal = "Research Square preprint",
  url = {https://www.researchsquare.com/article/rs-7101280/v1},
  doi = {10.21203/rs.3.rs-7101280/v1},
}

@article{dirr2009lie,
  title={Lie-semigroup structures for reachability and control of open quantum systems: Kossakowski-Lindblad generators form Lie wedge to Markovian channels},
  author={Dirr, G and Helmke, U and Kurniawan, I and Schulte-Herbr{\"u}ggen, T},
  journal={Reports on Mathematical Physics},
  volume={64},
  number={1-2},
  pages={93--121},
  year={2009},
  publisher={Elsevier}
}

@article{lindblad1976GKSL,
  title={On the generators of quantum dynamical semigroups},
  author={Lindblad, Goran},
  journal={Communications in mathematical physics},
  volume={48},
  pages={119--130},
  year={1976},
  publisher={Springer}
}

@article{gorini197GKSL,
  title={Completely positive dynamical semigroups of N-level systems},
  author={Gorini, Vittorio and Kossakowski, Andrzej and Sudarshan, Ennackal Chandy George},
  journal={Journal of Mathematical Physics},
  volume={17},
  number={5},
  pages={821--825},
  year={1976},
  publisher={AIP Publishing}
}

@misc{githubrepo,
  author = {Scala, Francesco},
  title = {{Improving Quantum Neural Networks exploration by Noise-Induced Equalization}},
  howpublished = {\url{https://github.com/fran-scala/public-noise-induced}},
  year = {2025}
}

\newpage
\appendix
\onecolumngrid

\section{Generalization and overfitting}
\label{appendix:Gen and overfit}

In this appendix, we provide a concise introduction to generalization, the 
overfitting problem, and theoretical approaches to addressing these 
issues. This section is not meant to be exhaustive, but rather intended as 
a gentle outline for readers new to the subject.

In supervised machine learning, the goal is to learn a function that maps input data $\mathbf{x}_i\in\mathbb{R}^m$ to corresponding outputs $y_i$, by minimizing a loss function over a training dataset. However, evaluating the model solely on the training data is not sufficient: what ultimately matters is the model's performance on unseen data drawn from the same underlying distribution $\mathcal{D}$. This ability is referred in the field as \emph{generalization}, a crucial property for machine learning because in real-world applications, models will be exposed to new data not seen before~\cite{mohri2018foundations}. This motivates the division of data into a \emph{training set} and a \emph{test set}, where the training set is used to optimize the model parameters, and the test set serves as a proxy to estimate the so-called \emph{true risk}, defined as the expected loss over the entire distribution:
\begin{equation}
R(f) = \mathbb{E}_{(\mathbf{x}, y) \sim \mathcal{D}} \left[ \mathcal{L}(f(\mathbf{x}), y) \right].
\end{equation}
Since $\mathcal{D}$ is unknown, we estimate $R(f)$ using the \emph{empirical risk} on a finite sample:
\begin{equation}
R_S(f) = \frac{1}{M} \sum_{i=1}^M \mathcal{L}(f(\mathbf{x}_i), y_i),
\end{equation}
with $(\mathbf{x}_i, y_i)$ belonging either to the training set or the test set, depending on the context.

In the case of Mean Squared Error (MSE) loss, we can better understand generalization via a theoretical framework called \emph{bias-variance decomposition}. As a matter of fact, a central issue in machine learning is to reach a tradeoff between \emph{bias}, accounting for limitation of the learning algorithm, and \emph{variance}, accounting for sensitivity to fluctuations in the training data. For regression tasks, the expected squared error at a test point $\mathbf{x}$ can be decomposed as:
\begin{equation}
\mathbb{E}\left[ (f(\mathbf{x}) - y)^2 \right] = \underbrace{( \mathbb{E}[f(\mathbf{x})] - y )^2}_{\text{Bias}^2} + \underbrace{\mathbb{E}\left[ (f(\mathbf{x}) - \mathbb{E}[f(\mathbf{x})])^2 \right]}_{\text{Variance}} + \underbrace{\sigma^2}_{\text{Irreducible noise}},
\end{equation}
where the expectations are taken over the randomness in the training set. A highly complex model typically exhibits low bias but high variance, meaning it can closely fit the training data but may perform poorly on unseen data—a phenomenon known as \emph{overfitting}~\cite{hawkins2004problem, ying2019overview}.

Overfitting occurs when a model fits the training data too closely, including its noise or spurious patterns, rather than capturing the underlying structure of the data distribution. Expressive models are particularly susceptible to this. Model expressiveness increases with the number of parameters and the nonlinearity of the function class. For instance, in quantum machine learning, parameterized quantum circuits with many layers, entangling gates and data re-uploading may become increasingly expressive and prone to overfitting.

Classical machine learning model's capacity was studied in terms of the \emph{interpolation threshold}. It refers to the regime in which the model has enough parameters to perfectly fit (i.e., interpolate) the training data, i.e. $P=M$ where $P$ is the number of parameters in the model and $M$ is the training set size. Classical results suggest that generalization should degrade when $P>M$, but recent empirical and theoretical developments have shown that models can generalize well even beyond this threshold, a phenomenon called \emph{double-descent}~\cite{nakkiran2021deep}. Nonetheless, for classical machine learning, the interpolation threshold is often associated with the onset of overfitting, particularly when the dataset is small or noisy. For what concerns quantum models, quantum neural networks were shown to be unable to reach double-descent~\cite{du2023problem} while recently Kempkes et al.~\cite{kempkes2025doubledescentquantummachine} demonstrated that this is achievable in quantum kernel methods. For this reason, overfitting remains a longstanding challenge in QML, with ongoing research focusing on developing novel methods to mitigate or prevent it.

To theoretically approach the model's performances on unseen data, one can make use of \emph{generalization bounds}, which provide probabilistic guarantees on how close $R_S(f)$ is to $R(f)$ for a given model (hypotesis) class $\mathcal{F}$ formalizing the relationship between empirical and true risk. A typical form of such bounds is:
\begin{equation}
R(f) \leq R_S(f) + \mathcal{C}(\mathcal{F}, M, \delta),
\end{equation}
which holds with high probability $1 - \delta$, where $\mathcal{C}$ is a complexity term that depends on the richness of $\mathcal{F}$, the number of training examples $M$, and the confidence level $\delta$. It is worth highlighting that the function $\mathcal{C}$ approaches $0$ as the number of samples $M$ tends to infinity.

One way to quantify the complexity of a hypothesis class is through the \emph{Rademacher complexity}, which measures how well functions in $\mathcal{F}$ can fit random noise~\cite{mohri2018foundations}. Given a sample $S = \{(\mathbf{x}_i, y_i)\}_{i=1}^M$, the empirical Rademacher complexity is defined as:
\begin{equation}
\hat{\mathfrak{R}}_S(\mathcal{F}) = \mathbb{E}_{\boldsymbol{\sigma}} \left[ \sup_{f \in \mathcal{F}} \frac{1}{M} \sum_{i=1}^M \sigma_i f(\mathbf{x}_i) \right],
\end{equation}
where $\sigma_i \in \{-1, +1\}$ are independent Rademacher variables. Intuitively, a high Rademacher complexity indicates a model class capable of fitting arbitrary labels, suggesting a high risk of overfitting. We point out that the bound in Eq.~\eqref{eq:gen_bound} belongs to the Rademacher complexity type.

In both classical and quantum settings, bounding the generalization gap via Rademacher complexity or related tools (such as VC dimension or covering numbers~\cite{mohri2018foundations}) is common practice for designing architectures that generalize well, even if this bounds have been shown to be somewhat vacuous~\cite{zhang2021understanding}. 
In this work, we study the effect of quantum noise on generalization performances and refer to the generalization bound given in Ref.~\cite{Khanal2025_data_JSuperC}, showing that, in contrast with our apporoach, this does not allow to accurately estimate a good noisy operating regime.

\section{Noise channels}
\label{appendix:noise chan}

A useful tool allowing us to describe the evolution of quantum mechanical systems is the quantum operation (channel) formalism~\cite{nielsen00}, which is particularly effective for characterizing quantum noise sources.
Here, we summarize the essential features of depolarizing, phase damping, and amplitude damping noise, which are the key types of quantum noise channels employed in the following of this work.

Depolarizing noise arises from random unitary rotations on the quantum state. This type of noise tends to isotropically reduce the coherence of the quantum state, effectively spreading the information uniformly across all possible outcomes, and reducing the original state to the completely mixed state, i.e., $\rho_{\mathrm{mix}}=\mathds{1}/2^d$. For a single qubit, this type of quantum noise can be mathematically described by the depolarizing channel: 
\begin{align}
    \mathcal{N}_{depol}(\rho,p)&= (1-p)\rho + \frac{p}{3}\left(X\rho X+Y\rho Y+Z\rho Z\right) \nonumber\\ 
    \label{eq:depolarizing}
    & =p\frac{\mathds{1}}{2}+(1-p)\rho \, ,
\end{align}
which is quantitatively characterized by the probability that a depolarizing event occurs, $p$. On the Bloch sphere, depolarizing noise can be visualized as the isotropic shrinking of the sphere towards the origin.

Phase damping (or dephasing) noise, on the other hand, is associated with the random introduction of phase errors in the quantum state. Unlike depolarizing noise, phase damping preserves the amplitude information but disrupts the relative phase relationships between different components of the quantum state. The dephasing channel, describing the action of the homonym noise, can be represented with a Kraus notation for a single qubit as:
\begin{align}
    \label{eq:dephasing1}
    &\mathcal{N}_{deph}(\rho,p)= \sum_i E_{i, deph} \rho E_{i, deph}^\dagger \quad , \quad \text{where} \\ 
    & E_{0, deph}=    \begin{bmatrix}
                1 & 0\\
                0 & \sqrt{1-p}
              \end{bmatrix} 
    \quad , \quad  
    E_{1, deph}=    \begin{bmatrix}
                0 & 0\\
                0 & \sqrt{p}
              \end{bmatrix} \, , \nonumber
\end{align}
with the degree of dephasing determined by the probability parameter, $p$. In Eq.~\eqref{eq:dephasing1}, $E_0$ leaves the state  $\ket{0}$ unchanged, but it reduces the amplitude of state $\ket{1}$, while $E_1$ destroys $\ket{0}$ and reduces the amplitude of state $\ket{1}$. Dephasing noise can be visualized as shrinking the Bloch sphere into an ellipsoid, in which the $z$-axis is left unchanged while the other two axes are contracted. 

Amplitude noise, also defined amplitude damping channel, represents a different facet of quantum noise. This noise source involves the loss of amplitude information, leading to the decay of the quantum state. The amplitude-damping channel is particularly relevant in describing processes in which the quantum system interacts with its environment, causing the loss of energy and coherence. For a single qubit, it can be described by the following Kraus decomposition:
\begin{align}
    \label{eq:amplitude1}
    &\mathcal{N}_{amp}(\rho,p)= \sum_i E_{i, amp} \rho E_{i, amp}^\dagger \quad , \quad \text{where} \\ 
    & E_{0, amp}=    \begin{bmatrix}
                1 & 0\\
                0 & \sqrt{1-p}
              \end{bmatrix} 
    \quad , \quad  
    E_{1, amp}=    \begin{bmatrix}
                0 & \sqrt{p}\\
                0 & 0
              \end{bmatrix} \, . \nonumber
\end{align}
From Eq.~\eqref{eq:amplitude1} it is possible to understand why amplitude noise is associated to energy loss: $E_1$ in Eq.~\eqref{eq:amplitude1} turns the state $\ket{0}$ into $\ket{1}$, which corresponds to the process of losing energy due to the interaction with an environment; $E_0$ leaves the state  $\ket{0}$ unchanged, but it reduces the amplitude of state $\ket{1}$, as for the dephasing noise.

In this work, we will consider noisy quantum learning models where noise acts after each quantum gate, be it a single qubit gate or an entangling gate (in this latter case, two single qubit noise channels act on the considered qubits). As all these noise channels depend on the parameter $p$, we will refer to it as the noise level, thus assuming all the noise channels to be characterized by the same parameter $p$, with obvious generalization.

\section{Growth of the Dynamical Lie Algebra in Noisy Quantum Circuits}
\label{appendix:noise dla}

In this appendix, we analyze how quantum noise can affect the dynamics of a quantum system through the Dynamical Lie Algebra (DLA) associated with the quantum circuit of interest and potentially lead to the noise-induced equalization (NIE). 
Different effects may occur depending on whether there is commutation or not between Hamiltonian dynamics and noise-induced dissipative generators and they can be rigorously understood by studying the evolution under the Lindblad master equation and applying the Baker-Campbell-Hausdorff (BCH) expansion~\cite{dirr2009lie}. We also discuss how the dimension of the DLA relates to the accessible directions in the system’s evolution.

\subsection*{The Dynamical Lie Algebra in the Noiseless Case}

In the absence of noise, a closed quantum system evolves under the Schrödinger equation. When the evolution is driven by a finite set of time-independent Hamiltonians $\{H_j\}$, the time evolution operators (quantum gates) take the form
\begin{equation}
U_j = e^{-i H_j t_j}, \quad j = 1, 2, \dots
\end{equation}
The Dynamical Lie Algebra $\mathfrak{g}$ is defined as the smallest Lie algebra closed under commutators and containing the skew-Hermitian generators $i H_j$. It determines the set of effective Hamiltonians that can be synthesized through combinations of the available gates.
If the generators $iH_j$ do not commute, then their products generate additional directions in $\mathfrak{g}$ through the BCH formula:
\begin{equation}
e^A e^B = e^{A + B + \frac{1}{2}[A,B] + \frac{1}{12}[A,[A,B]] - \frac{1}{12}[B,[A,B]] + \cdots} \quad .
\end{equation}
The nested commutators imply that the reachable set of unitaries expands beyond the span of the original Hamiltonians, depending on the algebraic structure of their commutators.
The \emph{dimension of the DLA} corresponds to the number of linearly independent skew-Hermitian operators generated from $\{iH_j\}$ and their nested commutators. 
Each independent direction in this algebra represents a possible trajectory in the system’s unitary evolution space. 
For a $n$-qubit system, the maximum DLA is $\mathfrak{su}(2^n)$, which has dimension $4^n - 1$; 
a lower-dimensional DLA means limited controllability and expressiveness.

In the unitary case, the evolution of a quantum state is restricted to the unitary orbit of the initial pure state:
\begin{equation}
\rho(t) = U(t) \rho(0) U^\dagger(t) \quad, 
\end{equation}
where $U=\prod_j U_j$.
This evolution preserves the eigenvalues of $\rho$, so the reachable set is confined to a lower-dimensional manifold.
The number of independent real parameters in a pure state (modulo global phase) is:
\begin{equation}
\dim_{\mathbb{R}}(\mathbb{CP}^{2^n - 1}) = 2^{n+1} - 2.
\end{equation}
This is far smaller than the DLA dimension $4^{n}-1$, highlighting that the DLA describes the possible dynamics, not the static configuration space.

\subsection*{Open-System Evolution: The Lindblad Master Equation}

When the system interacts with an environment, the dynamics are no longer unitary. Instead, the time evolution of the density matrix $\rho$ is governed by the Lindblad master equation~\cite{gorini197GKSL, lindblad1976GKSL}:
\begin{equation}
\frac{d\rho}{dt} = -i [H, \rho] + \sum_k \left( L_k \rho L_k^\dagger - \frac{1}{2} \left\{ L_k^\dagger L_k, \rho \right\} \right) \quad .
\end{equation}
Here, $H$ is the system Hamiltonian, and the Lindblad operators $L_k$ describe the dissipative interaction with the environment (e.g., depolarizing, dephasing, amplitude damping). The solution of this equation for time-independent generators is given by:
\begin{equation}
\rho(t) = e^{\mathcal{L} t}[\rho(0)]
\end{equation}
where $\mathcal{L}$ is the Liouvillian superoperator, which acts linearly on the space of density matrices:
\begin{equation}
\mathcal{L}[\rho] = \mathcal{L}_H+ \mathcal{L}_\mathcal{D} = -i [H, \rho] + \sum_k \left( L_k \rho L_k^\dagger - \frac{1}{2} \left\{ L_k^\dagger L_k, \rho \right\} \right)
\end{equation}
The Liouvillian defines a semigroup of completely positive, trace-preserving maps. While the dynamics are no longer represented by Lie groups of unitaries, the structure of $\mathcal{L}$ still allows for algebraic analysis via Lindbladian algebras, which extend the concept of DLAs to open systems.
Unlike unitary evolution, the Lindbladian can change the eigenvalues of $\rho$, enabling transitions from pure to mixed states and expanding the reachable set of states beyond unitary orbits~\cite{dirr2009lie}. As a matter of fact, the space $\mathfrak{D}_{2^n}$ of all density matrices (trace-one, positive semidefinite $2^n\times 2^n$  matrices) has real dimension 
\begin{equation}
\dim_{\mathbb{R}}(\mathfrak{D}_{2^n}) = 4^n - 1\quad ,
\end{equation} 
matching that of $\mathfrak{su}(2^n)$, including both pure and mixed states.

\subsection*{Dynamical Lie Algebra Growth Induced by Noise}

In the noisy setting, new generators emerge from the dissipative Lindblad terms. 
When the noise-induced generators $\mathcal{L}_\mathcal{D}$ do not commute with the original Hamiltonian generators $\mathcal{L}_H$, the algebra of effective dynamical generators expands through their nested commutators. This is analogous to the noiseless case, where consecutive quantum gates correspond to a product of exponentials of non-commuting generators and are combined via the Baker-Campbell-Hausdorff (BCH) formula. Specifically, if we consider two noisy gates modeled by superoperators $\mathcal{L}_1$ and $\mathcal{L}_2$, their consecutive application corresponds to:
\begin{equation}
e^{\mathcal{L}_1 t_1} e^{\mathcal{L}_2 t_2} = \exp\left( t_1 \mathcal{L}_1 + t_2 \mathcal{L}_2 + \frac{t_1 t_2}{2} [\mathcal{L}_1, \mathcal{L}_2] + \cdots \right),
\end{equation}
where we used the BCH expansion.
The DLA is then generated by both the Hamiltonian part and the dissipative part of the Liouvillian:
\begin{equation}
\mathfrak{g}_{\text{noisy}} = \text{Lie} \left( \{iH_j\} \cup \{ \mathcal{D}_k \} \right)
\end{equation}
where $\mathcal{D}_k$ denotes the superoperators associated with the dissipators:
\begin{equation}
\mathcal{D}_k[\rho] := L_k \rho L_k^\dagger - \frac{1}{2} \left\{ L_k^\dagger L_k, \rho \right\}
\end{equation}
At this stage, two cases may arise: the generators of the unitary dynamics either commute with the noise superoperators or they do not.
Commutation occurs, for instance, when the jump operators are eigenoperators of the Hamiltonian’s adjoint action, that is, $[H, L_k] = \omega_k L_k$, $\omega_k \in \mathbb{R}$,
which includes the special case $[H, L_k] = 0$ for all $k$. In the next section, we will show analytically tractable toy examples where, in the case of $[H, L] = 0$, NIE takes place.
On the other hand, the non-commutativity between Hamiltonian and noise superoperators allows for additional directions to emerge through commutators such as $[\mathcal{L}_H, \mathcal{D}_k]$, $[\mathcal{D}_k, \mathcal{D}_l]$, etc. This leads to a growth of the DLA, enriching the space of reachable operations.
The generators now include non-Hermitian and non-unitary elements, acting on the space of operators. The effective DLA becomes a subset of the space of superoperators on Hermitian matrices, which has real dimension:
\begin{equation}
\dim_{\mathbb{R}}(\text{End}(\mathfrak{su}(2^n))) = (4^n - 1)^2 \quad ,
\end{equation}
where the notation $\text{End}(\mathfrak{su}(2^n))$ stands for the space of endomorphisms, i.e. superoperators that map traceless skew-Hermitian operators to other such operators.

As in the unitary case, the dimension of the noisy DLA reflects the number of independent directions in the Liouvillian evolution space. A higher-dimensional DLA implies that the system can explore a larger portion of the operator or state space, potentially enabling more complex transformations, even in the presence of decoherence. In some cases, noise can paradoxically \emph{enhance controllability}, allowing the system to reach dynamical regimes that would not be accessible with Hamiltonian evolution alone. We argue that the noise-induced equalization stems from the balancing of the dissipative dynamics and the increased controllability provided by quantum noise.

\section{Analytical toy demonstrations of NIE}
\label{appendix:analytical toys}

In this appendix, leveraging the insights from 
Ref.~\cite{garcia2025noise}, we provide two illustrative examples that demonstrate the onset of noise-induced equalization (NIE). The derivation relies on the assumption that the generators of the unitary dynamics commute with the noise superoperators. 
The first example investigates a system with a single tunable parameter, revealing the existence of an optimal level of noise for the NIE. 
Furthermore, we compute the eigenvalues of the QFIM for systems with multiple variational parameters analytically uncovering the equalization process.

\subsection{State Evolution under Noise: $[H, L] = 0$}

We start by analyzing the time evolution of a quantum state subject to decoherence in the case where the system Hamiltonian $H$ and the Lindblad operator $L$ commute, i.e. $[H, L] = 0$. 
This commutation relation ensures that $H$ and $L$ share a common eigenbasis, which we denote by $\{\ket{E_0}, \ket{E_1}\}$, with corresponding eigenvalues $E_j$ and $\ell_j$
\begin{equation}
    H\ket{E_j}=E_j\ket{E_j},\quad L\ket{E_j}=\ell_j\ket{E_j} \, .
\end{equation}
We focus on an initial state given by the nontrivial superposition of $\ket{E_0}$, $\ket{E_1}$, a two-level state, whose density operator is
\begin{equation}
  \rho(0)=\ketbra{\psi(0)}{\psi(0)}
           = \tfrac12\bigl(\ketbra{E_0}{E_0}+\ketbra{E_1}{E_1}+\ketbra{E_0}{E_1}+\ketbra{E_1}{E_0}\bigr).
\end{equation}
Under unitary evolution $U(t)=e^{-iHt}$ and pure dephasing via $L$ at rate $\gamma$, the state at time $t$ becomes
\begin{align}
  \rho(t)&=e^{\mathcal{L}t}\rho(0)=e^{(-i[H,\cdot]-\frac{1}{2}\gamma \{L^\dagger L,\cdot\}+\gamma L\cdot L^\dagger )t}\rho(0)=\\
  &=\tfrac12\Bigl(\ketbra{E_0}{E_0}+\ketbra{E_1}{E_1}
  + e^{-i\Delta E\,t}e^{-\Gamma(t)}\ketbra{E_0}{E_1}
  + e^{i\Delta E\,t}e^{-\Gamma(t)}\ketbra{E_1}{E_0}\Bigr),
\end{align}
where
$ \Delta E=E_1-E_0$, and $ \Gamma(t)=\tfrac{\gamma t}{2}(\ell_1-\ell_0)^2\, . $

We now analyze the spectral properties of the time-evolved density matrix $\rho(t)$.
Its normalized eigenstates can be written as
\begin{equation}
\label{eq:normalized eigenstates}
  \ket{\phi_\pm(t)} =
  \frac{1}{\sqrt{2}}\Bigl(e^{-i\frac{\Delta E}{2}t}\ket{E_0}
  \pm e^{i\frac{\Delta E}{2}t}\ket{E_1}\Bigr),
\end{equation}
which represent, respectively, the in-phase ($+$) and out-of-phase ($-$) superpositions of the energy eigenstates, each evolving with opposite phase factors due to the energy splitting $\Delta E$.
With this basis, one finds that $\rho(t)$ has two nonzero eigenvalues given by
\begin{equation}
  \lambda_{\pm}(t) = \tfrac{1}{2}\bigl(1 \pm |r(t)|\bigr)
  = \tfrac{1}{2}\bigl(1 \pm e^{-\Gamma(t)}\bigr),
\end{equation}
where, for convenience, we introduced the complex decoherence factor
$
  r(t) = e^{-i\Delta E\,t} e^{-\Gamma(t)},
$
encoding both the unitary phase evolution due to the energy difference $\Delta E = E_1 - E_0$ and the exponential damping induced by the dephasing rate $\Gamma(t)$.

To investigate the parameter dependence of these quantities, we compute their derivatives, as these will be needed for deriving the Quantum Fisher Information (QFI). 
Differentiating the eigenvalues with respect to time yields
\begin{align*}
  \partial_t \lambda_{\pm}
  &= \pm \tfrac{1}{2}(-\Gamma') e^{-\Gamma}
   = \mp \tfrac{1}{2}\Bigl(\tfrac{\gamma}{2}(\ell_1 - \ell_0)^2\Bigr)e^{-\Gamma(t)},
\end{align*}
where $\Gamma' = d\Gamma/dt = \tfrac{\gamma}{2}(\ell_1 - \ell_0)^2$.
Similarly, differentiating the eigenstates gives
\begin{align*}
  \partial_t \ket{\phi_\pm(t)}
  = \frac{\pm i \Delta E}{2}\ket{\phi_\mp(t)},
\end{align*}
showing that the instantaneous rate of change of each eigenstate is proportional to
the energy splitting and points along the orthogonal superposition.

The QFI for single parameter $t$ is of a generic density matrix $\rho=\sum_k \lambda_k\ketbra{\psi_k}{\psi_k}$ is given by the following expression
\begin{align}
\mathcal{F}(t) &= \sum_{\substack{k\\\lambda_k\neq0}} \left[\frac{\partial_t \lambda_k }{\lambda_k} +4\lambda_k\langle  \partial_t\psi_k| \partial_t \psi_k  \rangle\right]- \sum_{\substack{k,l\\\lambda_k,\lambda_l\neq0}} \frac{8 \lambda_k \lambda_l}{\lambda_k + \lambda_l} |\langle  \psi_k| \partial_t \psi_l  \rangle |^2 \, .
\end{align} 
Plugging in the evolved state $\rho(t)$ yields
\begin{equation}
  \mathcal{F}(t)=\frac{(\Gamma'e^{-\Gamma})^2}{1-e^{-2\Gamma}}+(\Delta E)^2e^{-2\Gamma}=\left(\frac{\Gamma'^2}{1-e^{-2\Gamma}}+(\Delta E)^2\right)e^{-2\Gamma}
.
\end{equation}
In the limit $\gamma\to0$, $\Gamma\to0$, one recovers the noiseless value $\mathcal{F}(0)=(\Delta E)^2$. 
Here, we can notice that the presence of noise may enhance parameter sensitivity. 
This enhancement originates from the interplay between the coherent phase evolution and the decay
of off-diagonal terms: 
dephasing redistributes information between populations and coherences,
and the derivative of the damping factor $\Gamma'(t)$ contributes a positive term to the QFI.
Physically, this means that \emph{moderate noise} can increase the rate at which the state changes with
respect to $t$. 
However, for \emph{strong noise}, the exponential suppression $e^{-2\Gamma}$ dominates, leading to the expected decay of $\mathcal{F}(t)$.

At this point, we can try to find the optimal value of $\gamma$ maximazing the QFI in the limit of $\gamma\ll 1$. In order to do that we need to rewrite $\mathcal{F}$ as:
\begin{align}
    \mathcal{F}(t)&=\left(\frac{\Gamma'^2}{1-e^{-2\Gamma}}+(\Delta E)^2\right)e^{-2\Gamma}=\\
        &= \left(\frac{\gamma^2 A^2}{1-e^{-2\gamma A t}}+(\Delta E)^2\right)e^{-2\gamma A t}
\end{align}
where $\Delta \ell = (\ell_1-\ell_0)$ and $A=\tfrac{\Delta\ell^2}{2}$. Expanding to first order in the noise rate $\gamma$, the QFI takes the approximate form:
\begin{align}
    \mathcal{F}(t)&\approx \left(\frac{\gamma^2 A^2}{2\gamma A t}+(\Delta E)^2\right)(1-2\gamma A t)=\\
    &= -A^2 \gamma^2 +\left(\frac{A}{2t}-2(\Delta E)^2 A t\right)\gamma+(\Delta E)^2
\end{align}
which is a downward parabola in $\gamma$. Now, we can find the coordinate of the maximum $\gamma^*$ by imposing $d\mathcal{F}/d\gamma=0$
\begin{equation}
    \gamma^* = -\frac{\frac{A}{2t}-2(\Delta E)^2 A t}{-2A^2}=\frac{1-4(\Delta E)^2  t^2}{4At}=\frac{1-4(\Delta E)^2  t^2}{2\Delta\ell^2t} \, .
\end{equation}
Here we can see that $\gamma^*$ depends on the spectrum of the Hamiltonian generator ($\Delta E$), the spectrum of the noise generator ($\Delta \ell$) and the parameter ($t$). 
This could mean that in the context of QML where we have many parameters (even if some generators are shared), if the parameters $\theta$ are different, the optimal noise level would be, in general, different. 
We must come up with a collective measure that takes into account everything. 

The quadratic dependence on $\Delta E$ of the negative term might explain why equalization kills high eigenvalues first and help low ones: $\gamma^*$ for high eigenvalues would be too small, or even negative (not physically achievable).

\subsection{Multi-parameter Hamiltonian and State}

We now switch to a context more similar to the one of QML, where we have many different parameters associated with multiple generators. We again consider a two-level system with orthonormal eigenstates $\ket{E_0}$ and $\ket{E_1}$ of a family of commuting Hamiltonians $\{H_j\}_{j=1}^M$, and a Lindblad operator $L$ that also commutes with each $H_j$:
\begin{equation}
  H_j\ket{E_k}=E_k^{(j)}\ket{E_k},
  \quad L\ket{E_k}=\ell_k\ket{E_k},
  \quad [H_j,L]=0,\ \ \forall j,k.
\end{equation}
Given the multi-parameter vector $\mathbf t = (t_1\dots t_M)$We then define the multi-parameter Hamiltonian $H(\mathbf t)=\sum_{j=1}^M t_j\,H_j,$
and the initial state
\begin{equation}
  \rho(0)=\ketbra{\psi(0)}{\psi(0)}
           = \tfrac12\bigl(\ketbra{E_0}{E_0}+\ketbra{E_1}{E_1}+\ketbra{E_0}{E_1}+\ketbra{E_1}{E_0}\bigr).
\end{equation}

Proceeding in similarly to the what done for the single parameter case, the evolution of the initial state is given by
\begin{equation}
  \rho(\mathbf t)=e^{\sum_j\left(-i t_j[H_j,\cdot]+\gamma t_j\mathcal D\right)}\rho(0).
\end{equation}
with  $\mathcal{D}[\rho]=L\rho L-\tfrac12\{L^2,\rho\}$, leading to
\begin{equation}
  \rho(\mathbf t)=\frac12\Bigl(\ketbra{E_0}{E_0}+\ketbra{E_1}{E_1}
  +e^{-i\Delta(\mathbf t)}e^{-\Gamma(t)}\ketbra{E_0}{E_1}
  +\text{h.c.}\Bigr),
\end{equation}
where
\begin{equation}
  \Delta(\mathbf t)=\sum_{j=1}^M t_j\,\delta E^{(j)},
  \quad \delta E^{(j)}=E_1^{(j)}-E_0^{(j)}, \quad \Gamma(t)=\tfrac{\gamma t}{2}(\ell_1-\ell_0)^2, \quad t=\sum_j t_j.
\end{equation}
After generalizing the normalized eigenbasis defined in Eq.~\eqref{eq:normalized eigenstates} substituting $\Delta E$ with $\Delta(\mathbf t)$, we need to compute derivatives with respect to each parameter $t_j$. First, we define the following quantities for convenience
\begin{align}
  \partial_{t_j}\Delta=\delta E^{(j)},
  \quad\quad \partial_{t_j}\Gamma=\tfrac{\gamma}{2}(\ell_1-\ell_0)^2=\tfrac{\gamma}{2}\Delta\ell ^2=\Gamma', \quad\quad \partial_{t_j}r=\left(-i\delta E^{(j)}+\Gamma'\right)r
\end{align}
where $\Gamma'=\partial_t\Gamma$. Then for the eigenvalues we obtain
\begin{align}
  \partial_{t_j}\lambda_{\pm}&=\pm\tfrac12 \Re\bigl(r^*\partial_tr\bigr)
   =\mp\tfrac12\Gamma'e^{-\Gamma(t)} \, ,
\end{align}
and for the eigenstates:
\begin{equation}
  \partial_{t_j}\ket{\phi_\pm}=\frac{\pm i\delta E^{(j)}}{2}\ket{\phi_\mp}.
\end{equation}
The multi-parameter QFI matrix (QFIM) is given by
\begin{equation}
  \mathcal{F}_{ij}(\mathbf t) = \sum_{\substack{k\\\lambda_k\neq0}} \left[\frac{(\partial_i \lambda_k )(\partial_j \lambda_k)}{\lambda_k} +4\lambda_k\mathrm{Re}\left\{\langle  \partial_i\psi_k| \partial_j \psi_k  \rangle\right\}\right]- \sum_{\substack{k,l\\\lambda_k,\lambda_l\neq0}} \frac{8 \lambda_k \lambda_l}{\lambda_k + \lambda_l} \mathrm{Re}\left\{\langle \partial_i \psi_l | \psi_k \rangle \langle  \psi_k| \partial_j \psi_l  \rangle \right\} \, .
\end{equation}
This leads to a QFIM with the following elements:
\begin{equation}
  \mathcal{F}_{ij}=\left(\delta E^{(i)}\delta E^{(j)}+\frac{\Gamma'^2}{1-e^{-2\Gamma}}\right)e^{-2\Gamma}
\end{equation}
where on the diagonal we retrieve the single parameter case.
We now consider the approximate QFIM for $\gamma \ll 1$
\begin{equation}
\mathcal{F}\;\approx\;(1-2\Gamma)\,\delta\,\delta^{T}+\alpha\,u\,u^{T},
\end{equation}
with
$
\delta_{i}=\delta E^{(i)},
\quad u=(1\dots 1),
\quad \Gamma=\tfrac{\gamma t}{2}(\Delta\ell)^{2}\ll1,
\quad \alpha=\frac{\Gamma'^{2}}{1-e^{-2\Gamma}}\approx\frac{\gamma(\Delta\ell)^{2}}{4t}.
$
Here $\delta\,\delta^{T}$ and $u\,u^{T}$ are rank-1 matrices, so $F$ has at most two nonzero eigenvalues while the remaining $M-2$ are zero. Recall that since $H_j$ commute wiwth each other in the noiseless setting, the QFIM would have rank 1, already implying that noise is transforming one zero eigenvalue to a non-zero one. 
The nonzero eigenvalues $\lambda_{1},\lambda_{2}$ satisfy
$
\lambda_{1}+\lambda_{2}=\Tr\mathcal{F}=T,$ and $\lambda_{1}^{2}+\lambda_{2}^{2}=\Tr\mathcal{F}^{2}=S$. Then we can write
\begin{equation}
\mathcal{F}^2=(1-2\Gamma) \delta\delta^T \delta\delta^T+2(1-2\Gamma)\alpha\delta\delta^ T uu ^T +\alpha ^2uu^Tuu^T
\end{equation}
by defining the scalars
$$
a=\delta^{T}\delta=\sum_{i}(\delta E^{(i)})^{2},
\quad b=\delta^{T}u=\sum_{i}\delta E^{(i)},
\quad c=u^{T}u=M.
$$
Computing $T$ and $S$ allows to obtain the product $\lambda_{1}\lambda_{2}=D$ with which we can write the simple quadratic equation $\lambda^{2}-T\lambda+D=0$:
\begin{align*}
T&= (1-2\Gamma)\,\mathrm{Tr}(\delta\delta^{T}) + \alpha\,\mathrm{Tr}(u\,u^{T}) = (1-2\Gamma)\,a + \alpha\,c,\\[6pt]
S&= (1-2\Gamma)^{2}\,\mathrm{Tr}(\delta\delta^{T}\delta\delta^{T}) +2(1-2\Gamma)\alpha\,\mathrm{Tr}(\delta\delta^{T}u\,u^{T}) + \alpha^{2}\,\mathrm{Tr}(u\,u^{T}u\,u^{T}) \\
  &= (1-2\Gamma)^{2}a^{2} + 2(1-2\Gamma)\alpha\,b^{2} + \alpha^{2}c^{2}.
\end{align*}
Then
$D=\tfrac12\bigl[T^{2}-S\bigr]
  =(1-2\Gamma)\,\alpha\,(a\,c-b^{2})$
and
\begin{equation*}
\lambda_{\pm}=\frac{T\pm\sqrt{T^{2}-4D}}{2}
= \frac{(1-2\Gamma)\,a+\alpha\,c \;\pm\; \sqrt{(1-2\Gamma)^{2}a^{2} -2(1-2\Gamma)\alpha(a\,c-2
b^{2}) +\alpha^{2}c^{2}}}{2}\,.
\end{equation*}
We now expand to first order in the small quantities $\Gamma$ and $\alpha$ (i.e.\ $\Gamma,\alpha\sim \mathcal{O}(\gamma)\ll1$) neglecting all second order terms (like $\Gamma^2, \, \alpha^2, \, \alpha\Gamma$). 
Let $
\Delta_Q = T^{2}-4D$
\begin{equation*}
T^{2}= (a-2\Gamma a+\alpha c)^{2}
\approx a^{2} -4\Gamma a^{2} +2\alpha c a\,,\quad
4D \approx 4\alpha(a\,c-b^{2}),
\end{equation*}
hence
\begin{equation*}
\Delta_Q=T^{2}-4D 
\approx 
a^{2} -4\Gamma a^{2} -2\alpha a c +4\alpha b^{2}.
\end{equation*}
Thus since $\sqrt{1-x}\approx1-\tfrac{x}{2}$ for $x\ll1$ we obtain
$
\sqrt{\Delta_Q}
\approx a\sqrt{1 -4\Gamma -2\alpha\tfrac{c}{a} +4\alpha\tfrac{b^{2}}{a^{2}}}
\approx a\Bigl(1 -2\Gamma -\alpha\tfrac{c}{a} +2\alpha\tfrac{b^{2}}{a^{2}}\Bigr),
$
leading to the following eigenvalues expressions
\begin{align*}
\lambda_{+}&\approx\tfrac12\bigl[(a-2\Gamma a+\alpha c)+a(1-2\Gamma-\alpha\tfrac{c}{a}+2\alpha\tfrac{b^2}{a^2})\bigr]\\
&=a-2\Gamma a+\alpha\tfrac{b^2}{a},\\
\lambda_{-}&\approx\tfrac12\bigl[(a-2\Gamma a+\alpha c)-a(1-2\Gamma-\alpha\tfrac{c}{a}+2\alpha\tfrac{b^2}{a^2})\bigr]\\
&=\alpha\Bigl(c-\tfrac{b^2}{a}\Bigr).
\end{align*}
So in conclusion we have
\begin{align}
\lambda_{+}&\approx (1-\gamma t \Delta \ell^2)\sum_i(\delta E^{(i)})^2 + \frac{\gamma \Delta\ell^2}{4t} \frac{\left(\sum_i\delta E^{(i)}\right)^2}{\sum_i(\delta E^{(i)})^2}, \\
\lambda_{-}&\approx \frac{\gamma \Delta\ell^2}{4t}\left(M-\frac{\left(\sum_i\delta E^{(i)}\right)^2}{\sum_i(\delta E^{(i)})^2}\right),
\end{align}
and all other eigenvalues remain zero. Here is interesting to notice that the principal QFI mode ($\lambda_+$) starts at $\sum_i(\delta E^{(i)})^2$ when $\gamma=0$, then is suppressed by $\gamma t(\Delta\ell)^2$, but partly rescued by a “noise‐induced” boost.
The second non-zero eigenvalue ($\lambda_-$) is \emph{purely noise‐induced}, it vanishes for $\gamma=0$ and grows linearly in $\gamma$. 
This eigenvalue can also vanish if all the $\delta E^{(i)}$ are the same, i.e. all the generators $H_j$ are the same, since the ratio $\tfrac{\left(\sum_i\delta E^{(i)}\right)^2}{\sum_i(\delta E^{(i)})^2}$ can be related to a signal to noise ratio:
\begin{equation}
    \frac{\left(\sum_i\delta E^{(i)}\right)^2}{\sum_i(\delta E^{(i)})^2}=M\left(\frac{1}{\tfrac{\mathrm{Var}[\delta E]}{\mathbb{E}[\delta E]^2}+1}\right)\, ,
\end{equation}
this implies that if there is no variance in the $\delta E^{(i)}$ this ratio will just reduce to $M$ yielding $\lambda_-=0$. 
Instead, if some variability is allowed, the ratio will always be smaller than $M$.
The applied approximations also come with some drawback: the analytical expressions for $\lambda_\pm$ are linear in $\gamma$, hiding the possibility to have an optimal noise level for the equalization process and not showing the decay of $\lambda_-$.

In this simple setting we were able to derive the noise-induced equalization: noise enable a zero eigenvalue to become non-zero whereas the highest eigenvalue is damped. Avoiding neglecting the higher order terms, one could arrive at a cumbersome equation with also $\gamma^2$ and find the best noise level.

\section{Explicit generalization bound}
\label{appendix:explicit bound}

In this appendix, we state an adapted version of the theorem given in Ref.~\cite{Khanal2025_data_JSuperC} and then briefly present the derivation of our rewriting of the generalization bound.

\begin{theorem*}[Adapted from Ref.~\cite{Khanal2025_data_JSuperC}]
    Let $P,M\in \mathbb{N}$, $\delta\in[0,1)$ and $D=\{x_i, y_i\}_{i=1}^M$ an i.i.d. collection of data samples and target labels coming from the distribution $\mathcal{D}=\mathcal{X}\times \mathcal{Y}$. 
    Consider a $P$-dimensional parameter space $\Theta\subset\mathbb{R}^P$ and a class of quantum machine learning model $\mathcal{M}_\Theta=\{f_{\theta, p}(x):\theta\in\Theta\}$ subject to quantum noise of intensity $p\in[0,1)$. Assuming that:
    \begin{itemize}
        \item [-] The single samples loss $l:\mathcal{Y}\times\mathbb{R}\rightarrow[0,1]$ is Lipschitz continuous in its second argument with constant $0<L<1$.
        \item [-] The gradient of the model $\nabla_\theta f_{\theta,p}(x)$ is bounded by the Lipschitz constant $L_f$: $\|\nabla_\theta f_{\theta,p}(x)\|\leq L_f$, i.e. the model is Lipschitz continuous w.r.t the parameters $\theta$.
        \item [-] Let $\mathcal{F}(\theta)$ denote the quantum Fisher Information Matrix associated with the model. Suppose there exists $m > 0$ such that: $\sqrt{\det (\mathcal{F}(\theta))} \geq m>0 $ $\forall \theta \in \Theta$.
    \end{itemize}
    Then, for any $\delta >0$, with probability at least $1-\delta$ over the random draw of the i.i.d training set $D$, the following generalization bound holds uniformly for all $\theta\in\Theta$:
    \begin{equation}
        \label{eq:gen_bound}
        R(\theta)-R_S(\theta)\leq \frac{24\pi}{\sqrt{M}}B(p)+3\sqrt{\frac{\log(2/\delta)}{2M}} \, ,
    \end{equation}
    where $R(\theta)=\mathbb{E}_{(x,y)\sim \mathcal{D}}[l(y,f_{\theta,p}(x)]$ is the expected risk, $R_S(\theta)=\frac{1}{M}\sum_{i=1}^Ml(y,f_{\theta,p}(x))$ is the empirical risk and
    \begin{equation}
        \label{eq:noise term gen bound}
        B(p)=\sqrt{d_{eff}}\left[\Gamma\left(\frac{d_{eff}}{2}+1\right)\frac{1}{m}\right]^{1/d_{eff}}L_f \, ,
    \end{equation}
    is a term taking into account the effects of quantum noise with $\Gamma(\cdot)$ being the gamma function. In particular, the noise dependence is given by $d_{eff}$, $m$ and $L_f$.
\end{theorem*}

The original generalization bound is of the following form:
\begin{equation}
        \label{eq:gen_bound_original}
        R(\theta)-R_S(\theta)\leq \frac{12\sqrt{\pi d_{eff}}\exp{\left(\frac{C'}{d_{eff}}\right)}}{\sqrt{M}}+3\sqrt{\frac{\log(2/\delta)}{2M}} \, ,
\end{equation}
with 
\begin{equation}
C'=\log \left(\frac{V_\Theta L_f^{d_{eff}}}{V_{d_{eff}}m}\right)
\end{equation}
where $V_\Theta$ is the volume of the parameter space $\Theta$ and $V_{d_{eff}}$ is the volume of a unit ball in $\mathbb{R}^{d_eff}$. Also in this version of the generalization bound we stress the dependence on $d_{eff}$ instead of the total number of parameters $P$. This is motivated by two facts:
\begin{itemize}
    \item even in a noiseless setting, the effective dimension of a quantum model is different from the number of parameters (see overparametrization);
    \item noise can change the effective role of parameters via NIE.
\end{itemize}
The definition of 
\begin{equation}
    B(p)=\sqrt{d_{eff}}\left[\Gamma\left(\frac{d_{eff}}{2}+1\right)\frac{1}{m}\right]^{1/d_{eff}}L_f \, ,
\end{equation}
follows from giving the explicit form of $V_\Theta$, $V_{d_{eff}}$ in
\begin{equation}
    2\sqrt{\pi}B(p)=\sqrt{d_{eff}}\exp{\left(\frac{C'}{d_{eff}}\right)} \, .
\end{equation} 
In what follows we will use the notation $d=d_{eff}$ for brevity:
\begin{align}
    &V_\Theta = (2\pi)^d \quad , \quad V_{d} = \frac{\pi^{d/2}}{\Gamma\left(\frac{d}{2}+1\right)}\\
    &\frac{V_\Theta}{V_{d}}=2^d\pi^{d/2}\Gamma\left(\frac{d}{2}+1\right)
\end{align}
then
\begin{align}
    2\sqrt{\pi}B(p)&=\sqrt{d}\exp{\left(\frac{\log \left(2^d\pi^{d/2}\Gamma\left(\frac{d}{2}+1\right)\frac{L_f^{d}}{m}\right)}{d_{eff}}\right)} =\nonumber \\
    &= \sqrt{d} \left(2^d\pi^{d/2}\Gamma\left(\frac{d}{2}+1\right)\frac{L_f^{d}}{m}\right)^{1/d} =\nonumber \\
    &= 2\sqrt{\pi}\sqrt{d} \left(\Gamma\left(\frac{d}{2}+1\right)\frac{1}{m}\right)^{1/d}L_f 
\end{align} 

\section{Datasets}
\label{appendix:datasets}
\begin{figure*}
    \centering
    \includegraphics[width=0.325\linewidth]{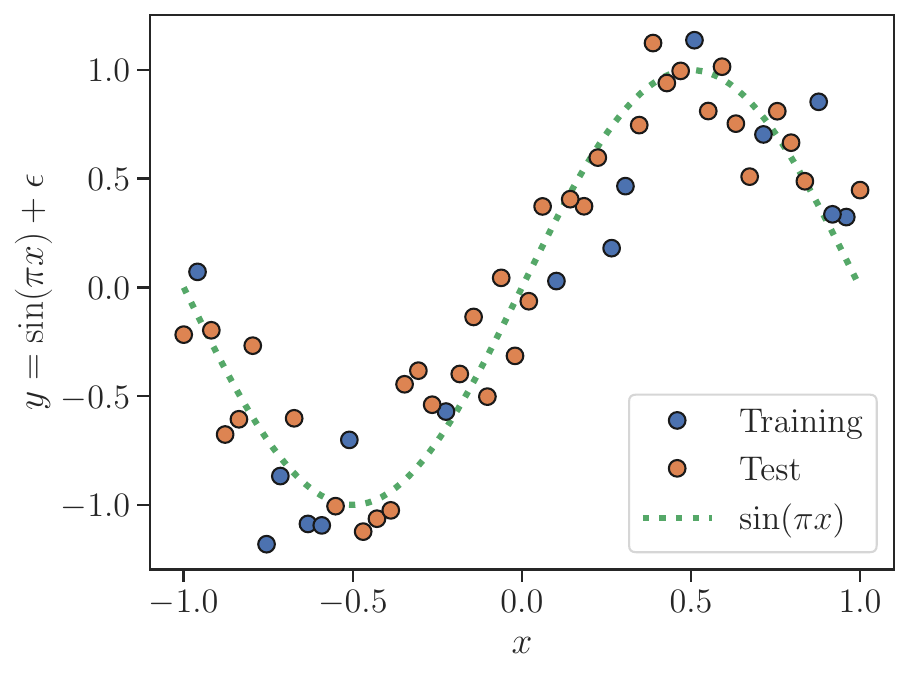}
    \includegraphics[trim={1.cm .cm .cm .cm}, clip,width=0.305\linewidth]{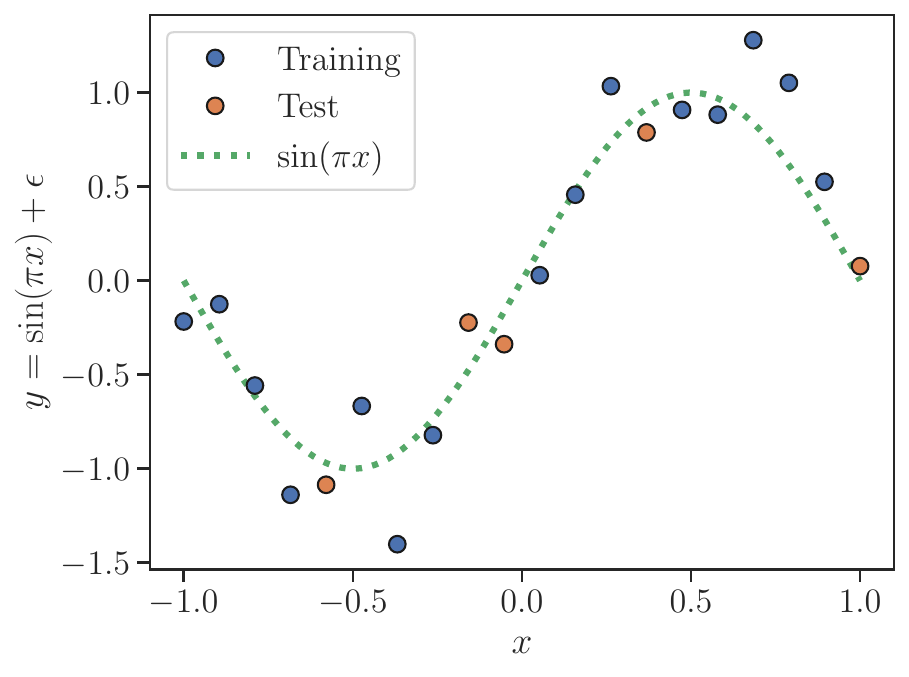}
    \includegraphics[width=0.325\linewidth]{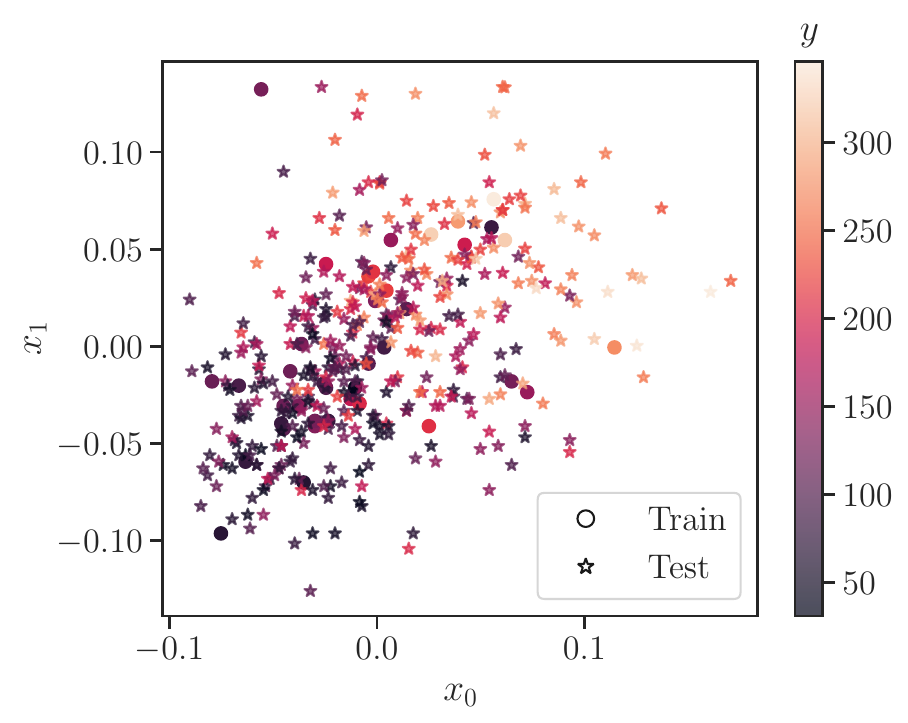}   
    \caption{Datasets: sinusoidal, sinusoidal 2, diabetes. These are raw data before classical preprocessing.}
    \label{fig:datasets}
\end{figure*}
In this section, we briefly describe the datasets under study which we also schematically report in Fig.~\ref{fig:datasets}. The first dataset analysed is a synthetic sinusoidal dataset. In particular, we generate two different datasets: the first one is composed of 50 points drawn with uniform probability in the interval $[-1,1]$ and then divided into 30\% training and 70\% test samples (sinusoidal), while the second one has 20 samples divided into 75\% training and 25\% test (sinusoidal2). The analytical expression describing the label that we assign to these points is the following:
\begin{equation}
    y = \sin(\pi x) + \epsilon\,,
\end{equation}
where $x \in [-1,1]$ and $\epsilon$ is an additive white Gaussian noise with amplitude equal to 0.4, zero mean and a standard deviation of 0.5. In order to properly fit the function, the $y$ variable is rescaled with a \texttt{MinMaxScaler} fitted on training data only to span the range $[-1,1]$.

The second dataset we tackle is a well-known benchmark dataset provided by Scikit-learn~\cite{sklearn}, with real medical data related to diabetes. It consists of physiological variables measured in patients, which are used to predict a quantitative measure of diabetes progression one year after baseline. It contains ten features, including age, sex, body mass index (BMI), blood pressure, total serum cholesterol, low-density lipoproteins, high-density lipoproteins, total cholesterol to HDL ratio, log of serum triglycerides level (LTG), and glucose level. The target variable represents a numerical value indicating the progression of diabetes. In this case, only BMI and LTG are used as input features. Then, the dataset is divided into 40 train and 400 test samples. Input features are rescaled to fit the range of angles of rotation gates, i.e. $[-\pi,\pi]$, with a \texttt{MinMaxScaler} fitted on training data only. Analogously, the target variable is rescaled within $[-1,1]$.

\section{Quantum Neural Network models}
\label{appendix: qnn models}

In this Appendix, we provide a detailed description of the quantum neural network (QNN) architectures employed in our study. The QNN employed to analyse the sinusoidal dataset is the same as in Ref.~\cite{scala2023dropout}, while for the diabetes dataset, we employ the same model as Ref.~\cite{somogyi2024niose_reg} to show that our procedure is capable of predicting the best noise level in agreement with previous results. 

The first QNN model consists of five qubits, all initialized in the computational $\ket{0}$ state. The classical features are encoded through two layers of single-qubit rotations $R_Y$ and $R_Z$. Since the dataset consists of single-feature data, all qubits encode the same value. The trainable part of the circuit is composed of three sublayers of single-qubit rotations, $R_X$, $R_Z$, and $R_X$, each followed by a sequence of CNOT gates that linearly entangle all qubits. With this elementary layer, we build an underparameterized QNN with $L=4$ layers and an overparameterized model with $L=10$ layers, resulting in $P=60$ and $P=150$ trainable parameters, respectively. The output of both models is the expectation value of the $Z$ Pauli operator on the first qubit.

The second QNN is used to study the diabetes dataset. The model consists of four qubits, also initialized in the computational $\ket{0}$ state. The encoding process applies two RX gates to the first and third qubits, embedding two classical features into the quantum state. The subsequent variational structure consists of a layer of single-qubit $R_Y$ gates and a ring of symmetric $RXX$ Ising gates that establish entanglement. We alternate this structure $L=3$ and $L=5$ times to obtain an underparameterized and an overparameterized QNN, respectively. The output of both models is the expectation value of the $Z^{\otimes 4}$ operator.

In Appendix~\ref{appendix:additional exp}, we cross-validate the architectures on the other dataset.

\section{Optimal noise level for diabetes dataset}
\label{appendix:diabetes dataset}

\begin{figure*}
    \centering
    \includegraphics[width=\linewidth]{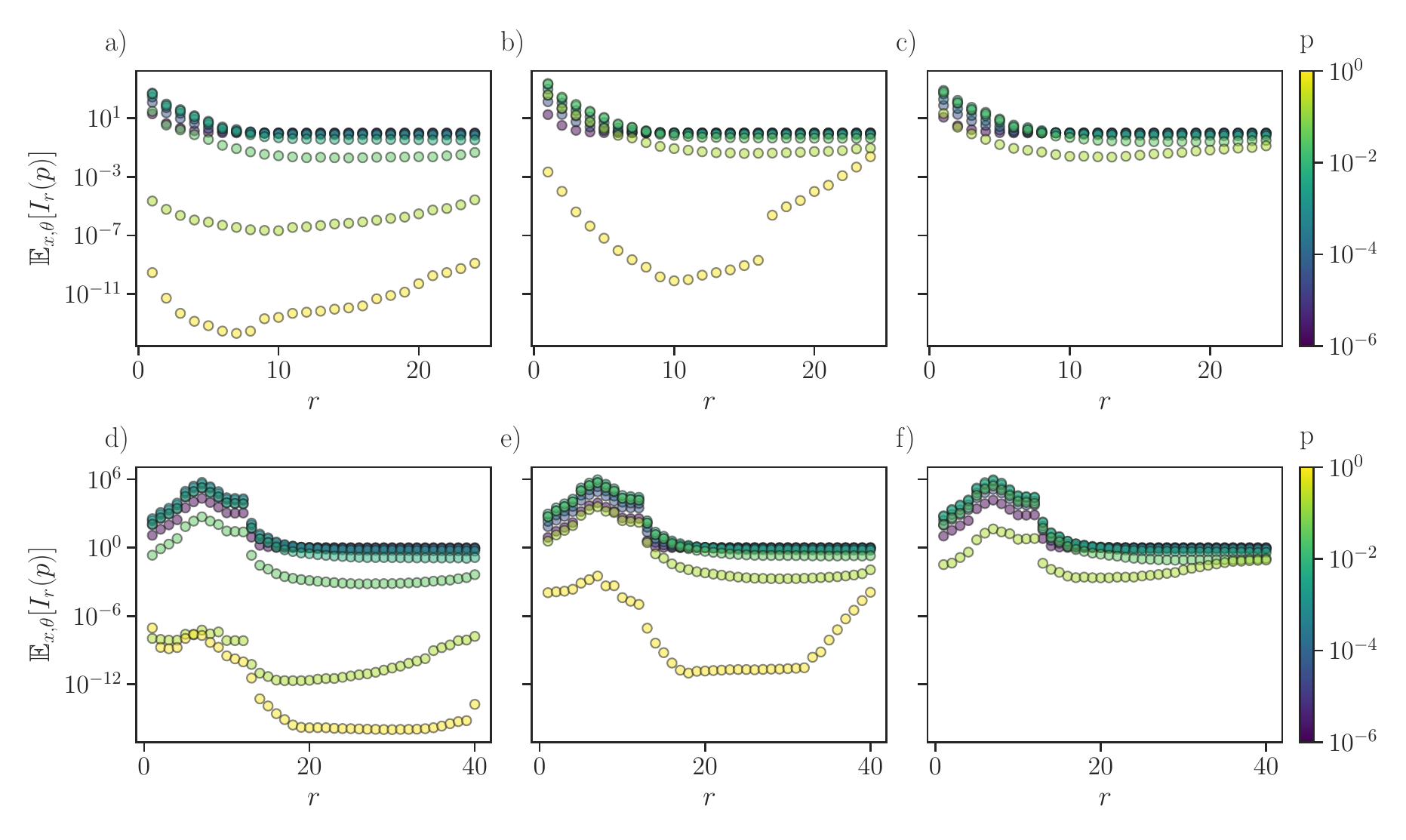}
    \caption{Relative change of the averaged QFIM eigenvalues $\lambda_m$ under different levels of noise $p$ with respect to the noiseless case ($p=0$) for a), d) depolarizing, b), e) dephasing and c), f) amplitude-damping noise. The first row represents the relative change for an underprametrized model ($24$ parameters), while the second for an overparameterized one ($40$ parameters).
    The average is computed over all the inputs of the training set (diabetes dataset) and 5 different parameter vectors.}
    \label{fig:relative change diabetes}
\end{figure*}

\begin{figure*}
    \centering
    \includegraphics[trim={.5cm .5cm .5cm .5cm},clip,width=0.95\linewidth]{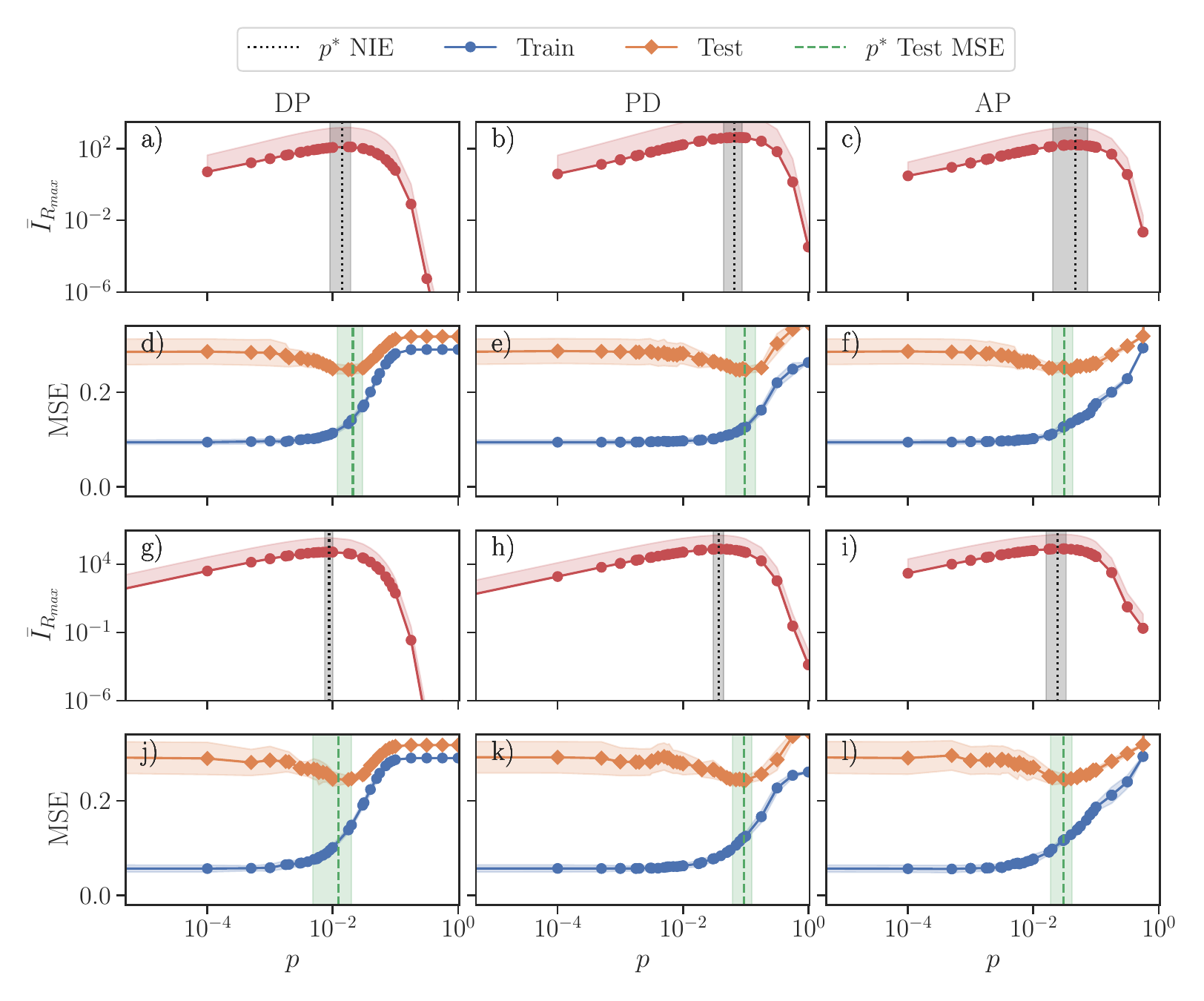}
    \caption{\textbf{Diabetes} Comparison of optimal noise level $p^*$ obtained from NIE and test MSE on diabetes dataset. Columns correspond to different types of noise: depolarizing (DP), phase damping (PD) and amplitude damping (AD). First and second rows refer to the underparameterized model, while third and fourth to the overparameterized one. Shaded regions represent one standard deviation above and below the mean, computed over 10 independent runs, and are plotted throughout, though not always visible (except for $\bar{I}_{R_{\max}}$, where only the upper standard deviation is shown for clarity).}
    \label{fig:generalization p* diabetes}
\end{figure*}

\begin{table}[]
\centering
\begin{tabular}{l|c|c|c|}
    \cline{2-4}
       & \multicolumn{1}{c|}{DP} & \multicolumn{1}{c|}{PD} & \multicolumn{1}{c|}{AD} \\ \hline
    \multicolumn{1}{|l|}{NIE}                                              & $(8.8 \pm 1.3)10^{-3}$                   & $(3.76 \pm 0.74)10^{-2}$                   & $(2.45 \pm 0.84)10^{-2}$         \\ \hline
    \multicolumn{1}{|l|}{Test MSE}                                         & $(1.24 \pm 0.76)10^{-2}$                  & $(9.38 \pm 3.20)10^{-2}$                  & $(3.01 \pm 1.12)10^{-2}$                 \\ \hline
    \multicolumn{1}{|l|}{Test MSE~\cite{somogyi2024niose_reg}} & 0.010                   & 0.056                   & 0.018                   \\ \hline
    \multicolumn{1}{|l|}{$B(p)$}                                           & $(1.78 \pm 0.00)10^{-1}$                       & $(1.90 \pm 0.39)10^{-1}$                      & $(6.97 \pm 0.19)10^{-2}$                      \\ \hline
    \multicolumn{1}{|l|}{Gen. gap}                                         & $(5.38 \pm 0.74)10^{-1}$    & $(3.66 \pm 0.98)10^{-1}$       & $(5.62 \pm 0.00)10^{-1}$   \\ \hline
\end{tabular}

\caption{Comparison of different values of $p^*$ for depolarizing (DP), phase damping (PD) and amplitude damping (AD) for diabetes dataset estimated with different methods for overparameterized QNN. }
    \label{tab:diabetes over p*}
\begin{tabular}{l|c|c|c|}
\cline{2-4}
    & \multicolumn{1}{c|}{DP} & \multicolumn{1}{c|}{PD} & \multicolumn{1}{c|}{AD} \\ \hline
\multicolumn{1}{|l|}{NIE} 
    & $(1.41 \pm 0.51)10^{-2}$ 
    & $(6.62 \pm 2.20)10^{-2}$ 
    & $(4.72 \pm 2.68)10^{-2}$ \\ \hline
\multicolumn{1}{|l|}{Test MSE} 
    & $(2.10 \pm 0.91)10^{-2}$ 
    & $(9.60 \pm 4.73)10^{-2}$ 
    & $(3.10 \pm 1.14)10^{-2}$ \\ \hline
\multicolumn{1}{|l|}{$B(p)$} 
    & $(3.71 \pm 1.86)10^{-1}$ 
    & $(4.04 \pm 1.31)10^{-1}$ 
    & $(1.00 \pm 0.00)10^{-1}$ \\ \hline
\multicolumn{1}{|l|}{Gen. gap} 
    & $(7.81 \pm 2.19)10^{-1}$ 
    & $(6.25 \pm 3.25)10^{-1}$ 
    & $(5.62 \pm 0.00)10^{-1}$ \\ \hline
\end{tabular}

\caption{Comparison of different values of $p^*$ for depolarizing (DP), phase damping (PD) and amplitude damping (AD) for diabetes dataset estimated with different methods for underparameterized QNN. }
    \label{tab:diabetes under p*}
\end{table}
In this section, we show that we are capable of approximating the optimal level of noise found in Ref.~\cite{somogyi2024niose_reg} and we extend the analysis also to underparameterized QNNs with the same architecture. We study the diabetes dataset with the second QNN model described in Appendix~\ref{appendix: qnn models}.
The analysis of the NIE is presented in Fig.~\ref{fig:relative change diabetes}, while results concerning the estimation of the optimal noise level $p^*$ are reported in Fig.~\ref{fig:generalization p* diabetes} and Tabs.~\ref{tab:diabetes over p*}-\ref{tab:diabetes under p*}.
The NIE can be appreciated via the analysis of $I_r(p)$ also for this second architecture and dataset. A non-trivial increase for the least important eigenvalues is observed at non-zero noise levels for both underparameterized and overparameterized models under the action of different kinds of quantum noise. Notice that Fig.~\ref{fig:generalization p* diabetes}a-b-c-i are missing the noise level $p=10^{-6}$ due to numerical issues arising when computing the QFIM.
Moving to the optimal noise level estimation, we can notice a good agreement between the two values found with the NIE-based procedure and the MSE. The only case in which we find a slight mismatch is for the overparameterized QNN when phase damping noise is present. This might be only a fluctuation due to the particular training-test splitting as the value of $p^*$ given by the NIE-based estimation is close to the noise level found in Ref.~\cite{somogyi2024niose_reg} while the MSE estimation is quite far (see values in Tab.~\ref{tab:diabetes over p*}).

\section{Additional numerical experiments}
\label{appendix:additional exp}
\begin{figure*}
    \centering
    \includegraphics[trim={0cm 1.0cm 0.cm 0.2cm}, clip,width=0.45\linewidth]{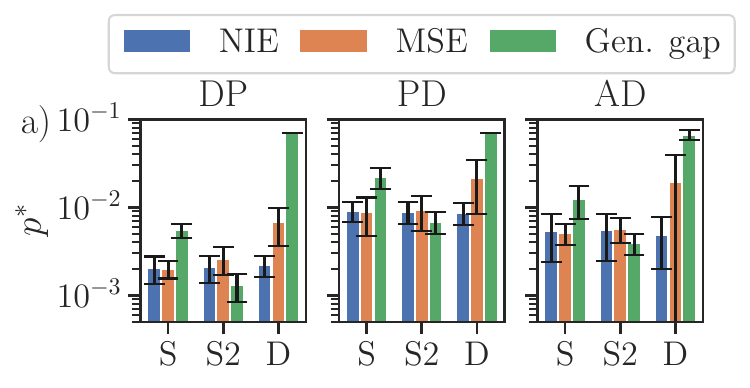}
    \includegraphics[trim={0cm 1.0cm 0.cm 0.2cm}, clip,width=0.45\linewidth]{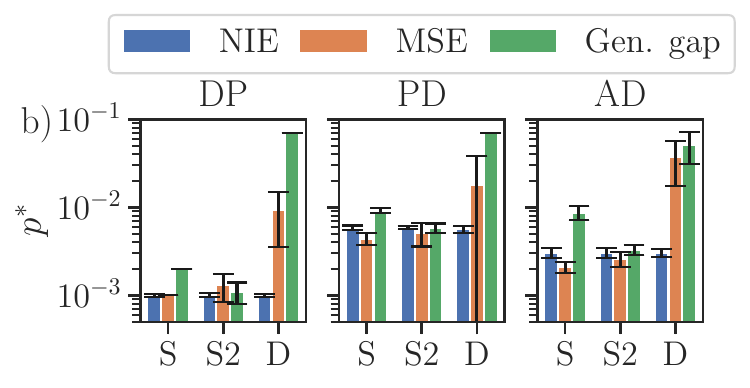}
    \includegraphics[trim={0cm 0.0cm 0.cm 1.3cm}, clip,width=0.45\linewidth]{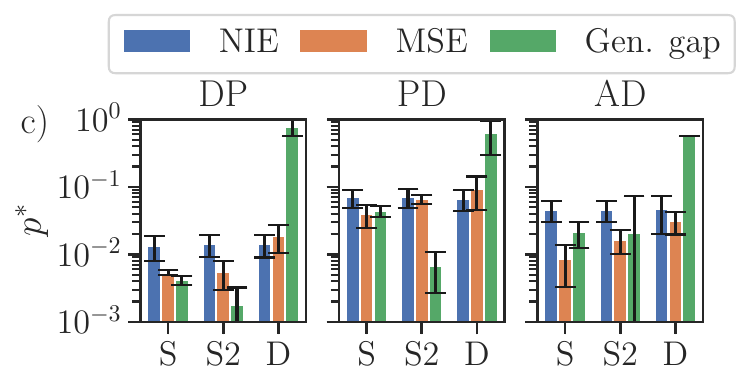}
    \includegraphics[trim={0cm 0.0cm 0.cm 1.3cm}, clip,width=0.45\linewidth]{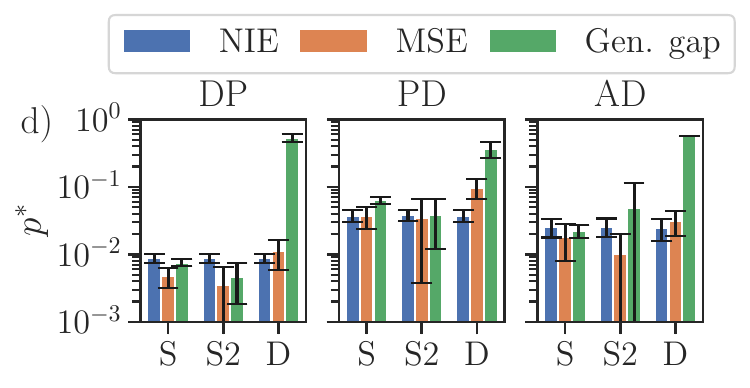}
    \caption{Comparison of different values of $p^*$ for depolarizing (DP), phase damping (PD) and amplitude damping (AD) for sinusoidal (S, S2) and diabetes (D) datasets estimated with different  methods and QNN architectures: a) first QNN underparameterized, b) first QNN overparameterized, c) second QNN underparameterized, d) second QNN overparameterized. The errorbars are one standard deviation above and below the mean, computed over 10 independent runs.}
    \label{fig:comp opt p}
\end{figure*}
In Fig.~\ref{fig:comp opt p}, we summarize the results of the cross-validation of architectures and datasets when estimating $p^*$ for different types of quantum noise and different estimation methods. 
In particular, we compare first (first row) and second (second row) QNNs in both under- (left column) and over-parameterized (right column) regimes on sinusoidal and diabetes datasets. 
The NID estimation appears almost always consistent with the MSE estimation. A substantial discrepancy is observed only for the overparameterized version of the first QNN on the diabetes dataset, highlighting that our pre-training analysis based on averaged quantities on the optimization landscape may fail to find the best regularizing regime. A local approach could maybe lead to better performance in such cases. For what concerns the estimation via generalization gap, as already seen, this is in general not a good estimation method.

\section{Numerical analysis of the generalization bound}
\label{appendix:numerical genbound}

We now report a numerical analysis for the generalization bound. In order to estimate $p^*$ from the generalization bound reported in Eq.~\eqref{eq:gen_bound}, we focus on the noise-dependent term $B(p)$ (see Eq.~\eqref{eq:noise term gen bound}). In particular, we also study the noise dependence of the single components affected by quantum noise, namely the Lipschitz constant of the quantum model $L_f$, the (square root of the) determinant of the QFIM and the effective dimension ($d_{eff}$) of the the quantum model. In particular, we remind the definition of $B(p)$
$$B(p)=\sqrt{d_{eff}}\left[\Gamma\left(\frac{d_{eff}}{2}+1\right)\frac{1}{m}\right]^{1/d_{eff}}L_f \, .$$
This quantity is evaluated for 5 random parameter vectors per each one of the $M$ training samples (for a total of $5M$) as a function of noise $p$. Specifically, in Figs.~\ref{fig:gen bound sin 4 layer}-~\ref{fig:gen bound diabetes 5 layer}, we plot $B(p)$ in panels a,~e,~i, the Lipschitz constant in panels b,~f,~j, the square root of the determinant of the QFIM in panels c,~g,~k and the effective dimensions in panel d,~h,~l.

The Lipschitz constant $L_f$ is estimated as the maximal gradient over different samplings (5 random parameter vectors per each one of the $M$ training samples) given that 
\begin{equation}
    \|\nabla_\theta f_{\theta,p}(x)\|\leq L_f\implies \forall p \quad L_f\approx \max_{x,\theta} \|\nabla_\theta f_{\theta,p}(x)\| \, .
\end{equation}
This estimation of the Lipschitz function is quite loose, a more accurate estimate would require exponentially many samples. This highlights how our method is more applicable and less demanding in terms of computational resources.

For what concerns the determinant of the QFIM $\mathcal{F}$, we compute the full eigenspectrum, and then we take the product of all the eigenvalues:
\begin{equation}
    \det{\mathcal{F}(x, \theta, p)}=\prod_{r=1}^P \lambda_r \, .
\end{equation}
Since the QFIM depends on the specific input, parameter vector and noise level ($\mathcal{F}=\mathcal{F}(x,\theta,p)$), to analyze its behaviour as a function of the noise only, we compute its average and standard deviation with respect to our finite sampling (5 random parameter vectors per each one of the $M$ training samples). It is worth pointing out that we numerically see many of the eigenvalues being smaller than $1$. Consequently, for models with many parameters, this leads to determinants extremely close to $0$ or considered equivalent to $0$ numerically. 
In such situations, it will be impossible to estimate the generalization bound of Eq.~\eqref{eq:gen_bound}, as this has an inverse dependence on the square root of the determinant, implying a diverging bound. This is represented by red areas in the Figs.~\ref{fig:gen bound sin 4 layer}-~\ref{fig:gen bound diabetes 5 layer}.

For what concerns the effective dimension $d_{eff}$, it is determined as the number of non-trivial direction in the parameter space. This is measured in terms of the number of non-zero eigenvalues of the QFIM, i.e. its rank. In particular, the numerical precision in this case is the one set by the numerical precision of the machine ($\epsilon=2.22\cdot 10^{-16}$) times the total number of parameters in the model $P$. 

\begin{figure*}
    \centering
    \includegraphics[trim={0cm 1.0cm 0.cm 0.cm}, clip,width=0.825\linewidth]{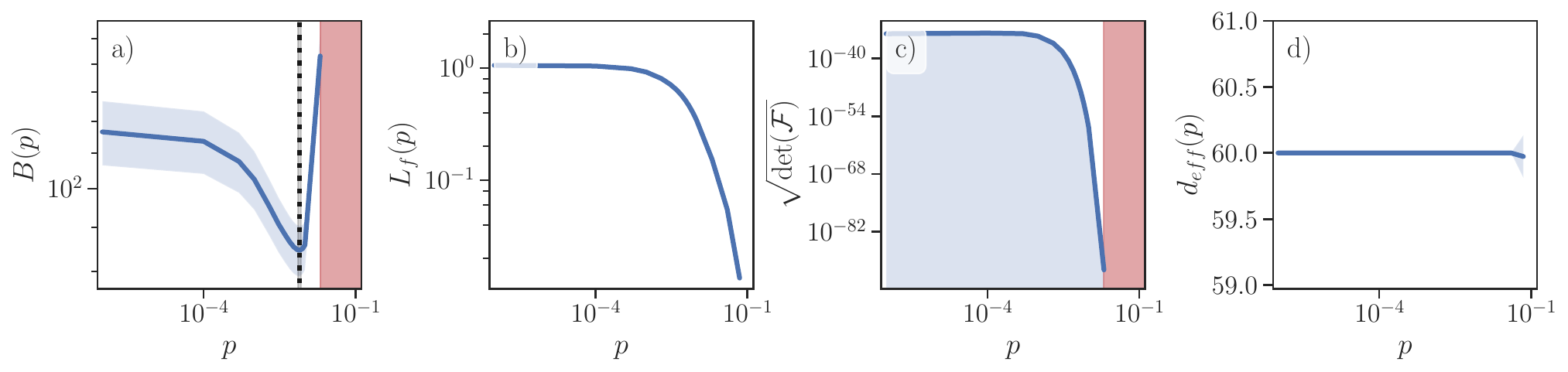}
    \includegraphics[trim={0cm 1.0cm 0.cm 0.cm}, clip,width=0.825\linewidth]{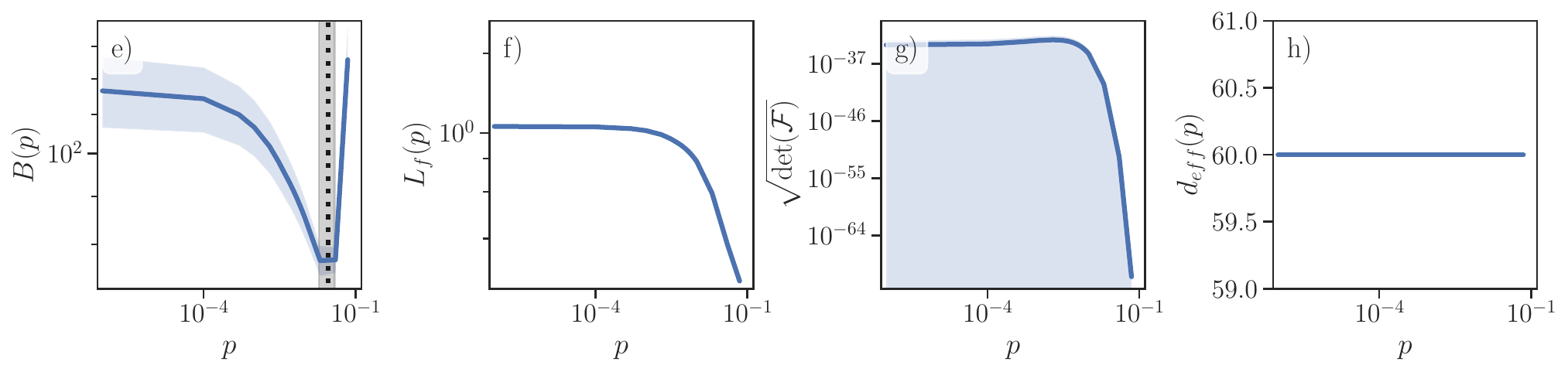}
    \includegraphics[width=0.825\linewidth]{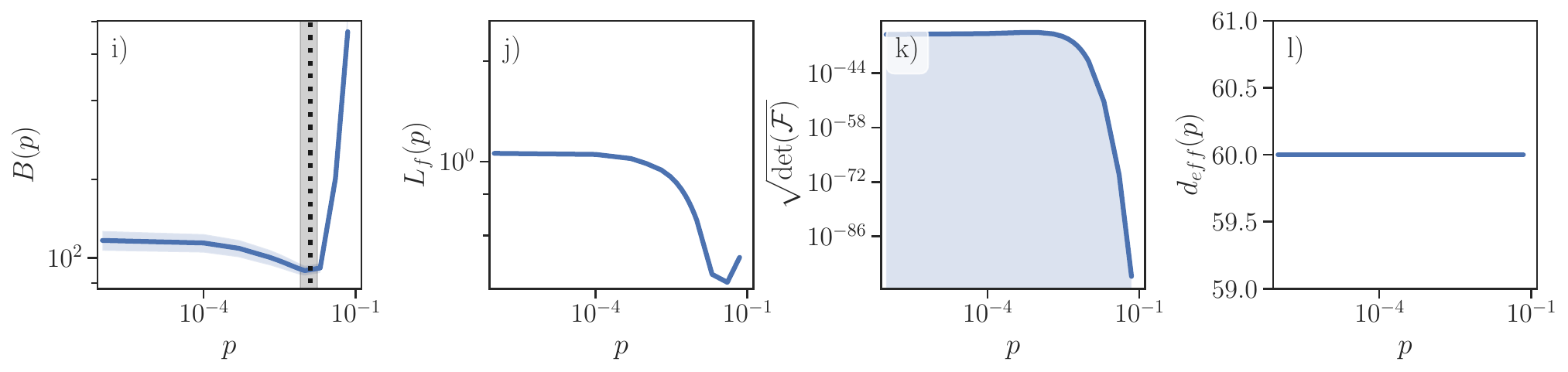}
    \caption{Generalization bound for the first QNN, in underparameterized regime with sinusoidal dataset, as a function of the noise level $p$ is reported in panel a). The other panels analogously report specific contributions to the generalization bound: b) Lipschitz constant $L_f$, c) $\sqrt{\det(\mathcal{F})}$, d) effective dimension $d_{eff}$. The vertical dotted line represents the level of noise minimizing the bound. The shaded areas are one standard deviation above and below the mean, calculated from 5 independent runs. Red areas highlight where the bound is not computable.
    }
    \label{fig:gen bound sin 4 layer}
\end{figure*}

\begin{figure*}
    \centering
    \includegraphics[trim={0cm 1.0cm 0.cm 0.cm}, clip,width=0.825\linewidth]{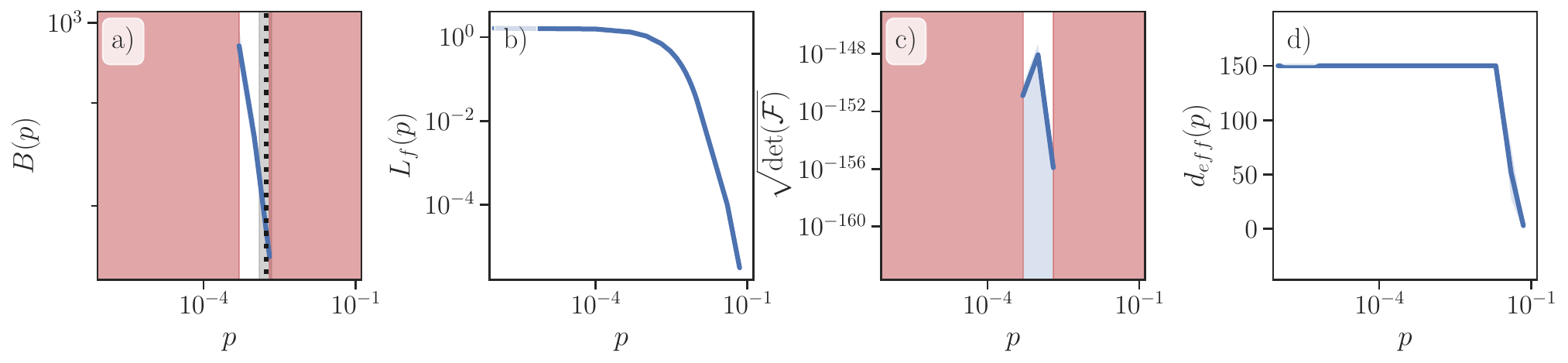}
    \includegraphics[trim={0cm 1.0cm 0.cm 0.cm}, clip,width=0.825\linewidth]{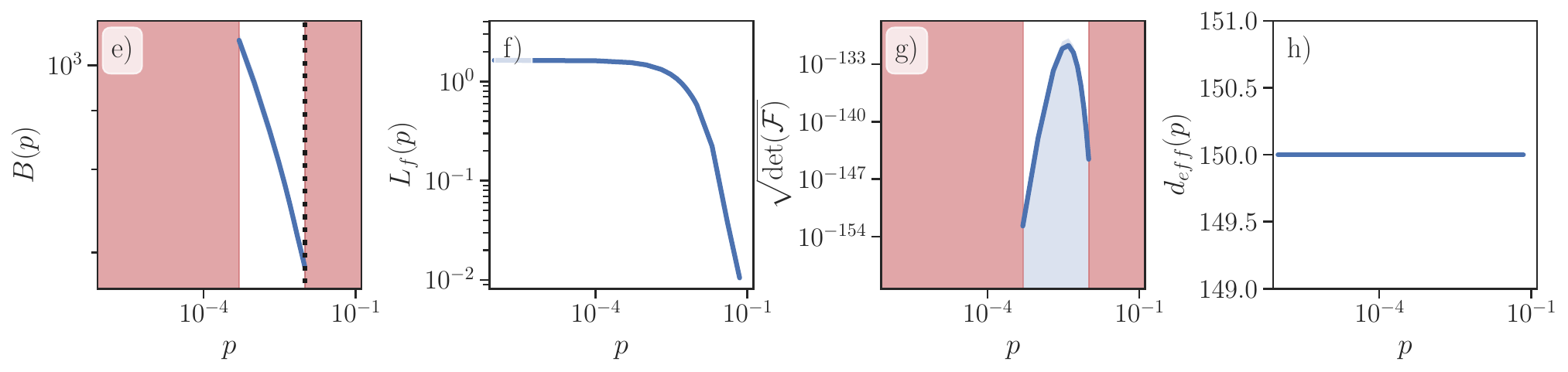}
    \includegraphics[width=0.825\linewidth]{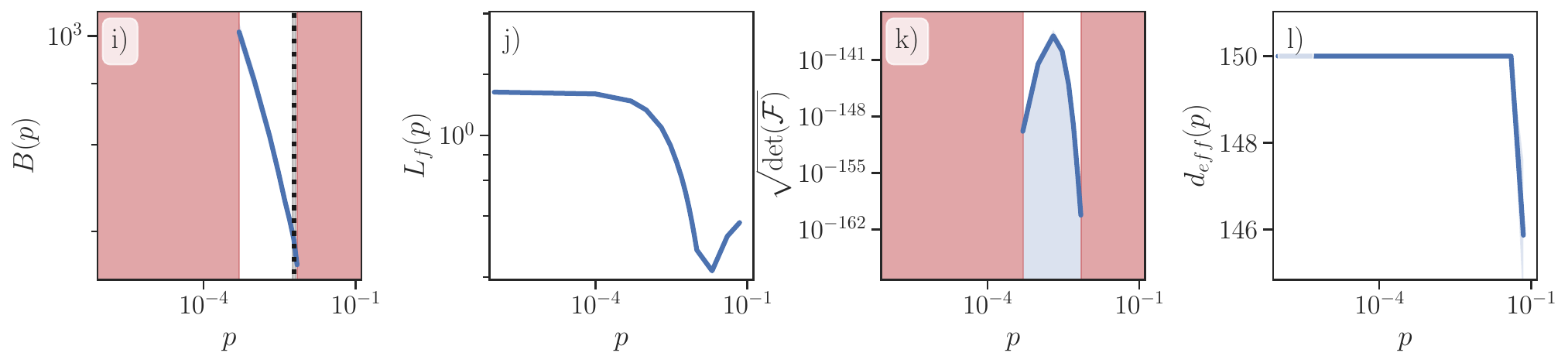}
    \caption{Generalization bound for the first QNN, in overparameterized regime with sinusoidal dataset, as a function of the noise level $p$ is reported in panel a). The other panels analogously report specific contributions to the generalization bound: b) Lipschitz constant $L_f$, c) $\sqrt{\det(\mathcal{F})}$, d) effective dimension $d_{eff}$. The vertical dotted line represents the level of noise minimizing the bound. The shaded areas are one standard deviation above and below the mean, calculated from 5 independent runs. Red areas highlight where the bound is not computable.}
    \label{fig:gen bound sin 10 layer}
\end{figure*}

\begin{figure*}
    \centering
    \includegraphics[width=0.825\linewidth]{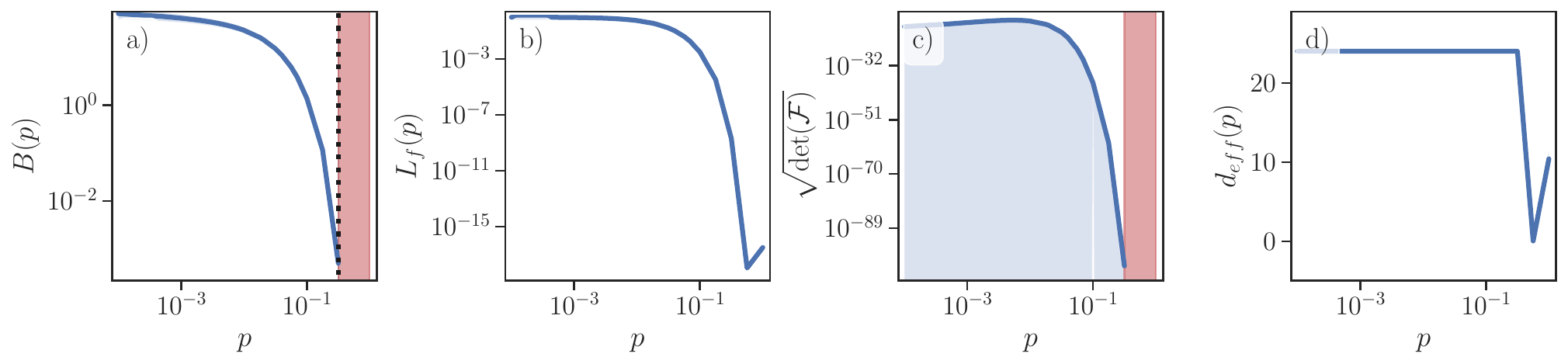}
    \includegraphics[width=0.825\linewidth]{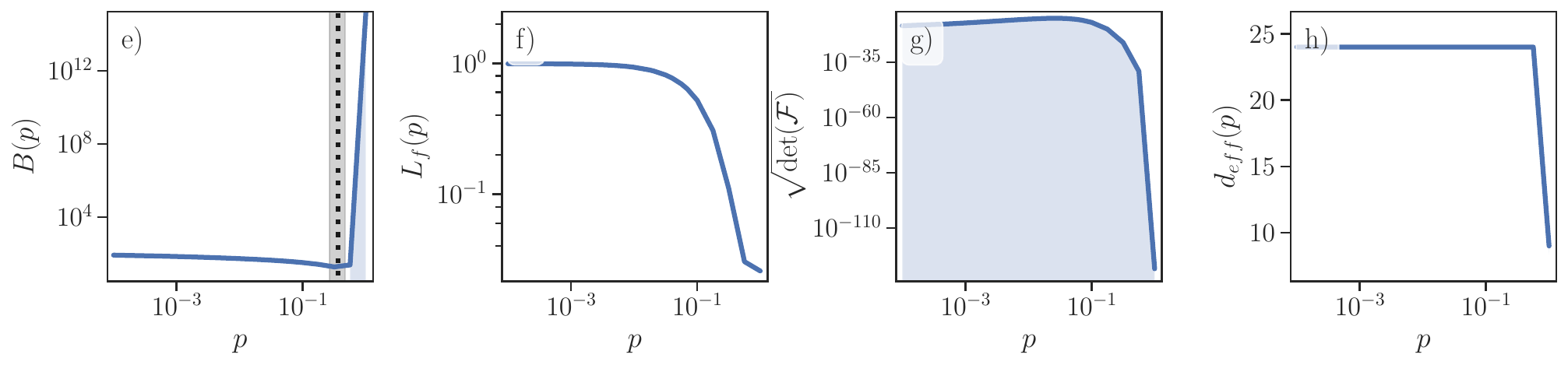}
    \includegraphics[width=0.825\linewidth]{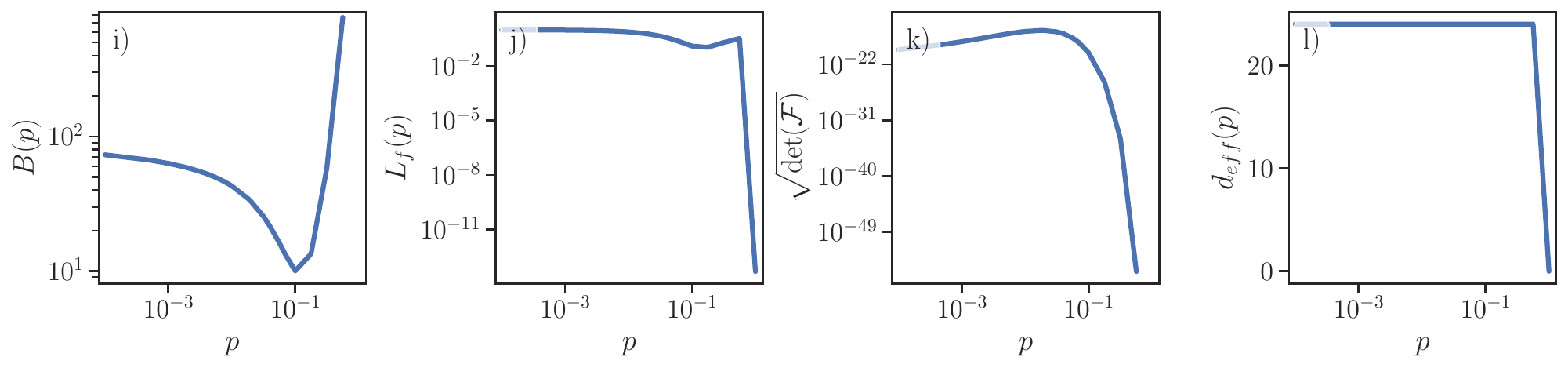}
    \caption{Generalization bound for the first QNN, in underparameterized regime with diabetes dataset, as a function of the noise level $p$ is reported in panel a). The other panels analogously report specific contributions to the generalization bound: b) Lipschitz constant $L_f$, c) $\sqrt{\det(\mathcal{F})}$, d) effective dimension $d_{eff}$. The vertical dotted line represents the level of noise minimizing the bound. The shaded areas are one standard deviation above and below the mean, calculated from 5 independent runs. Red areas highlight where the bound is not computable.}
    \label{fig:gen bound diabetes 3 layer}
\end{figure*}

\begin{figure*}
    \centering
    \includegraphics[width=0.825\linewidth]{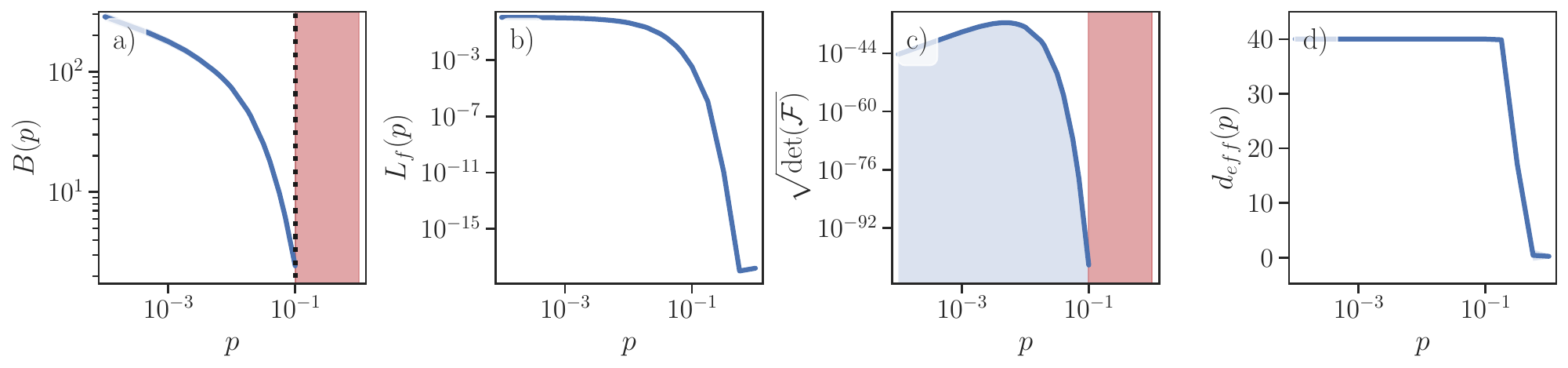}
    \includegraphics[width=0.825\linewidth]{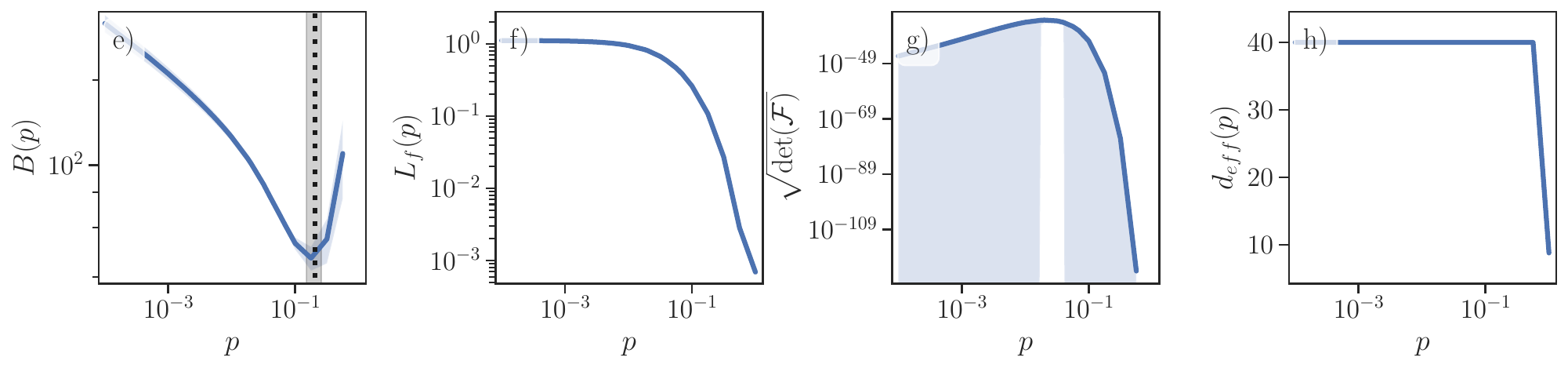}
    \includegraphics[width=0.825\linewidth]{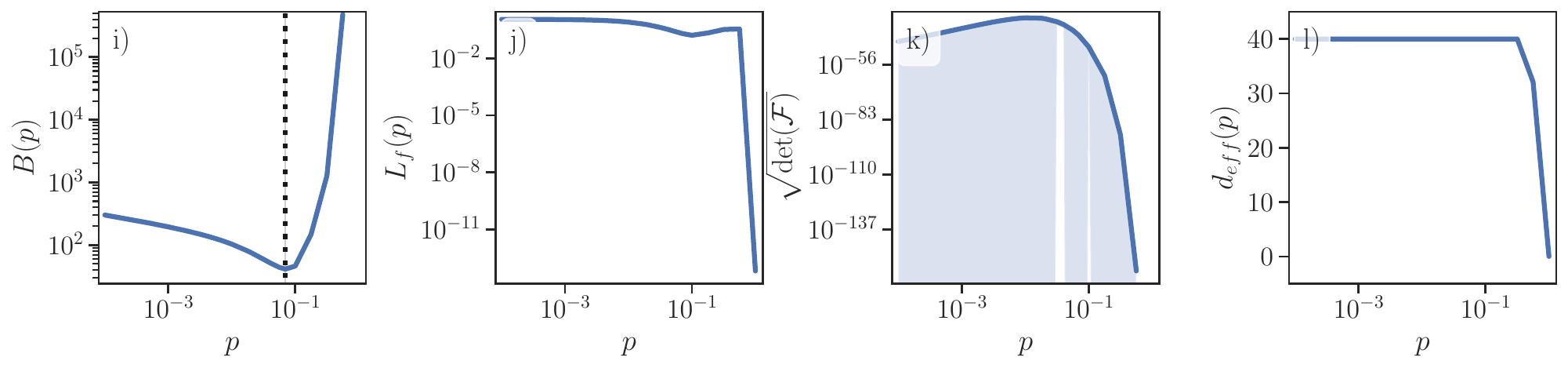}
    \caption{Generalization bound for the first QNN, in overparameterized regime with diabetes dataset, as a function of the noise level $p$ is reported in panel a). The other panels analogously report specific contributions to the generalization bound: b) Lipschitz constant $L_f$, c) $\sqrt{\det(\mathcal{F})}$, d) effective dimension $d_{eff}$. The vertical dotted line represents the level of noise minimizing the bound. The shaded areas are one standard deviation above and below the mean, calculated from 5 independent runs. Red areas highlight where the bound is not computable.}
    \label{fig:gen bound diabetes 5 layer}
\end{figure*}

\section{Dependence on input dataset}

Here we test the dependence of the method with respect to the input dataset. We estimate the optimal level of noise via synthetic datasets that are not related to the learning tasks described above. In particular, we create two datasets for each learning model with 15 samples drawn from a uniform distribution in the interval $[-\pi,\pi]$ and from a Gaussian distribution with zero mean and standard deviation equal to 1. For the first QNN architecture, the datasets are single-feature, to reflect the structure of the sinusoidal dataset, while for the second one, the datasets have 2 input features. The estimates of $p^*$ with the NIE procedure for different datasets and models are gathered in Tabs.~\ref{tab:2design independence under}-\ref{tab:paper model independence over}. It is possible to see that the $p^*$ is almost the same when varying the input dataset. This is most likely due to the fact that the average eigenspectrum is the same when changing the dataset. 

\begin{table}[]
\begin{tabular}{l|c|c|c|}
\cline{2-4}
 & DP  & PD & AD   \\ \hline
\multicolumn{1}{|l|}{Sinusoidal}           & $(2.1\pm 0.7)10^{-3}$ & $(9.1\pm 2.3)10^{-3}$ & $(5.4\pm 3.0)10^{-3}$ \\ \hline
\multicolumn{1}{|l|}{Sinusoidal2}          & $(2.1\pm 0.7)10^{-3}$ & $(8.9\pm 2.5)10^{-3}$ & $(5.5\pm 3.0)10^{-3}$ \\ \hline
\multicolumn{1}{|l|}{Diabetes}             & $(2.2\pm 0.6)10^{-3}$ & $(8.7\pm 2.4)10^{-3}$ & $(4.9\pm 2.9)10^{-3}$ \\ \hline
\multicolumn{1}{|l|}{Uniform $[-\pi,\pi]$} & $(2.1\pm 0.7)10^{-3}$ & $(8.8\pm 2.1)10^{-3}$ & $(5.0\pm 2.9)10^{-3}$ \\ \hline
\multicolumn{1}{|l|}{Gaussian $(0,1)$}     & $(2.1\pm 0.7)10^{-3}$ & $(8.7\pm 2.4)10^{-3}$ & $(5.2\pm 2.8)10^{-3}$ \\ \hline
\end{tabular}
\caption{Comparison of different values of $p^*$ for depolarizing (DP), phase damping (PD) and amplitude damping (AD) with respect to various input datasets for the first QNN in underparameterized regime. Values are reported as the mean together with standard deviation computed over 10 independent runs.}
    \label{tab:2design independence under}
\end{table}

\begin{table}
\centering
\begin{tabular}{l|c|c|c|}
\cline{2-4}
     & DP  & PD & AD   \\ \hline
\multicolumn{1}{|l|}{Sinusoidal}           & $(1.0\pm 0.0)10^{-3}$ & $(5.9\pm 0.4)10^{-3}$ & $(3.1\pm 0.4)10^{-3}$ \\ \hline
\multicolumn{1}{|l|}{Sinusoidal2}          & $(1.0\pm 0.0)10^{-3}$ & $(5.9\pm 0.3)10^{-3}$ & $(3.1\pm 0.4)10^{-3}$ \\ \hline
\multicolumn{1}{|l|}{Diabetes}             & $(1.0\pm 0.0)10^{-3}$ & $(5.6\pm 0.5)10^{-3}$ & $(3.0\pm 0.3)10^{-3}$ \\ \hline
\multicolumn{1}{|l|}{Uniform $[-\pi,\pi]$} & $(1.0\pm 0.0)10^{-3}$ & $(5.6\pm 0.5)10^{-3}$ & $(3.0\pm 0.3)10^{-3}$ \\ \hline
\multicolumn{1}{|l|}{Gaussian $(0,1)$}     & $(1.0\pm 0.0)10^{-3}$ & $(5.8\pm 0.4)10^{-3}$ & $(3.0\pm 0.3)10^{-3}$ \\ \hline
\end{tabular}
\caption{Comparison of different values of $p^*$ for depolarizing (DP), phase damping (PD) and amplitude damping (AD) with respect to various input datasets for the first QNN in overparameterized regime. Values are reported as the mean together with standard deviation computed over 10 independent runs.}
    \label{tab:2design independence over}
\end{table}

\begin{table}
\centering
\begin{tabular}{l|c|c|c|}
\cline{2-4}
 & DP  & PD & AD   \\ \hline
\multicolumn{1}{|l|}{Sinusoidal}           & $(1.41\pm 0.53)10^{-2}$ & $(6.97\pm 2.06)10^{-2}$ & $(4.60\pm 1.56)10^{-2}$ \\ \hline
\multicolumn{1}{|l|}{Sinusoidal2}          & $(1.42\pm 0.51)10^{-2}$ & $(7.06\pm 2.16)10^{-2}$ & $(4.56\pm 1.53)10^{-2}$ \\ \hline
\multicolumn{1}{|l|}{Diabetes}             & $(1.41\pm 0.51)10^{-2}$ & $(6.62\pm 2.20)10^{-2}$ & $(4.72\pm 2.68)10^{-2}$ \\ \hline
\multicolumn{1}{|l|}{Uniform $[-\pi,\pi]$} & $(1.46\pm 0.47)10^{-2}$ & $(6.15\pm 2.54)10^{-2}$ & $(4.64\pm 2.67)10^{-2}$ \\ \hline
\multicolumn{1}{|l|}{Gaussian $(0,1)$}     & $(1.46\pm 0.46)10^{-2}$ & $(6.59\pm 2.03)10^{-2}$ & $(4.68\pm 2.02)10^{-2}$ \\ \hline
\end{tabular}
\caption{Comparison of different values of $p^*$ for depolarizing (DP), phase damping (PD) and amplitude damping (AD) with respect to various input datasets for the second QNN in underparameterized regime. Values are reported as the mean together with standard deviation computed over 10 independent runs.}
    \label{tab:paper model independence under}
\end{table}

\begin{table}[]
\centering
\begin{tabular}{l|c|c|c|}
\cline{2-4}
     & DP  & PD & AD   \\ \hline
\multicolumn{1}{|l|}{Sinusoidal}           & $(8.8\pm 1.3)10^{-3}$ & $(3.75\pm 0.74)10^{-2}$ & $(2.56\pm 0.79)10^{-2}$ \\ \hline
\multicolumn{1}{|l|}{Sinusoidal2}          & $(8.8\pm 1.3)10^{-3}$ & $(3.79\pm 0.72)10^{-2}$ & $(2.59\pm 0.80)10^{-2}$ \\ \hline
\multicolumn{1}{|l|}{Diabetes}             & $(8.8\pm 1.3)10^{-3}$ & $(3.76\pm 0.74)10^{-2}$ & $(2.45\pm 0.84)10^{-2}$ \\ \hline
\multicolumn{1}{|l|}{Uniform $[-\pi,\pi]$} & $(8.8\pm 1.3)10^{-3}$ & $(3.71\pm 0.70)10^{-2}$ & $(2.47\pm 1.02)10^{-2}$ \\ \hline
\multicolumn{1}{|l|}{Gaussian $(0,1)$}     & $(8.7\pm 1.3)10^{-3}$ & $(3.73\pm 0.67)10^{-2}$ & $(2.51\pm 0.73)10^{-2}$ \\ \hline
\end{tabular}
\caption{Comparison of different values of $p^*$ for depolarizing (DP), phase damping (PD) and amplitude damping (AD) with respect to various input datasets for the second QNN in overparameterized regime. Values are reported as the mean together with standard deviation computed over 10 independent runs.}
    \label{tab:paper model independence over}
\end{table}

\end{document}